\documentclass[aps,twocolumn,pra,reprint,amsmath,amssymb,floatfix,footinbib,superscriptaddress]{revtex4}
\usepackage[utf8]{inputenc}

\usepackage{amsmath}	
\usepackage{gensymb}
\usepackage{hyperref}
\hypersetup{colorlinks=true}
\usepackage{upgreek}
\usepackage{graphics}
\usepackage{hyperref}
\usepackage{epsfig}
\usepackage{color}
\usepackage{bm}
\usepackage{float}
\usepackage{ulem} 
\usepackage{blindtext}
\usepackage{appendix} %

\usepackage{CJK}        
\usepackage{url}
\usepackage{multirow}

\newcommand{\upperRomannumeral}[1]{\uppercase\expandafter{\romannumeral#1}}

\usepackage{soul} 
\usepackage{color,xcolor}
\soulregister\cite7
\begin{document}

\title{Modeling Bond-Dependent Kitaev-like interaction in 2D Edge-Sharing Tetrahedral Magnets: FeX (X=Te, Se)}

\author{Mengdong Li}
\affiliation{College of Physics, Key Laboratory of Aerospace Information Materials and Physics (NUAA), MIIT, Nanjing University of Aeronautics and Astronautics, Nanjing 210016, China}
\author{Can Huang}
\email[]{canhuang@mail.usts.edu.cn}
\affiliation{Jiangsu Key Laboratory of Micro and Nano Heat Fluid Flow Technology and Energy Application, School of Mathematics and Physics, Suzhou University of Science and Technology, Suzhou 215009, China}
\affiliation{Advanced Technology Research Institute of Taihu Photon Center, School of Physical Science and Technology, Suzhou University of Science and Technology, Suzhou, 215009, China}
\author{Bingjie Liu}
\affiliation{College of Physics, Key Laboratory of Aerospace Information Materials and Physics (NUAA), MIIT, Nanjing University of Aeronautics and Astronautics, Nanjing 210016, China}
\author{Zhixin Liu}
\affiliation{College of Physics, Key Laboratory of Aerospace Information Materials and Physics (NUAA), MIIT, Nanjing University of Aeronautics and Astronautics, Nanjing 210016, China}
\author{Yanfei Pan}
\affiliation{College of Physics, Key Laboratory of Aerospace Information Materials and Physics (NUAA), MIIT, Nanjing University of Aeronautics and Astronautics, Nanjing 210016, China}
\author{Jiyu Fan}
\affiliation{College of Physics, Key Laboratory of Aerospace Information Materials and Physics (NUAA), MIIT, Nanjing University of Aeronautics and Astronautics, Nanjing 210016, China}
\author{Chunlan Ma}
\email[]{wlxmcl@mail.usts.edu.cn}
\affiliation{Jiangsu Key Laboratory of Micro and Nano Heat Fluid Flow Technology and Energy Application, School of Mathematics and Physics, Suzhou University of Science and Technology, Suzhou 215009, China}
\affiliation{Advanced Technology Research Institute of Taihu Photon Center, School of Physical Science and Technology, Suzhou University of Science and Technology, Suzhou, 215009, China}
\author{Daning Shi}
\affiliation{College of Physics, Key Laboratory of Aerospace Information Materials and Physics (NUAA), MIIT, Nanjing University of Aeronautics and Astronautics, Nanjing 210016, China}
\author{Yan Zhu}
\email[]{yzhu@nuaa.edu.cn}
\affiliation{College of Physics, Key Laboratory of Aerospace Information Materials and Physics (NUAA), MIIT, Nanjing University of Aeronautics and Astronautics, Nanjing 210016, China}

\date{\today}

\begin{abstract}
  Bond-dependent magnetic interactions, exemplified by the Kitaev model, are known to arise from the interplay between spin-orbit coupling (SOC) and specific coordination geometries, but have so far been almost exclusively identified in edge-sharing octahedral systems. Whether such interactions persist in edge-sharing tetrahedral environments—characteristic of the parent compounds of iron-based superconductors—remains an open question. Here, we construct a Kitaev-like model for monolayer FeTe and FeSe and demonstrate the presence of a previously unrecognized bond-dependent Ising-type interaction, induced jointly by chalcogen-mediated SOC and the tetrahedral crystal-field geometry. A microscopic spin model for these bond-dependent interactions is derived via strong-coupling perturbation theory, and the strengths of the individual exchange terms are extracted by partitioning the magnetic anisotropy energy calculated using density functional theory across various collinear magnetic orders. We reveal that the Kitaev-like interaction dominates the magnetic anisotropy in FeTe, whereas in FeSe, it strongly competes with a single-ion anisotropy of opposite sign. The resulting noncollinear local anisotropy axes generate intrinsic single-site spin frustration, providing a microscopic mechanism for magnetic disorder that transcends isotropic exchange models. Our results establish edge-sharing tetrahedral magnets as a new platform for bond-dependent interactions and extend the scope of Kitaev physics beyond octahedral coordination. 
\end{abstract}


\maketitle

\noindent \textit{Introduction.}---The exploration of novel quantum states, such as quantum spin liquids (QSLs) \cite{re1,re2,re3,re4,re5}, relies heavily on understanding anisotropic magnetic interactions beyond the Heisenberg model. Historically, the theoretical framework for such bond-dependent anisotropy was established by the quantum compass model \cite{re6,re7,re8,re9}. This concept has found its most profound realization in the Kitaev model on a honeycomb lattice \cite{re10,re11,re12,re13}, which has become increasingly important as a universal quantum interaction mechanism. The most representative example is the Kitaev interaction in two-dimensional (2D) honeycomb lattices constructed from edge-sharing octahedral coordination, \textit{e.g.}, $\alpha\text{-RuCl}_3$ \cite{re5,re14,re15,re16,re17}, which is closely related to frontier states such as QSLs and Majorana fermions \cite{re18,re19,re20}. However, in another significant class of systems possessing strong electron correlation and rich quantum phenomena—edge-sharing tetrahedral coordination structures like 2D iron-based superconductors—whether similar critical bond-dependent interactions exist and how to characterize them remains an underexplored topic. This is closely related to whether the magnetic disorder in iron-based superconductor parent compounds originates from, or partially involves, bond-dependent interactions, necessitating urgent research into such interactions.

Generalizing bond-dependent interactions from octahedral to tetrahedral coordination structures, and ultimately to other lattice geometries, has been hindered by the lack of a unified calculation scheme to quantitatively characterize their underlying rules. The ``Energy mapping combined with magnetic anisotropy energy (MAE)'' calculation scheme \cite{re21,re22,re23} developed by our group provides a basis for establishing bond-dependent interaction models; its principle is that differences in the MAE of various collinear magnetic configurations can reflect the existence and laws of bond-dependent interactions.

Bond-dependent interactions are inherently tied to spin-orbit coupling (SOC) \cite{re24,re25,re26,re27}, yet conventional theories often underestimate the influence of SOC on magnetism in iron-based superconductors \cite{re28,re29,re30,re31}. However, the iron-based superconductor parent compound FeTe incorporates the heavy ligand element Te, which induces substantial SOC \cite{re21,re32,re33,re34,re35}; similarly, the ligand element Se in FeSe also imparts non-negligible SOC \cite{re22,re36,re37,re38}. This suggests that novel bond-dependent interaction mechanisms driven by SOC, which have not yet been fully appreciated, may lie hidden within FeTe and FeSe parent compounds, with their strength and anisotropy inherently governed by the local crystal structure (especially bond directionality).

In this Letter, we demonstrate the presence of Kitaev-like bond-dependent interactions induced by chalcogen ligands ($\mathrm{Te/Se}$) in the edge-sharing tetrahedral environments of 2D iron-based superconductors FeX (X=Te, Se). Such distinctive interactions are intrinsically tied to specific bond directions, offering a framework to understand their unique magnetic and superconducting phenomena. We first construct a Kitaev-like model based on the characteristics of the edge-sharing tetrahedral geometry and the physics of the Ising model, and derive a minimal microscopic spin model—incorporating Heisenberg, Dzyaloshinskii-Moriya (DM), Kitaev-like, and off-diagonal exchange interactions—via strong-coupling perturbation theory. Through first-principles calculations combining energy mapping with MAE \cite{re21,re22,re23}, we evaluate the angle-dependent MAE for various collinear magnetic configurations and parameterize this bond-dependent anisotropic model. By isolating the contribution from SIA, we accurately extract the Kitaev-like interaction parameters. Despite their similar tetragonal layered structures, FeTe and FeSe exhibit fundamentally distinct microscopic origins of MAE: the Kitaev-like interaction dominates in FeTe, whereas in FeSe it becomes comparable to SIA. Furthermore, we demonstrate that these Kitaev-like interactions in iron-based superconductor parent compounds can induce single-site spin frustration at Fe sites. This finding highlights the universality of bond-dependent models in tetragonal systems, providing a fresh perspective toward addressing key questions in high-temperature superconductivity.
\begin{figure}[!ht]
\centering
\includegraphics[width=0.99\columnwidth, clip]{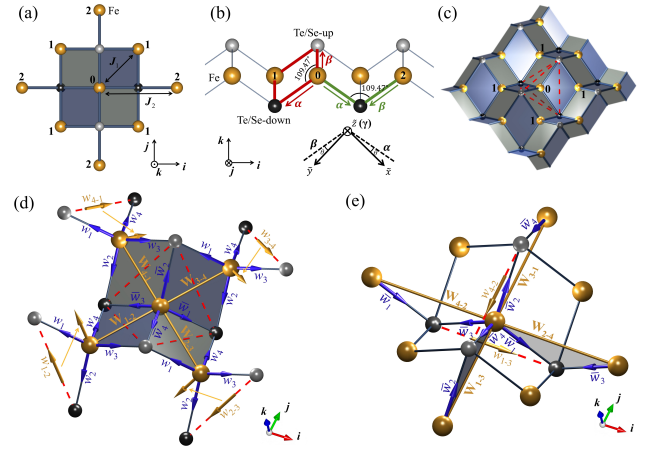}\\
\caption{Structure of monolayer FeX(X=Te, Se) and schematic of Kitaev-like interactions. Fe is represented by brown atoms, upper-layer X by gray atoms, and lower-layer X by black atoms. (a) Top view of monolayer FeX and schematic of first- and second-nearest-neighbor (1NN and 2NN) Fe atoms. The unit cell is a square formed by 1NN connections. (b) Side view of monolayer FeX, where $\alpha, \beta$, and $\gamma$ denote the Kitaev-like basis vectors for the corresponding planes, while $\tilde{x}, \tilde{y}$, and $\tilde{z}$ represent the local spin quantization axes. (c) Structural diagram of monolayer FeX. (d), (e) 1NN and 2NN Kitaev-like interactions, respectively. $w_p$ ($\bar{w}_p$) ($p=1, 2, 3, 4$) represent the Fe--X bond directions (where $w_p$ and $\bar{w}_p$ only distinguish different sites within the same FeX unit cell). Taking 1NN as an example, the Kitaev-like interaction basis directions in different planes are along the two in-plane bond directions $w_p, w_{p+1}$ ($\alpha$ or $\beta$) and the vector difference of the two out-of-plane bond directions $w_{(p+2)-(p+3)}=w_{p+2} - w_{p+3}$ ($\gamma$, the Ising axis direction, indicated by yellow arrows); the Fe--Fe bond direction $W_{(p+2)-(p+3)}$ is perpendicular to the Ising axis direction. The subscript $p+q$ ($q=1, 2, 3$) denotes a shift in the index, with cyclic symmetry (looping back to 1 when exceeding 4). 
}\label{FIG-1}
\end{figure}

The core issue of this work is how to construct and describe bond-dependent interactions in monolayer FeX (X=Te, Se) in the 2D edge-sharing tetrahedral coordination structure. As shown in Fig.~\ref{FIG-1}(a) and~\ref{FIG-1}(b), the basic structural unit of monolayer FeX consists of one Fe atomic layer sandwiched between two X atomic layers. As illustrated in Fig.~\ref{FIG-1}(c), each Fe atom is located at the center of a tetrahedron formed by four surrounding X atoms, forming Fe--X--Fe--X planes with the X atoms. Adjacent planes without shared edges [faces of the same color in Fig.~\ref{FIG-1}(c)] are mutually perpendicular, while those with shared edges form a specific angle. These mutually perpendicular planes give rise to single-site spin frustration induced by bond-dependent interactions. However, the lack of a natural orthogonal basis along the bond directions within the structure complicates the construction of a bond-dependent model. Nevertheless, we find that for any two of the four Fe--X bonds formed by an Fe atom and its four surrounding X atoms, their vector difference is perpendicular to the other two bonds as well as to the Fe--X--Fe(--X) plane between 1NN (2NN) sites [Figs.~\ref{FIG-1}(d) and~\ref{FIG-1}(e)], providing a crucial prerequisite for constructing bond-dependent interactions.

In the following, combining second-order strong-coupling perturbation theory, we formulate the Kitaev-like bond-dependent interaction model and derive its microscopic spin Hamiltonian in an ideal 2D edge-sharing tetrahedral structure \cite{re25,re26,re27,re39,re40,nre1,nre2}. Unlike the conventional Kitaev model where the Kitaev local spin basis $\alpha, \beta$, and $\gamma$ on a given plane are mutually orthogonal, we denote the two non-orthogonal Fe--X bonds in the Fe--X--Fe(--X) plane corresponding to a given Fe--Fe bond as $\alpha$ and $\beta$ [Fig.~\ref{FIG-1}(b)], while $\gamma$ corresponds to  direction perpendicular to this plane, which can be described by the vector difference between two out-of-plane Fe--X bonds; specific setups for different bonds are detailed in the captions of Fig.~\ref{FIG-1} and Sec.~\textcolor{red}{S1} of the Supplemental Material (SM) \cite{re41}. Consequently, we can model the bond-dependent Kitaev-like interactions by leveraging the inherent bond orientations of the 2D edge-sharing tetrahedral geometry. 

However, as shown in Figs.~\ref{FIG-1}(d) and~\ref{FIG-1}(e), the normal vectors of the planes between 1NN and 2NN differ, making it impossible to describe the superexchange processes between different Fe--Fe atom pairs within a single global coordinate system and its permutations. Furthermore, with the non-normality, the Kitaev-like local basis $\alpha, \beta$, and $\gamma$ cannot be directly treated as local quantization axes. Therefore, we establish local coordinate systems $(\tilde{x}, \tilde{y}, \tilde{z})$ for different Fe--X--Fe(--X) planes to unify the superexchange processes across all Fe--Fe bonds [Fig.~\ref{FIG-1}(b)] (see SM Sec.~\textcolor{red}{S2} for details \cite{re41}). In the microscopic derivation, we consider the Kanamori interaction \cite{re40} for Fe in the hole representation alongside crystal-field splitting to construct the tight-binding Hamiltonian. Under strong coupling, treating the tight-binding hopping integrals as perturbations, we analyze the indirect superexchange process between Fe atoms mediated by ligand X-atom SOC. Finally, mapping the spin operators from local coordinates back to the global Cartesian coordinate system yields the minimal spin Hamiltonian originating from 1NN and 2NN superexchange in the 2D edge-sharing system:
\begin{equation}
\begin{aligned}
H_s =& \sum_{\langle i,j \rangle_{1,2}} \frac{1}{2} \left[ J_{ij} (\boldsymbol{S}_i \cdot \boldsymbol{S}_j)\right] \\
&+ \sum_{\langle i,j \rangle_{1,2}} \frac{1}{2} \left[K_{ij} S_i^\gamma S_j^\gamma + \Gamma_{ij} (S_i^\alpha S_j^\beta + S_i^\beta S_j^\alpha)\right]  \\
&+\sum_{\langle i,j \rangle_2} \frac{1}{2} D_{ij} \left[ (\mathbf{z} \times \mathbf{d}_{ij}) \cdot (\boldsymbol{S}_i \times \boldsymbol{S}_j) \right] \\
&+ \sum_i A_k \left[ 1 - (\boldsymbol{S}_i \cdot \mathbf{k})^2 \right],
\end{aligned}
\label{Eq-1}
\end{equation}
where the terms represent the Heisenberg interaction $J_{ij}$, the Kitaev-like interaction $K_{ij}$, the off-diagonal exchange $\Gamma_{ij}$, the DM interaction $D_{ij}$. In addition, a single-ion anisotropy (SIA) term $A_k$ is incorporated~\cite{re27,re37}. Here, $\langle i,j \rangle$ represent Fe atom neighbor pairs, where subscripts 1 and 2 denote the 1NNs and 2NNs, respectively, and $\boldsymbol{S}$ is the normalized spin operator. The local Kitaev-like basis $\{\alpha, \beta, \gamma\}$ are the spin component directions of the planar basis in Figs.~\ref{FIG-1}(b),~\ref{FIG-1}(d), and~\ref{FIG-1}(e). The explicit form of the Hamiltonian formulated using the bond directions is given in Eq.~(\textcolor{red}{S61}) in the Supplemental Material~\cite{re41}. Here, $\alpha$ and $\beta$ form an angle of $109.47^\circ$ and are not perpendicular. $\mathbf{i}, \mathbf{j}, \mathbf{k}$ are the basis vectors of the Cartesian coordinate system. Specifically, the Kitaev-like and off-diagonal interactions originate from spin-flip processes involving ligand SOC during virtual hoppings along both $\mathrm{Fe}_i \rightarrow \mathrm{Fe}_j$ and $\mathrm{Fe}_j \rightarrow \mathrm{Fe}_i$ paths~\cite{re27}; the DM interaction arises from superexchange processes where SOC acts during only a single virtual hopping step; and the Heisenberg interaction receives contributions from both spin-flip processes and spin-conserving processes without SOC.
 \begin{figure}[!ht]
\centering
\includegraphics[width=0.99\columnwidth, clip]{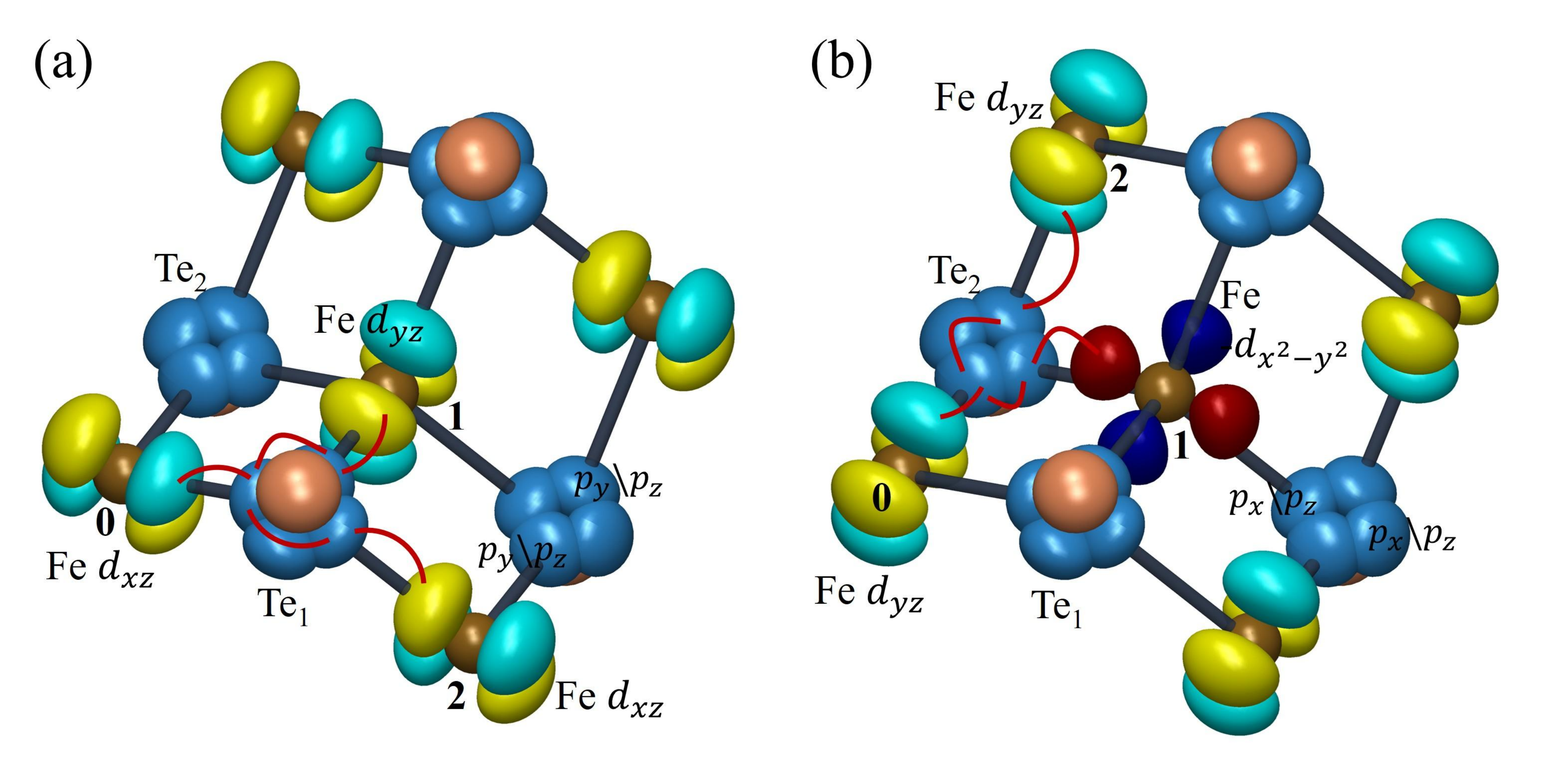}\\
\caption{Superexchange spin-flip paths in the global coordinate system. Paths connecting different atoms correspond to hopping integrals, while paths on the same Te atom represent $p$-orbital transformations under SOC. In the schematic of Te atomic orbitals, the blue regions denote hybrid orbitals simultaneously incorporating $p_x/p_z$ or $p_y/p_z$ interactions. Spin-flip superexchange processes between (a) $d_{xz}$ and $d_{yz}$ orbitals for 1NN, $d_{xz}$ and $d_{xz}$ orbitals for 2NN; (b) $d_{yz}$ and $d_{x^2-y^2}$ orbitals for 1NN, $d_{yz}$ and $d_{yz}$ orbitals for 2NN.
  }\label{FIG-2}
\end{figure}

In the edge-sharing tetrahedral coordination structure, the Fe atom is at the center of the crystal field of the tetrahedron formed by Te atoms, and the five degenerate $d$ orbitals split into higher-energy triply degenerate $t_2$ and lower-energy doubly degenerate $e$ orbitals. Analysis of the superexchange processes reveals that, in the global coordinate system, the $d_{xz}$ ($d_{yz}$) orbitals of the central $\mathrm{Fe}_0$ atom and the $d_{yz}$ ($d_{x^2-y^2}$) orbitals of $\mathrm{Fe}_1$ mediate the 1NN Kitaev-like interaction via the $p$ orbitals of Te. Meanwhile, the $d_{xz}$ ($d_{yz}$) orbitals of $\mathrm{Fe}_0$ and the $d_{xz}$ ($d_{yz}$) orbitals of $\mathrm{Fe}_2$ mediate the 2NN Kitaev-like interaction via the $p$ orbitals of Te [Figs.~\ref{FIG-2}(a) and ~\ref{FIG-2}(b)]. Here, only representative spin-flip superexchange paths involving $\sigma$ channels are presented; additional processes are detailed in Sec.~\textcolor{red}{S5} of the SM~\cite{re41}. 

Naturally, how to intuitively and concisely extract and parameterize the edge-sharing tetrahedral bond-dependent Kitaev-like interactions in DFT calculations and decompose the interaction parameters in Eq.~\ref{Eq-1} becomes the focus of our subsequent discussion.
To further analyze and validate our model via DFT calculations~\cite{re42,re43,re44,re45,re46}, we obtained the electronic structures of monolayer FeX  (see Sec.~\textcolor{red}{S.3} of the SM~\cite{re41}). The band structure shows that in the presence of SOC, band gaps open at crossing points that exist without SOC. This indicates that the Te/Se ligands induce strong SOC~\cite{re21,re32}, which serves as the entry point for Kitaev-like interactions.

 \begin{figure}[!ht]
\centering
\includegraphics[width=0.99\columnwidth, clip]{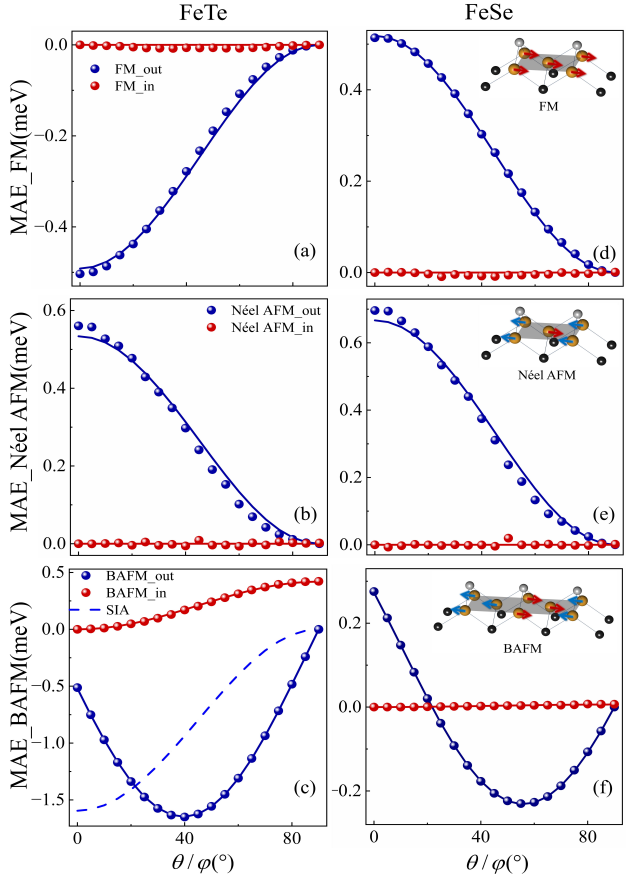}\\
\caption{Calculation and formula fitting results of $\mathrm{MAE}(\theta,\varphi)$ for monolayer FeTe and FeSe under various magnetic configurations. (a)--(c) show $\mathrm{MAE}(\theta,\varphi)$ for FeTe under FM, Néel AFM, and BAFM magnetic orders, respectively. (d)--(f) show the corresponding results for FeSe. Blue and red discrete points correspond to DFT calculation results (taken every $5^\circ$) for out-of-plane and in-plane $\mathrm{MAE}(\theta,\varphi)$, respectively; blue and red curves are the fitted curves obtained after substituting the coefficients.
  }\label{FIG-3}
\end{figure}

Based on previous work, Kitaev interactions can be directly reflected in the differences of MAE among various collinear magnetic orders~\cite{re21,re22,re23}. Because the magnetic moments are parallel or antiparallel in these collinear configurations, isotropic Heisenberg and DM interactions do not contribute to the collinear MAE. Therefore, to verify the existence of Kitaev-like interactions in monolayer FeX, we need to calculate diverse and reasonable magnetic order structures. Previous DFT studies on FeTe have primarily focused on typical collinear configurations such as FM, Néel AFM, and BAFM~\cite{re47,re48,re49}, as shown in Fig.~\ref{FIG-3}. Among them, BAFM order has been widely determined by calculations to be the magnetic ground state of the FeTe system~\cite{re47}. For the FeSe system, the magnetic ground state remains a subject of ongoing debate and theoretical challenge~\cite{re47,re50,re51}.

For specific FeX systems, we can obtain the parameters for Kitaev-like and SIA interactions by fitting the theoretically derived formulas to DFT calculation results using the least-squares method. Substituting these parameters into the formulas to generate MAE curves, self-consistency with calculation results would verify the correctness of the model~\cite{re21,re22,re23}.

Based on this, we performed first-principles DFT calculations to evaluate the angular dependence of both out-of-plane and in-plane collinear MAE$(\theta, \varphi)$ for monolayer FeX in FM, Néel AFM, and BAFM orders (discrete points in Fig.~\ref{FIG-3}). If the anisotropy were governed solely by SIA, the profiles would be strictly identical across all magnetic structures and confined to the out-of-plane component.

Therefore, simple SIA is far from sufficient to explain the complexity of the DFT results in Fig.~\ref{FIG-3}. Taking FeTe as an example, the calculated $\mathrm{MAE}(\theta, \varphi)$ reveals a strong dependence on the magnetic order: the out-of-plane trends for FM and Néel AFM are diametrically opposite [Figs.~\ref{FIG-3}(a) vs ~\ref{FIG-3}(b)]. Furthermore, the BAFM order [Fig.~\ref{FIG-3}(c)] exhibits a non-zero in-plane anisotropy that is strictly forbidden in a pure SIA model. These contradictions provide clear evidence that bond-dependent anisotropic interactions must exist beyond SIA in monolayer FeX.

Notably, a theoretical model restricted to SIA and 1NN interactions predicts a vanishing in-plane MAE for the BAFM order. This contrasts the in-plane calculation results for FeTe, which show a clear $\sin^2\varphi$ oscillation. This indicates that introducing 1NN Kitaev-like interactions is insufficient to capture the full magnetic anisotropy. Based on the results from perturbation theory, the superexchange interaction between 2NN Fe ions holds equal status to that of the 1NN. Therefore, we naturally introduced 2NN Kitaev-like interactions, which resolved this issue. 

\begin{figure}[!ht]
\centering
\includegraphics[width=0.99\columnwidth, clip]{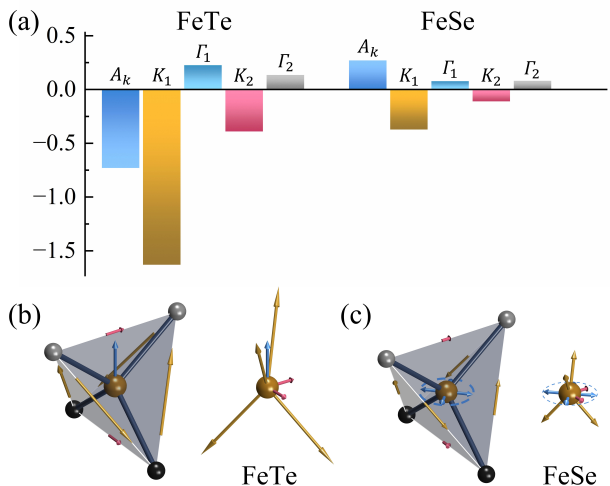}\\
\caption{(a) Fitted magnetic interaction parameters derived from the MAE mapping for monolayer FeTe (left) and FeSe (right). (b), (c) Schematic visualization of competing anisotropy terms acting on the central Fe atom for (b) FeTe and (c) FeSe. The vectors represent the local easy axes induced by different mechanisms: SIA (blue), 1NN Kitaev interaction (yellow), and 2NN Kitaev interaction (pink). Arrow lengths indicate the relative strength of the influence on spin orientation; spins tend to align with the dominantly longer arrows.
  }\label{FIG-4}
\end{figure}

By fitting the calculation results of collinear MAE under different magnetic configurations to the theoretical formulas using the least-squares method, as shown in Fig.~\ref{FIG-4}(a), we extract the full set of magnetic interaction parameters. As shown in Fig.~\ref{FIG-3}, the fitted curves utilizing these parameters perfectly match the calculated values, demonstrating the correctness of our proposed Kitaev-like model.

For FeTe, the magnetic anisotropy is dominated by the 1NN Kitaev-like interaction, $K_1 = -1.63~\text{meV/Fe}$. This term is significantly larger than both the SIA ($A_k = -0.73~\text{meV/Fe}$) and the off-diagonal exchange terms. In contrast, for monolayer FeSe, the lighter Se ligand induces a weaker SOC, resulting in generally reduced anisotropy scales. Nevertheless, $K_1$ retains the leading role with a value of $-0.37~\text{meV/Fe}$. Crucially, however, this term faces strong competition from the $A_k$ term, which possesses a comparable magnitude of $0.27~\text{meV/Fe}$ but acts in the opposite sign. 
Furthermore, the coefficients of the Kitaev-like terms are generally larger than those of the off-diagonal interaction terms, indicating that the Kitaev-like interactions in monolayer FeTe and FeSe are primarily Ising-type. 

Additionally, the parameters for FeTe are generally larger than those for FeSe [Fig.~\ref{FIG-4}(a)], consistent with the SOC effects of Te and Se, confirming that SOC plays a key role in bond-dependent interactions. Furthermore, we examined the additional off-diagonal interaction $\Gamma'$~\cite{re11,re23,re27}, which are subleading relative to the dominate Kitaev interaction. Their inclusion changes the fitted $K$, $\Gamma$ and $A_k$ parameters by less than ($1\%$) and does not alter our main conclusions (see Sec.~\textcolor{red}{S3.3.4} of the SM~\cite{re41}.

To understand the physical implications of these extracted parameters, we visualize their spatial arrangement with respect to the lattice [Figs.~\ref{FIG-4}(b) and~\ref{FIG-4}(c)]. As discussed in Fig.~\ref{FIG-1}, the directions of the Kitaev-like interaction for 1NN and 2NN are perpendicular to the Fe-X-Fe-X planes and match the directions of the tetrahedron edges formed by the four X atoms surrounding the central Fe atom [indicated by red dashed lines in Figs.~\ref{FIG-1}(d) and~\ref{FIG-1}(e)]. Therefore, these directions are mutually perpendicular or form angles in space [Figs.~\ref{FIG-4}(b) and ~\ref{FIG-4}(c)], with each favoring the central spin alignment along its own axis. This results in the spin being unable to simultaneously satisfy the energy minimization of all interactions, giving rise to ground-state degeneracy and spin frustration at single lattice sites. Such spin frustration induced by structure-compatible Kitaev-like bond-dependent interactions provides a plausible microscopic mechanism for understanding the magnetic disorder observed in parent compounds of iron-based superconductors, offering a new perspective beyond conventional exchange models.

Notably, the 2NN DM interaction parameters in FeTe and FeSe reach $D_2 = -7.56$ and $-4.98~\text{meV/Fe}$, respectively. When considering DM interactions alone, the spins on each Fe sublattice tend to form short-period noncollinear helical orders propagating along the $(1\bar{1}0)$ and equivalent directions, indicating potential chiral magnetic competition (see SM Sec.~\textcolor{red}{S4} of the SM~\cite{re41}). This adds to the magnetic complexity and provides a symmetry foundation for multi-$q$ chiral magnetic states driven by external magnetic fields, magnetic anisotropy, or higher-order exchange interactions~\cite{re52,re53,re54,re55}, making these materials promising candidate platforms for exploring topological spin textures under external fields, strain, or interface engineering.

In summary, we have established a universal bond-dependent magnetic interaction model and derived its Hamiltonian for non-honeycomb, 2D edge-sharing tetrahedral iron-based superconductors, exemplified by monolayer FeTe and FeSe. Our work points out that magnetic interactions related to bond directionality universally exist in lattices with SOC. This not only lays the foundation for extending Kitaev physics to 2D tetrahedral and even more general lattice structures, but also opens a promising research platform for iron-based superconducting materials: their unique crystal structure and electronic properties provide rich possibilities for exploring novel interaction mechanisms deeply linked to superconducting pairing, quantum magnetism, and topological states, offering a fresh perspective toward understanding the mechanism of high-temperature superconductivity.

\noindent \textit{Acknowledgments.}---This work was supported by the National Natural Science Foundation of China (NSFC) under Grant Nos.~11204131, 11974181, and 12474155, and the Natural Science Foundation of the Jiangsu Higher Education Institutions of China (Grant No.~25KJB140017). We thank Associate Professors Qiang Luo from NUAA and Hongjun Xiang from Fudan University for their assistance and helpful discussions.


%

\clearpage

\onecolumngrid

\newpage

\newcounter{sectionSM}
\newcounter{equationSM}
\newcounter{figureSM}
\newcounter{tableSM}

\setcounter{section}{0}
\setcounter{subsection}{0}
\setcounter{subsubsection}{0}
\setcounter{equation}{0}
\setcounter{figure}{0}
\setcounter{table}{0}
\setcounter{page}{1}

\makeatletter

\@removefromreset{equation}{section}

\renewcommand{\thesection}{\textsc{S}\arabic{section}}
\renewcommand{\thesubsection}{\textsc{S}\arabic{section}.\arabic{subsection}}
\renewcommand{\thesubsubsection}{\textsc{S}\arabic{section}.\arabic{subsection}.\arabic{subsubsection}}

\renewcommand{\p@subsection}{}
\renewcommand{\p@subsubsection}{}

\renewcommand{\theequation}{\textsc{S}\arabic{equation}}
\renewcommand{\thefigure}{\textsc{S}\arabic{figure}}
\renewcommand{\thetable}{\textsc{S}\arabic{table}}

\makeatother


\begin{center}
{\large{\bf Supplemental Material for\\
``Modeling Bond-Dependent Kitaev-like interaction in 2D Edge-Sharing Tetrahedral Magnets: FeX (X=Te, Se)''}}
\end{center}
\begin{center}
Mengdong Li$^{1}$, Can Huang$^{2,\;3}$, Bingjie Liu$^{1}$, Zhixin Liu$^{1}$, Yanfei Pan$^{1}$, Jiyu Fan$^{1}$, Chunlan Ma$^{2,\;3}$, Daning Shi$^{1}$, Yan Zhu$^{1}$ \\
\quad\\
$^1$\textit{College of Physics, Key Laboratory of Aerospace Information Materials and Physics (NUAA), MIIT, Nanjing University of Aeronautics and Astronautics, Nanjing 210016, China}\\
$^2$\textit{Jiangsu Key Laboratory of Micro and Nano Heat Fluid Flow Technology and Energy Application, School of Mathematics and Physics, Suzhou University of Science and Technology, Suzhou 215009, China} \\
$^3$\textit{Advanced Technology Research Institute of Taihu Photon Center, School of Physical Science and Technology, Suzhou University of Science and Technology, Suzhou, 215009, China}\\
(Dated: August 21, 2026)
\quad\\
\end{center}

\vspace{-0.00cm}
\section{Basis-vector selection rules for the Kitaev(-like) model}\label{Sec-S1}
\subsection{Kitaev interactions in octahedral coordination}\label{Subsec-S1.1}
\begin{figure}[!ht]
\centering
\includegraphics[width=0.35\columnwidth, clip]{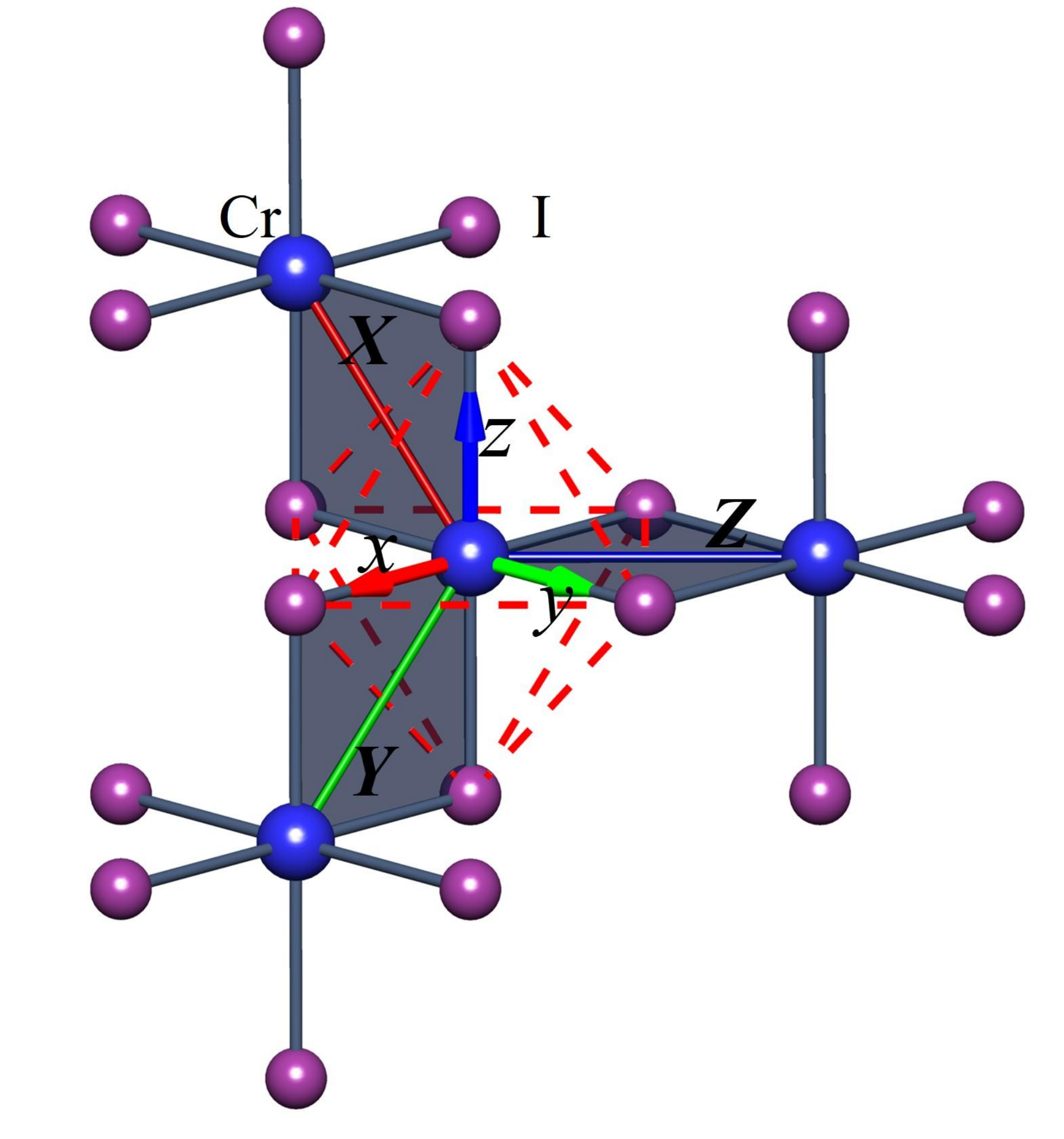}\\
\caption{ Monolayer 1T-CrI$_3$ structure and schematic illustration of the Kitaev model. Cr is represented by blue atoms, and I by purple atoms. $X$, $Y$, $Z$ and $x$, $y$, $z$ correspond to the red, green, and blue Cr-Cr bond directions and Kitaev basis directions, respectively, where $X\;(Y, Z)$ and $x\;(y, z)$ are mutually perpendicular.
  }\label{FIG-s1}
\end{figure}
Taking the hexagonal lattice 1T-CrI$_3$ as an exemplary system, the Kitaev spin model is intimately correlated with its distinctive crystal structure and spin-orbit coupling effects~[\textcolor{green}{24-27}]. As illustrated in Fig.~\ref{FIG-s1}, 1T-CrI$_3$ exhibits a characteristic sandwich-type layered architecture, wherein the Cr atomic layer is sandwiched between two I atomic layers, forming a Cr-I-Cr heterogeneous interlayer stacking configuration. From a crystallographic perspective, each Cr atom occupies the center of a near-perfect octahedral coordination environment constituted by six nearest-neighbor I atoms. Particularly noteworthy is that adjacent Cr atoms achieve magnetic coupling through nearly square Cr-I-Cr-I coplanar motifs, and the three Cr-I-Cr-I planes within the octahedral coordination field are mutually orthogonal. This unique geometric configuration breaks the spatial isotropy of spin exchange interactions, thereby inducing spin frustration at individual lattice sites—that is, the orientational preferences of neighboring spins cannot be simultaneously satisfied, leading to high ground-state degeneracy and strongly correlated quantum fluctuations in the system.

The Kitaev interaction in 1T-CrI$_3$ fundamentally originates from the strong spin-orbit coupling(SOC) effect induced by the heavy coordinating element I. Specifically, the heavy halogen element I, possessing a large atomic number, exhibits pronounced SOC effects that render the magnetic interactions between Cr-Cr pairs anisotropic, generating Ising-type exchange interactions perpendicular to the Cr-I-Cr-I planes (i.e., the Kitaev interaction term). Meanwhile, within the Cr-I-Cr-I planes, the spin magnetic moments of Cr ions produce in-plane off-diagonal exchange interaction terms (i.e., the $\Gamma$ term) mediated by the $p$ orbitals of I atoms along the Cr-I bond directions.
The geometric representation of the Kitaev interaction is depicted in Fig.~\ref{FIG-s1}, where the uppercase letters $X$, $Y$, and $Z$ correspond to the three types of Cr-Cr bond directions marked in red, green, and blue, respectively (representing the bond orientations of Kitaev interactions), while the lowercase letters $x$, $y$, and $z$ denote the local coordinate basis vectors of the Kitaev spins. These satisfy the orthogonality constraint that the $X\;(Y,\;Z)$ axes are perpendicular to the $x\;(y,\;z) $axes, respectively. Based on these definitions, the Hamiltonian of the Kitaev model in the 1T-CrI$_3$ system can be formulated as:
\begin{equation}
    H_{kit}=\sum\limits_{\langle i,j\rangle}{\frac{1}{2}}[K_{ij} S^{\gamma}_{i} S^{\gamma}_{j}+\Gamma_{ij}(S^{\alpha}_{i} S^{\beta}_{j}+S^{\beta}_{i} S^{\alpha}_{j})],
    \label{Eq-S1}
\end{equation}
where $\langle i,\;j\rangle$ denotes the nearest-neighbor Cr atom pairs, and $\alpha$, $\beta$, $\gamma$ correspond to the spin component directions along the $X$, $Y$, $Z$ bonds, respectively. For the three mutually orthogonal Cr-I-Cr-I planes, cyclic permutation symmetry exists among $\alpha$, $\beta$, and $\gamma$.

\subsection{Kitaev-like interactions in tetrahedral coordination}\label{Subsec-S1.2}
In monolayer FeX (X=Te, Se) systems, the Kitaev-like interactions are similarly dependent on the unique tetrahedral coordination structure. As illustrated in Figs.~(\textcolor{red}{1}))(a)-(c), the fundamental structural unit of monolayer FeX consists of a single Fe atomic layer sandwiched between two X atomic layers. Each Fe atom occupies the center of a tetrahedral coordination environment formed by four surrounding X atoms, giving rise to four distinct Fe-X-Fe-X planes. Among these planes, two adjacent planes without shared edges (indicated by identical colors) are mutually perpendicular, whereas planes sharing common edges are oriented at specific angles relative to each other. Analogous to the octahedral coordination configuration, this geometric arrangement induces spin frustration at individual lattice sites.

A brief geometric proof of the aforementioned orthogonality relation is provided as follows: considering the two blue Fe-X-Fe-X planes within the unit cell as an example, the three Fe atoms are collinear and lie within the planes, while being simultaneously perpendicular to the lines connecting the X atoms within each plane (indicated by red dashed lines in the Fig.~(\textcolor{red}{1})(c). According to the geometric properties of the regular tetrahedron, the lines connecting the X atoms in the two planes are also mutually perpendicular. By applying spatial geometry theorems, it can be deduced that these two planes are orthogonal to each other. The same reasoning can be applied to verify that the remaining pairs of adjacent planes without shared edges are likewise mutually perpendicular.

\begin{figure}[!ht]
\centering
\includegraphics[width=0.7\columnwidth, clip]{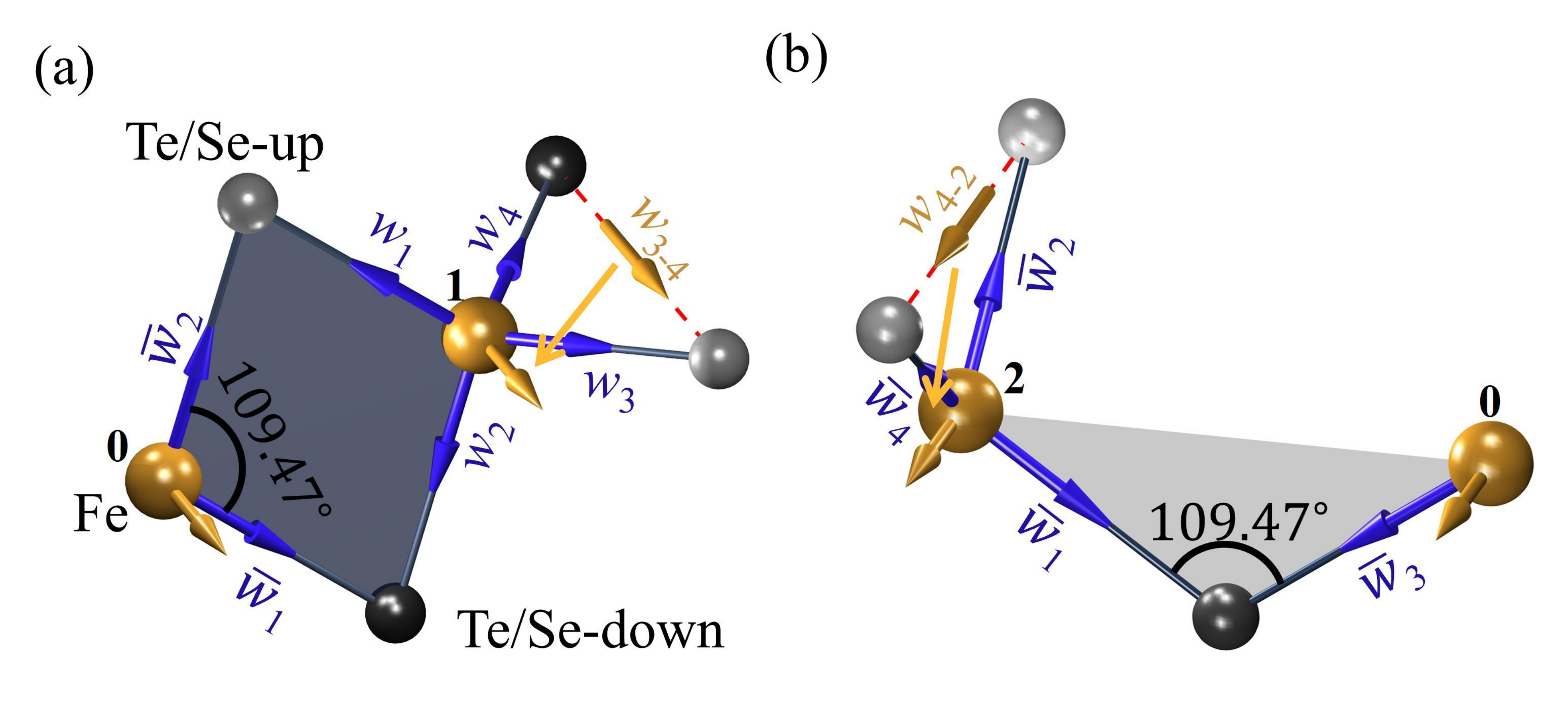}\\
\caption{ Schematic of local Kitaev-like interactions. Brown spheres denote Fe atoms, labeled $0$, $1$ and $2$ for the central, 1NN and 2NN sites, respectively. Black (gray) spheres denote lower- (upper-) layer X (X=Te, Se) atoms. Blue arrows indicate Fe–X bonds; yellow arrows indicate the vector differences of the two bonds, which lie perpendicular to the plane under consideration. (a) Kitaev-like interaction between the central Fe atom and the 1NN Fe atom at (1,1). (b) Kitaev-like interaction between the central Fe atom and the 2NN Fe atom at (-2,0).
  }\label{FIG-s2}
\end{figure}
Based on the unique lattice structure of the FeX system, we construct a corresponding bond-dependent interaction model. Under the SOC effect induced by the X atoms, the magnetic moment interactions between the central Fe atom and its four nearest-neighbor(1NN) Fe atoms generate components perpendicular to the Fe-X-Fe-X planes, constituting the Kitaev-like interaction term. Meanwhile, within the planes, the spin magnetic moments of Fe ions along the Fe-X bond directions are mediated by the orbitals of X ions, giving rise to in-plane off-diagonal exchange interaction terms. For convenience, we denote the directions from the central Fe atom along the four Fe-X bonds pointing toward the X atoms as $w_p$, $\bar{w}_p$($p$ = 1, 2, 3, 4). The local coordinate basis vectors of different Fe-X-Fe-X planes can be expressed as linear combinations of these bond directions. As shown in Fig.~\ref{FIG-s2}(a), the local geometry of the 1NN Kitaev-like interaction is depicted. The vector difference of the two bonds lying out of the Fe-X-Fe-X plane is perpendicular to that plane and is denoted $w_{3-4}$; it is identified as the $\gamma$ bond of the local frame. The remaining two vectors $w_1$ and $w_2$ correspond to the $\alpha$ and $\beta$ bonds, respectively. The triad {$\alpha,\;\beta,\;\gamma$} thus constitutes the basis of the Kitaev-like interaction of the plane. 

Kitaev-like interactions also exist between the central Fe atom and its second nearest-neighbor(2NN) Fe atoms, involving two mutually perpendicular Fe-X-Fe planes. Similarly, under the SOC effect induced by the X atoms, the magnetic moment interactions between the central Fe atom and its four 2NN Fe atoms generate components perpendicular to the planes, constituting the Kitaev-like interaction term. Meanwhile, the in-plane spin magnetic moments of Fe ions along the Fe-X bond directions are mediated by X ions, producing in-plane off-diagonal exchange interaction terms. The local basis vectors of different planes can likewise be expressed in terms of the bond directions. As shown in Fig.~\ref{FIG-s2}(b), the vector difference of the two bonds lying out of the Fe-X-Fe plane is perpendicular to that plane and is denoted $w_{4-2}$; it is identified as the $\gamma$ bond of the local frame. The remaining two vectors $w_3$ and $w_1$ correspond to the $\alpha$ and $\beta$ bonds, respectively. The triad {$\alpha,\;\beta,\;\gamma$} thus constitutes the Kitaev-like interaction basis of the plane. Note that the interactions of the central Fe atom with the pair of opposite 2NN Fe atoms lie in the same plane and therefore give identical Kitaev-like contributions; they are nevertheless treated separately for clarity.

The local basis vectors defined on the planes that contain the various Fe–Fe bonds are listed in the Table~\ref{Table-s1}; they correspond to the models shown in Figs.~(\textcolor{red}{1})(d) and (e).

\begin{table}[htbp]
  \centering
  \caption{Local Kitaev-like interaction basis-vector directions for the distinct Fe-X-Fe(-X) planes.}
  \label{Table-s1}
   \begin{ruledtabular}
 \label{Table-s1}
    \begin{tabular}{ c c c c c c}
      1NN Fe-Fe bond & 1NN Fe-atom positions & $p$ & 1NN $\alpha$ bond & 1NN $\beta$ bond  & 1NN $\gamma$ bond \\
      \hline
      W$_{3-4}$ & (1,1) & 1 & $\bar{w}_1$ & $\bar{w}_2$ & $w_{3-4}$\\W$_{4-1}$ & (-1,1) & 2 & $\bar{w}_3$ & $\bar{w}_2$ & $w_{4-1}$\\W$_{1-2}$ & (-1,-1) & 3 & $\bar{w}_3$ & $\bar{w}_4$ & $w_{1-2}$\\W$_{2-3}$ & (1,-1) & 4 & $\bar{w}_1$ & $\bar{w}_4$ & $w_{2-3}$\\
      \hline
      \hline
      2NN Fe-Fe bond & 2NN Fe-atom positions & $p$ & 2NN $\alpha$ bond & 2NN $\beta$ bond  & 2NN $\gamma$ bond \\ 
      \hline
      W$_{2-4}$ & (2,0) & 1 & $\bar{w}_1$ & $\bar{w}_3$ & $w_{2-4}$\\W$_{3-1}$ & (0,2) & 2 & $\bar{w}_4$ & $\bar{w}_2$ & $w_{3-1}$\\W$_{4-2}$ & (-2,0) & 3 & $\bar{w}_3$ & $\bar{w}_1$ & $w_{4-2}$\\W$_{1-3}$ & (0,-2) & 4 & $\bar{w}_2$ & $\bar{w}_4$ & $w_{1-3}$\\
    \end{tabular}
     \end{ruledtabular}
\end{table}
Taken together, the model conventions imply that, on each Fe-X-Fe(-X) plane under study, the $\gamma$ bond stands perpendicular to the plane while the $\alpha$ and $\beta$ bonds lie within it. Their in-plane angle is not $90^\circ$ but $109.47^\circ$, as illustrated in Fig.~\ref{FIG-s2}. A new coordinate frame must therefore be constructed when the effective spin Hamiltonian is analyzed; this construction is discussed in detail below. It should also be noted that the bond-direction expressions given here apply to only one of the two sublattices; the corresponding expressions on the other sublattice are modified accordingly.

\newpage
\section{Microscopic derivation of the Kitaev-like model}\label{Sec-S2}
In studies of the Kitaev model on ideal honeycomb lattices with edge-sharing octahedra, microscopic derivations are typically performed via the Kanamori interaction and a tight-binding model~[\textcolor{green}{24,40}]; the effective spin interactions are obtained by describing virtual hopping processes between magnetic atoms M$_1$ and M$_2$. The microscopic origin of the Kitaev interaction is the strong spin–orbit coupling (SOC) of ligands X$_1$ and X$_2$, which drives spin-flip virtual transitions and thereby yields bond-dependent anisotropic exchange. In conventional derivations for such ideal octahedral systems, however, only Heisenberg and Kitaev terms appear; off-diagonal interactions require further local symmetry lowering induced by lattice distortions.

For the two-dimensional edge-sharing tetrahedral structure investigated here, applying the same Kanamori interaction and tight-binding framework reveals that the Kitaev-like interaction also originates from ligand-SOC-induced spin-flip virtual processes. Crucially, even in the absence of any lattice distortion, off-diagonal exchange interactions emerge spontaneously in the ideal edge-sharing tetrahedral geometry. Below, taking FeTe as an example, we derive explicitly the effective magnetic exchange interactions for this ideal geometry based on the Kanamori interaction and tight-binding model.

\subsection{Kanamori interaction and tight-binding model}\label{Subsec-S2.1}
In monolayer FeTe, each Fe atom sits at the center of a regular tetrahedron formed by four surrounding Te atoms; the edge-sharing tetrahedral network is composed primarily of Fe $d$ orbitals and Te $p$ orbitals. The Fe $d$ shell is more than half-filled, corresponding to a $d^6$ configuration, whereas the Te $p$ orbitals are fully occupied ($p^6$). Thus, the total Hamiltonian of the system comprises the Kanamori interaction on the Fe sites and the tight-binding Hamiltonian.

However, because the Fe $d$ orbitals are more than half-filled, we adopt the hole picture to simplify the model. The Fe $d^6$ electron configuration is then mapped onto a $d^4$ hole configuration. Consequently, the local Hamiltonian on each Fe site, expressed in the hole basis, is given by the Kanamori interaction together with crystal-field splitting, and takes the following form~[\textcolor{green}{40}]:
\begin{equation}
    \begin{aligned}
	H_{ee}=&U\sum\limits_\alpha{n_{\alpha\uparrow}} n_{\alpha\downarrow}+\frac{U'}{2}\sum\limits_{\alpha\neq\beta,\sigma,\sigma'}{n_{\alpha\sigma}} n_{\beta\sigma'}-\frac{J_H}{2}\sum\limits_{\alpha\neq\beta,\sigma,\sigma'}{c^\dagger_{\alpha\sigma}} c^\dagger_{\beta\sigma'} c_{\beta\sigma} c_{\alpha\sigma'} \\
	+&J_H\sum\limits_{\alpha\neq\beta,}{c^\dagger_{\alpha\uparrow}} c^\dagger_{\alpha\downarrow} c_{\beta\downarrow} c_{\beta\uparrow}+\Delta_c\sum\limits_{\alpha\in e,\sigma}{c^\dagger_{\alpha\sigma}} c_{\alpha\sigma}.
\end{aligned}
\label{Eq-S2}
\end{equation}

The density operator satisfies the expression $n_{\alpha\sigma}=c^\dagger_{\alpha\sigma} c_{\alpha\sigma}$, where $c^\dagger_{\alpha\sigma}$ and $c_{\alpha\sigma}$ represent the creation and annihilation operators for orbital $\alpha$ and spin $\sigma$ on the Fe site, respectively. The terms involving $U$ and $U'$ correspond to the Hubbard interactions for intra-orbital and inter-orbital processes, respectively. The $J_H$ terms describe Hund’s coupling, where the third and fourth terms represent spin-exchange interactions between different orbitals and pair-hopping processes, respectively. The $\Delta c$ term denotes the tetrahedral crystal-field splitting acting on the Fe site. Under the tetrahedral crystal field, the Fe $d$ orbitals split into a lower-energy doubly degenerate $e$ level and a higher-energy triply degenerate $t_2$ level, with the electronic configuration denoted as $e^3t_2^3$. The Hund’s coupling selects the high-spin $S=2$ ground state for FeTe, and in the hole representation, the $t_2$ level is preferentially occupied, yielding the configuration $t_2^3e^1$. Consequently, the intermediate states for hole hopping exhibit multiple possibilities, with energy variations dependent on $U,\;U',\;J_H$, and $\Delta c$. Under the strong-coupling approximation, the tight-binding hopping integrals are treated as perturbations, as they are much smaller than the excitation energy changes.

In the edge-sharing tetrahedral lattice structure (as shown in Fig.~(\textcolor{red}{1})), each Fe atom forms four Fe-Fe bonds with its 1NNs, where each bond plane contains two coordinating Te atoms. Similarly, four Fe-Fe bonds are formed with 2NNs, but each bond plane contains only one coordinating Te atom. The magnetic exchange interactions between Fe atoms can be decomposed into two microscopic mechanisms: (1) direct exchange processes, where electrons undergo virtual hopping via direct Fe-Fe paths, and (2) indirect superexchange processes, where virtual hopping occurs via Fe-Te-Fe pathways, with the SOC of the Te $p$ orbitals playing a critical role. Based on this analysis, the tight-binding Hamiltonian describing the interaction between Fe$_i$ and Fe$_j$ can be constructed as follows~[\textcolor{green}{27}]:
\begin{equation}
H_{\mathrm{TB}_1}
=
\begin{pmatrix}
\boldsymbol{0}_{\mathbf{5}\times\mathbf{5}} &
\boldsymbol{T}_{\mathbf{M}_1\mathbf{M}_2} &
\boldsymbol{T}_{\mathbf{M}_1\mathbf{X}_1} &
\boldsymbol{T}_{\mathbf{M}_1\mathbf{X}_2}
\\
\boldsymbol{T}_{\mathbf{M}_1\mathbf{M}_2}^{\dagger} &
\boldsymbol{0}_{\mathbf{5}\times\mathbf{5}} &
\boldsymbol{T}_{\mathbf{M}_2\mathbf{X}_1} &
\boldsymbol{T}_{\mathbf{M}_2\mathbf{X}_2}
\\
\boldsymbol{T}_{\mathbf{M}_1\mathbf{X}_1}^{\dagger} &
\boldsymbol{T}_{\mathbf{M}_2\mathbf{X}_1}^{\dagger} &
\boldsymbol{0}_{\mathbf{3}\times\mathbf{3}} &
\boldsymbol{0}_{\mathbf{3}\times\mathbf{3}}
\\
\boldsymbol{T}_{\mathbf{M}_1\mathbf{X}_2}^{\dagger} &
\boldsymbol{T}_{\mathbf{M}_2\mathbf{X}_2}^{\dagger} &
\boldsymbol{0}_{\mathbf{3}\times\mathbf{3}} &
\boldsymbol{0}_{\mathbf{3}\times\mathbf{3}}
\end{pmatrix},
\qquad
H_{\mathrm{TB}_2}
=
\begin{pmatrix}
\boldsymbol{0}_{\mathbf{5}\times\mathbf{5}} &
\boldsymbol{T}_{\mathbf{M}_1\mathbf{M}_2} &
\boldsymbol{T}_{\mathbf{M}_1\mathbf{X}_1}
\\
\boldsymbol{T}_{\mathbf{M}_1\mathbf{M}_2}^{\dagger} &
\boldsymbol{0}_{\mathbf{5}\times\mathbf{5}} &
\boldsymbol{T}_{\mathbf{M}_2\mathbf{X}_1}
\\
\boldsymbol{T}_{\mathbf{M}_1\mathbf{X}_1}^{\dagger} &
\boldsymbol{T}_{\mathbf{M}_2\mathbf{X}_1}^{\dagger} &
\boldsymbol{0}_{\mathbf{3}\times\mathbf{3}}
\end{pmatrix},
\label{Eq-S3}
\end{equation}
where $\mathbf{0_{n\times n}}$ denotes an $n\times n$ null matrix, and $\mathbf{T^\dagger},\;\mathbf{T}$ represent hopping operators between atomic orbitals, with subscripts indicating the involved atoms. Here, M$_i$ and X$_m$($i,\;m$=1,2) denote Fe and Te atoms, respectively. Specifically, $\mathbf{T_{M_1 M_2}}$ describes direct hopping from the $d$ orbitals of Fe$_2$ to Fe$_1$, while $\mathbf{T_{M_i X_m}}$ represents indirect hopping via Te$_m$ $p$ orbitals to Fe$_i$ $d$ orbitals. In the global coordinate system shown in Fig.~(\textcolor{red}{1}), the five $d$-orbital basis states on Fe$_i$ are defined as: $C_{M_i, d}^{\dagger} = ( c_{i, d_{xy}}^{\dagger}, c_{i, d_{yz}}^{\dagger}, c_{i, d_{xz}}^{\dagger}, c_{i, d_{x^2-y^2}}^{\dagger}, c_{i, d_{3z^2-r^2}}^{\dagger} )$, and the three $p$-orbital basis states on Te$_m$ are defined as: $C_{X_m, p}^{\dagger} = ( c_{m, p_x}^{\dagger}, c_{m, p_y}^{\dagger}, c_{m, p_z}^{\dagger} )$. Since our focus is on Kitaev-like interactions, we only consider the $d$--$p$--$d$ indirect hopping processes.

\subsection{Effective d-d Hopping Model and SOC Effects}\label{Subsec-S2.2}
For the indirect $d$--$p$--$d$ hopping, we employ second-order perturbation theory to eliminate the $p$ orbitals, yielding an effective $d$--$d$ hopping model expression~[\textcolor{green}{27}], which can be recast into a Green's function projection form for clarity and computational~[\textcolor{green}{41,42}]:
\begin{equation}
    \bm{T}_{\bm{M}_j \bm{M}_i}^{eff} = - \sum_{a,m} \frac{\bm{T}_{\bm{M}_j \bm{X}_m} |a\rangle \langle a| \bm{T}_{\bm{X}_m \bm{M}_i}}{\Delta E_a} = \sum_m \bm{T}_{\bm{M}_j \bm{X}_m} \bm{G}_p \bm{T}_{\bm{X}_m \bm{M}_i},
\label{Eq-S4}
\end{equation}
where the summation runs over all intermediate states $|a\rangle$ on each ligand atom X$_m$. Summing over all intermediate states for a given ligand yields the propagator operator $G_p$, where $\Delta E_a=E_a-E_0$. In the hole representation, the physical process can be described as follows: a hole hops from atom M$_i$ to atom X$_m$, undergoes SOC effects, and then hops to atom M$_j$. The hole hopping from M$_i$ to X$_m$ is equivalent to $\mathbf{T_{X_m M_i}}$, the action on X$_m$ is equivalent to $G_p$, and the subsequent hopping from X$_m$ to M$_j$ is equivalent to $\mathbf{T_{M_j X_m}}$.

Below, we derive the explicit form of the propagator operator under the SOC effects of the ligand Te atoms.

The intermediate state on the Te atom contains only one hole, with orbital angular momentum $L = 1$ and spin angular momentum $S = \frac{1}{2}$. Under LS coupling with SOC, the states split into a fourfold degenerate manifold with $J = L + S = 3/2, M_j = \pm 3/2, \pm 1/2$, and a twofold degenerate manifold with $J = L - S = 1/2, M_j = \pm 1/2$. Including the energy contribution from SOC, the Hamiltonian is expressed as:
\begin{equation}
    H_{SOC} = \lambda_p (\bm{L} \cdot \bm{S}),
    \label{Eq-S5}
\end{equation}
where $\lambda_p$ represents the SOC strength of the Te $p$ orbitals, which takes a negative value in the hole representation. Under different total angular momenta, the corresponding SOC energy corrections are obtained as $H_{J=\frac{3}{2}} = \frac{1}{2}\lambda_p, H_{J=\frac{1}{2}} = -\lambda_p$. Consequently, the excitation energy of the intermediate state relative to the ground state is given by $\Delta E_a = \epsilon_p - \epsilon_d - \frac{1}{2}\lambda_p$ or $\Delta E_a = \epsilon_p - \epsilon_d + \lambda_p$. For subsequent calculations, we define the charge-transfer energy as $\Delta = \epsilon_p - \epsilon_d$. The projection operator onto a specific $J$ manifold can be expressed as $P_J = |a\rangle\langle a|$, yielding:
\begin{equation}
    P_{J=\frac{3}{2}} = \frac{2}{3}(I + \bm{L} \cdot \bm{S}), 
    P_{J=\frac{1}{2}} = \frac{1}{3}(I - 2\bm{L} \cdot \bm{S}),
    \label{Eq-S6}
\end{equation}
where $I$ denotes the $6\times 6$ identity matrix.

The propagator operator can then be expressed as:
\begin{equation}
    G_p = -\frac{P_{J=\frac{3}{2}}}{\Delta - \frac{1}{2}\lambda_p} - \frac{P_{J=\frac{1}{2}}}{\Delta + \lambda_p} = -\frac{1}{3}\left( \frac{2}{\Delta - \frac{\lambda_p}{2}} + \frac{1}{\Delta + \lambda_p} \right) I - \frac{2}{3}\left( \frac{1}{\Delta - \frac{\lambda_p}{2}} - \frac{1}{\Delta + \lambda_p} \right) (\bm{L} \cdot \bm{S}),
    \label{Eq-S7}
\end{equation}
where the two terms represent the spin-independent and SOC-dependent contributions, respectively. We define $A = -\frac{1}{3}\left( \frac{2}{\Delta - \lambda_p/2} + \frac{1}{\Delta + \lambda_p} \right)$ as the coefficient for the spin-independent part, and $B = -\frac{2}{3}\left( \frac{1}{\Delta - \lambda_p/2} - \frac{1}{\Delta + \lambda_p} \right)$ as the coefficient for the SOC part.

The contribution of SOC can be rewritten in the following form:
\begin{equation}
    \bm{L} \cdot \bm{S} = L_x \otimes S_x + L_y \otimes S_y + L_z \otimes S_z = \sum_{\gamma} L_{\gamma} \otimes \frac{1}{2}\sigma_{\gamma},
    \label{Eq-S8}
\end{equation}
where $L_{\gamma}$ and $\sigma_{\gamma}$ are the orbital angular momentum and Pauli operators for the respective components. The SOC process filters the allowed $d-p$ and $d-p$ hopping pathways through the action of the orbital angular momentum operators.

\subsection{Coordinate transformation and Slater-Koster transformation}\label{Subsec-S2.3}
\begin{figure}[!ht]
\centering
\includegraphics[width=0.7\columnwidth, clip]{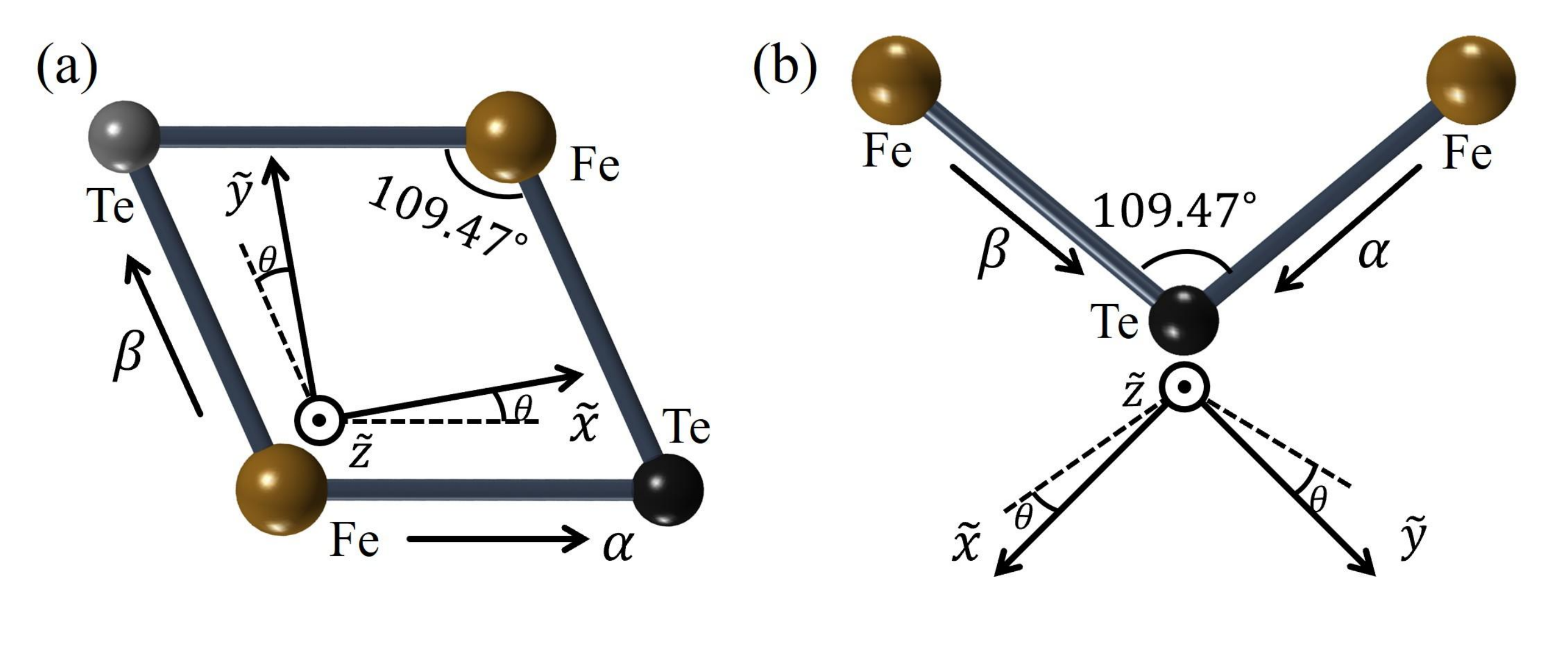 }\\
\caption{Schematic of the local coordinate configuration. Here, $\tilde{x}$, $\tilde{y}$ and $\tilde{z}$ denote the local coordinate axes, while $\alpha$ and $\beta$ represent the directions of the $\alpha$ and $\beta$ bonds of the plane in the global coordinate system. $\theta$ is the angle between the coordinate axes and the bond directions. (a) First nearest neighbor. (b) Second nearest neighbor.}
\label{FIG-s3}
\end{figure}

As shown in Fig.~(\textcolor{red}{1}), the Fe-Te-Fe-Te and Fe-Te-Fe planes formed by the Fe atoms and their neighbors via the coordinating Te atoms are not mutually perpendicular. Consequently, it is difficult to define a single unified coordinate system to describe both the orbital and spin bases in different plane. To address this, we define distinct local coordinate systems $(\tilde{x}, \tilde{y}, \tilde{z})$ for the different planes to perform the Slater-Koster transformation~[\textcolor{green}{39}], as illustrated in Fig.~\ref{FIG-s3}.
We construct the coordinate system according to the bond directions defined in the model, where the $\tilde{z}$ direction corresponds to the $w_{p-p'}$ bond direction perpendicular to the plane in Fig.~(\textcolor{red}{1}), matching the $\gamma$ direction in the corresponding Hamiltonian. Since the Fe-Te bond angles within the Fe-Te-Fe(-Te) exchange plane are not $90^{\circ}$, we choose the $\tilde{x}$ and $\tilde{y}$ directions within the plane at positions corresponding to $\alpha$ and $\beta$, as shown in Fig.~\ref{FIG-s3}, with $\tilde{z}$ pointing out of the page. When applying this to the real structure, we first fix $\tilde{z}$ along the $w_{p-p'}$ bond direction as the reference, ensuring compliance with the right-hand rule. Under this convention, the resulting local coordinate expressions are identical for all 1NN and 2NN pairs, respectively.

The relationship between the local and global coordinate systems is generally described by a rotation matrix. The local coordinate axes are expressed in the global coordinate system as:
\begin{equation}
    \tilde{x} = (x_1, x_2, x_3), \quad 
    \tilde{y} = (y_1, y_2, y_3), \quad 
    \tilde{z} = (z_1, z_2, z_3),
    \label{Eq-S9}
\end{equation}
then the transformation from local to global coordinates is given by:
\begin{equation}
    \begin{pmatrix} x_m \\ y_m \\ z_m \end{pmatrix} 
    = 
    \begin{pmatrix} 
    x_1 & y_1 & z_1 \\ 
    x_2 & y_2 & z_2 \\ 
    x_3 & y_3 & z_3 
    \end{pmatrix}
    \begin{pmatrix} \tilde{x}_m \\ \tilde{y}_m \\ \tilde{z}_m \end{pmatrix}
    = R 
    \begin{pmatrix} \tilde{x}_m \\ \tilde{y}_m \\ \tilde{z}_m \end{pmatrix},
    \label{Eq-S10}
\end{equation}
where $R$ is the rotation matrix.

For calculations performed in the local coordinate system, a new orbital basis must be established within that frame, namely $\bm{C}_{M_i, d'}^{\dagger}$ and $\bm{C}_{X_m, p'}^{\dagger}$, where $d'$ corresponds to $(d'_{xy}, d'_{yz}, d'_{xz}, d'_{x^2-y^2}, d'_{3z^2-r^2})$ and $p'$ corresponds to $(p'_x, p'_y, p'_z)$. Applying the Slater-Koster transformation in the local coordinate system yields the hopping matrix for transitions from the Fe $d$ orbitals to the Te $p$ orbitals~[\textcolor{green}{39}]:
\begin{equation}
    \bm{T}_{X_1 M_1} = - \bm{T}_{X_2 M_2} = 
    \begin{pmatrix} 
    t_1 & t_2 & 0 \\ 
    0 & 0 & t_3 \\ 
    0 & 0 & t_4 \\ 
    t_5 & t_6 & 0 \\ 
    t_7 & t_8 & 0 
    \end{pmatrix}, \quad
    \bm{T}_{X_2 M_1} = - \bm{T}_{X_1 M_2} = 
    \begin{pmatrix} 
    t_2 & t_1 & 0 \\ 
    0 & 0 & t_4 \\ 
    0 & 0 & t_3 \\ 
    -t_6 & -t_5 & 0 \\ 
    t_8 & t_7 & 0 
    \end{pmatrix}.
    \label{Eq-S11}
\end{equation}

The corresponding hopping integrals are expressed as:
\begin{equation}
    \begin{aligned}
    t_1 &= -\frac{2+\sqrt{2}}{12}V_{pd\sigma} &+\frac{2(\sqrt{6}-\sqrt{3})}{9}V_{pd\pi}&, \\
    t_2 &= \frac{2-\sqrt{2}}{12}V_{pd\sigma} &+\frac{2(\sqrt{6}+\sqrt{3})}{9}V_{pd\pi}&, \\
    t_3 &= &\frac{\sqrt{6}-2\sqrt{3}}{6}V_{pd\pi}, \\
    t_4 &= &\frac{\sqrt{6}+2\sqrt{3}}{6}V_{pd\pi}&, \\
    t_5 &= \frac{1+\sqrt{2}}{3}V_{pd\sigma} &+\frac{2\sqrt{3}-\sqrt{6}}{18}V_{pd\pi}&, \\
    t_6 &= \frac{1-\sqrt{2}}{3}V_{pd\sigma} &+ \frac{2\sqrt{3}+\sqrt{6}}{18}V_{pd\pi}&, \\
    t_7 &= -\frac{2\sqrt{3}+\sqrt{6}}{12}V_{pd\sigma}&&,  \\
    t_8 &= \frac{2\sqrt{3}-\sqrt{6}}{12}V_{pd\sigma}&&, 
    \end{aligned}
    \label{Eq-S12}
\end{equation}
where $V_{pd\sigma} < 0, V_{pd\pi} > 0$ and $|V_{pd\sigma}| > |V_{pd\pi}|$. The hopping integrals for $p \rightarrow d$ and $d \rightarrow p$ transitions along the same bond are equal in magnitude.

\subsection{Superexchange process and effective spin Hamiltonian}\label{Subsec-S2.4}
The complete superexchange process consists of a hole hopping virtually from $\text{Fe}_1$ to $\text{Fe}_2$ mediated by the $\text{Te}_1/\text{Te}_2$ atoms, followed by a virtual return hopping from $\text{Fe}_2$ back to $\text{Fe}_1$ via $\text{Te}_1/\text{Te}_2$. Accordingly, the Hamiltonian governing the entire virtual hopping process can be expressed as:
\begin{equation}
    H_s = -\sum_n \frac{\bm{T}_{M_i M_j}^{eff} |n\rangle \langle n| \bm{T}_{M_j M_i}^{eff}}{\Delta E_n},
    \label{Eq-S13}
\end{equation}
where $\Delta E_n = E_n - E_0$ represents the excitation energy of the virtual intermediate state $|n\rangle$ relative to the magnetic ground state.

Although the Te-induced SOC has been analyzed, we must consistently treat the SOC term $\tilde{\bm{L}} \cdot \tilde{\bm{S}}$ in the local coordinate frame, transforming it in the same manner as described above. By combining the local-frame Slater-Koster hopping matrices between the Fe $d$ and Te $p$ orbitals, we obtain the total superexchange interaction between $\text{Fe}_1$ and $\text{Fe}_2$. Throughout this section, all descriptions of the $d$ and $p$ orbitals are defined with respect to the local coordinate system established in Fig.~\ref{FIG-s3}.

\subsubsection{\texorpdfstring{$t_2$–$t_2$ superexchange process}{t2–t2 superexchange process}}\label{Subsubsec-S2.4.1}
\begin{figure}[!ht]
\centering
\includegraphics[width=0.85\columnwidth, clip]{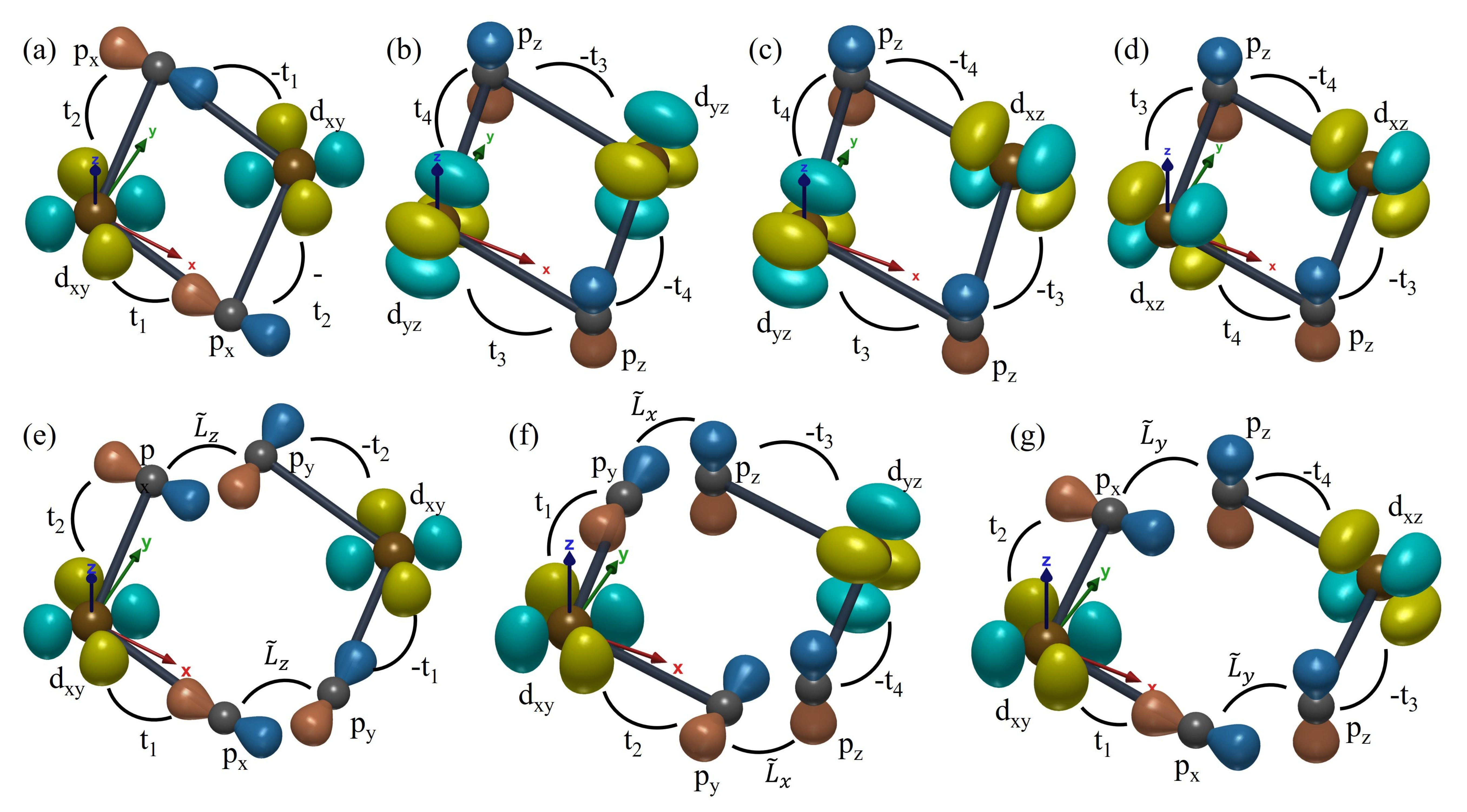}\\
\caption{Schematic of the $t_2$-$t_2$ virtual hopping processes in the hole representation. Brown and dark-gray spheres denote Fe and Te atoms, respectively. (a)-(d) Spin-conserving superexchange pathways between $\text{Fe}_1$ and $\text{Fe}_2$ mediated by the Te atoms. (e)-(g) Spin-flip superexchange pathways between $\text{Fe}_1$ and $\text{Fe}_2$ assisted by the SOC of the Te atoms.
  }\label{FIG-s4}
\end{figure}

We first consider the superexchange processes involving exclusively the t$_2$ orbitals. Utilizing the Slater-Koster formalism, several representative superexchange pathways in the local coordinate frame are illustrated in Fig.~\ref{FIG-s4}.
As visually demonstrated in it, several virtual hopping processes are directly visualized. Since the bond directions in the local coordinate system are not aligned along the basis vectors, the inter-orbital interactions exhibit considerable complexity. As a representative example, we present the complete calculation for the $\text{Fe}_1\ d_{xy}\text{--}\text{Fe}_2\ d_{xy}$ process. The total process of a hole hopping from $\text{Fe}_1$ to $\text{Fe}_2$, expressed in terms of operators and eigenstates, is given by:
\begin{equation}
    \langle d_{j,xy} | \bm{T}_{\bm{M}_2 \bm{M}_1}^{eff} | d_{i,xy} \rangle = \sum_{m,\alpha,\beta} \langle d_{j,xy} | c_{j,d_{xy}}^{\dagger} c_{m,p_{\beta}} | p_{m,\beta} \rangle \langle p_{m,\beta} | G_{p,m} | p_{m,\alpha} \rangle \langle p_{m,\alpha} | c_{m,p_{\alpha}}^{\dagger} c_{i,d_{xy}} | d_{i,xy} \rangle.
    \label{Eq-S14}
\end{equation}

The term $|p_{m,\beta}\rangle \langle p_{m,\beta} | G_{p,m} | p_{m,\alpha} \rangle \langle p_{m,\alpha}|$ describes a hole hopping onto the $\alpha$ orbital of the $\text{Te}_m$ site, evolving under $G_{p,m}$, and subsequently transitioning to the $\beta$ orbital. For the $\text{Fe}_1\ d_{xy}\text{--}\text{Fe}_2\ d_{xy}$ virtual hopping via $\text{Te}_m$, only four pathways exist: $d_{xy}\text{-}p_x\text{-}d_{xy}$ (spin-conserving), $d_{xy}\text{-}p_x\text{-}p_y\text{-}d_{xy}$ (spin-flip), $d_{xy}\text{-}p_y\text{-}p_x\text{-}d_{xy}$ (spin-flip), and $d_{xy}\text{-}p_y\text{-}d_{xy}$ (spin-conserving). The corresponding propagator actions on the p orbitals can be expressed as:
\begin{equation}
    \begin{aligned}
    \langle p_{m,x} | A I + B (\tilde{\bm{L}} \cdot \tilde{\bm{S}}) | p_{m,x} \rangle &= \langle p_{m,x} | A I | p_{m,x} \rangle = A \langle p_{m,x} | I | p_{m,x} \rangle = A = A \tilde{\sigma}_0, \\
    \langle p_{m,x} | A I + B (\tilde{\bm{L}} \cdot \tilde{\bm{S}}) | p_{m,y} \rangle &= \langle p_{m,x} | B (\tilde{L}_z \otimes \frac{1}{2}\tilde{\sigma}_z) | p_{m,y} \rangle = B \langle p_{m,x} | \tilde{L}_z | p_{m,y} \rangle \otimes \frac{1}{2}\tilde{\sigma}_z = -i \frac{1}{2} B \tilde{\sigma}_z, \\
    \langle p_{m,y} | A I + B (\tilde{\bm{L}} \cdot \tilde{\bm{S}}) | p_{m,x} \rangle &= \langle p_{m,y} | B (\tilde{L}_z \otimes \frac{1}{2}\tilde{\sigma}_z) | p_{m,x} \rangle = B \langle p_{m,y} | \tilde{L}_z | p_{m,x} \rangle \otimes \frac{1}{2}\tilde{\sigma}_z = i \frac{1}{2} B \tilde{\sigma}_z, \\
    \langle p_{m,y} | A I + B (\tilde{\bm{L}} \cdot \tilde{\bm{S}}) | p_{m,y} \rangle &= \langle p_{m,y} | A I | p_{m,y} \rangle = A \langle p_{m,y} | I | p_{m,y} \rangle = A = A \tilde{\sigma}_0,
    \end{aligned}\label{Eq-S15}
\end{equation}
where $\tilde{\sigma}_0, \tilde{\sigma}_x, \tilde{\sigma}_y,$ and $\tilde{\sigma}_z$ denote the identity spin operator and the Pauli matrices along the $x, y,$ and $z$ directions in the local coordinate system, respectively. The subscripts $x$ and $y$ correspond to the $p_x$ and $p_y$ orbitals. After evaluating the propagator operator, the $p$-orbital degrees of freedom can be integrated out, yielding:
\begin{equation}
    \begin{aligned}
    c_{m,p_{\beta}} |p_{m,\beta}\rangle &\left( [A \tilde{\sigma}_0]_{xx} + [-i\frac{1}{2}B \tilde{\sigma}_z]_{yx} + [i\frac{1}{2}B \tilde{\sigma}_z]_{xy} + [A \tilde{\sigma}_0]_{yy} \right) \langle p_{m,\alpha}| c_{m,p_{\alpha}}^{\dagger} \\
    &= [A \tilde{\sigma}_0]_{xx} + [-i\frac{1}{2}B \tilde{\sigma}_z]_{yx} + [i\frac{1}{2}B \tilde{\sigma}_z]_{xy} + [A \tilde{\sigma}_0]_{yy}.
    \end{aligned}\label{Eq-S16}
\end{equation}
This expression is incomplete because the orbital integrals arising from the actions of $c_{m,p_{\alpha}}^{\dagger} c_{i,d_{xy}}$ and $c_{j,d_{xy}}^{\dagger} c_{m,p_{\beta}}$ have not been included; the focus here is on demonstrating the elimination of the p-orbital integrals in the empty (filled) state representation. Assuming that the operator $c_{m,p_{\alpha}}^{\dagger}$ creates an electron with spin $\tilde{\sigma}_1$ that has hopped from the $\text{Fe}_1$ d orbital, and $c_{m,p_{\beta}}$ annihilates an electron with spin $\tilde{\sigma}_2$ that hops to the $\text{Fe}_2$ d orbital, the Pauli matrices in Eq.~(\ref{Eq-S16}) must be projected onto the matrix element $[\tilde{\sigma}_{\lambda}]_{\tilde{\sigma}_2 \tilde{\sigma}_1}$ according to the values of $\tilde{\sigma}_1$ and $\tilde{\sigma}_2$. Substituting the hopping integrals obtained from the Slater-Koster calculation yields:
\begin{equation}
    \begin{aligned}
    \langle d_{j,xy} | c_{j,d_{xy}}^{\dagger} &\left( T_{X_1 M_2, 11} [A \tilde{\sigma}_0]_{xx} T_{X_1 M_1, 11} + T_{X_1 M_2, 11} [-i \frac{1}{2} B \tilde{\sigma}_z]_{yx} T_{X_1 M_1, 12} \right.  \\
    &\quad \left. + T_{X_1 M_2, 12} [i \frac{1}{2} B \tilde{\sigma}_z]_{xy} T_{X_1 M_1, 11} + T_{X_1 M_2, 12} [A \tilde{\sigma}_0]_{yy} T_{X_1 M_1, 12} \right) c_{i,d_{xy}} | d_{i,xy} \rangle  \\
    &= \langle d_{j,xy} | c_{j,d_{xy}}^{\dagger} \left( -2t_1 t_2 A \tilde{\sigma}_0 - i \frac{1}{2} (t_1^2 - t_2^2) B \tilde{\sigma}_z \right) c_{i,d_{xy}} | d_{i,xy} \rangle,
    \end{aligned}
\label{Eq-S17}
\end{equation}
where $-2t_1 t_2 A \tilde{\sigma}_0 - i \frac{1}{2} (t_1^2 - t_2^2) B \tilde{\sigma}_z$ represents the effective hopping integral from the $\text{Fe}_1\ d_{xy}$ orbital to the $\text{Fe}_2\ d_{xy}$ orbital via $\text{Te}_1$. For notational simplicity, we omit the hole operators $c_{i,d_{xy}}$ and $c_{j,d_{xy}}^{\dagger}$ in this expression and retain only the corresponding coefficients; the hopping operators will be transformed into spin operators when considering the complete superexchange process. However, the contribution via $\text{Te}_2$ must also be calculated, and summing both contributions yields the total effective hopping integral from $\text{Fe}_1\ d_{xy}$ to $\text{Fe}_2\ d_{xy}$ as $-4t_1 t_2 A \tilde{\sigma}_0$.

Using the same method, the effective hopping integrals for the 1NN $t_2\text{-}t_2 (d_{xy}, d_{yz}, d_{xz})$ exchange can be computed as:
\begin{equation}
    \begin{matrix}
    \bm{T}_{\bm{M}_2\bm{M}_1}^{eff} (t_2 \otimes t_2) = \begin{pmatrix} 
    -4t_1 t_2 A \tilde{\sigma}_0 & \bm{T}_{\bm{M}_2\bm{M}_1, 12}^{eff} & \bm{T}_{\bm{M}_2\bm{M}_1, 13}^{eff} \\ 
    -i \frac{1}{2}(t_1 t_4 + t_2 t_3)B \tilde{\sigma}_y + i \frac{1}{2}(t_1 t_3 + t_2 t_4)B \tilde{\sigma}_x & -2t_3 t_4 A \tilde{\sigma}_0 & -(t_3^2 + t_4^2)A \tilde{\sigma}_0 \\ 
    -i \frac{1}{2}(t_1 t_3 + t_2 t_4)B \tilde{\sigma}_y + i \frac{1}{2}(t_1 t_4 + t_2 t_3)B \tilde{\sigma}_x & -(t_3^2 + t_4^2)A \tilde{\sigma}_0 & -2t_3 t_4 A \tilde{\sigma}_0 
    \end{pmatrix}, \\
    \bm{T}_{\bm{M}_2\bm{M}_1, 12}^{eff} (t_2 \otimes t_2) = i \frac{1}{2}(t_1 t_4 + t_2 t_3)B \tilde{\sigma}_y - i \frac{1}{2}(t_1 t_3 + t_2 t_4)B \tilde{\sigma}_x, \\
    \bm{T}_{\bm{M}_2\bm{M}_1, 13}^{eff} (t_2 \otimes t_2) = i \frac{1}{2}(t_1 t_3 + t_2 t_4)B \tilde{\sigma}_y - i \frac{1}{2}(t_1 t_4 + t_2 t_3)B \tilde{\sigma}_x.
\end{matrix}\label{Eq-S18}
\end{equation}

In the complete superexchange process, as shown in Eq.~(\ref{Eq-S13}), the reverse hopping from $\text{Fe}_2$ back to $\text{Fe}_1$ must also be considered. Here, the relation $\bm{T}_{\bm{M}_1\bm{M}_2}^{eff} = \bm{T}_{\bm{M}_2\bm{M}_1}^{eff\dagger}$ holds, thereby determining all effective hopping integrals between the $\text{Fe}_1\ t_2$ and $\text{Fe}_2\ t_2$ orbitals.

The effective hopping integrals for the 2NN must also be considered. Their hopping pathways are similar to those of the 1NN, but with only one coordinating atom and different bond directions in the local coordinate system. Denoting the 2NN Fe atom as $\bm{M}'_2$, the effective hopping integral can be expressed as:

\begin{equation}
    \begin{aligned}
    \bm{T}_{\bm{M}'_2\bm{M}_1}^{eff} (t_2 \otimes t_2) = \begin{pmatrix} 
    2t_1 t_2 A \tilde{\sigma}_0 + i \frac{1}{2}(t_2^2 - t_1^2)B \tilde{\sigma}_z & -i \frac{1}{2}t_2 t_3 B \tilde{\sigma}_y + i \frac{1}{2}t_1 t_3 B \tilde{\sigma}_x & -i \frac{1}{2}t_2 t_4 B \tilde{\sigma}_y + i \frac{1}{2}t_1 t_4 B \tilde{\sigma}_x \\ 
    i \frac{1}{2}t_1 t_4 B \tilde{\sigma}_y - i \frac{1}{2}t_2 t_4 B \tilde{\sigma}_x & t_3 t_4 A \tilde{\sigma}_0 & t_4^2 A \tilde{\sigma}_0 \\ 
    i \frac{1}{2}t_1 t_3 B \tilde{\sigma}_y - i \frac{1}{2}t_2 t_3 B \tilde{\sigma}_x & t_3^2 A \tilde{\sigma}_0 & t_3 t_4 A \tilde{\sigma}_0 
    \end{pmatrix}.
\end{aligned}\label{Eq-S19}
\end{equation}

This expression also satisfies the relation $\bm{T}_{\bm{M}_1\bm{M}'_2}^{eff} = \bm{T}_{\bm{M}'_2\bm{M}_1}^{eff\dagger}$.

The excitation energy change from the $d^4 d^4$ ground state to the $d^3 d^5$ intermediate state in the hole representation is calculated using the tight-binding model under the Kanamori interaction in Eq.~(\ref{Eq-S2}). For $t_2$-$t_2$ hopping, only one degenerate intermediate-state energy exists, with $\Delta E_n = U + 3J_H$. Substituting this into the complete superexchange process yields:

\begin{equation}
    \langle LS | H_s | GS \rangle = - \frac{1}{U + 3J_H} \langle LS | c_{i,d_{\alpha},\sigma_4}^{\dagger} \bm{T}_{\bm{M}_2\bm{M}_1}^{eff\dagger} c_{j,d_{\beta},\sigma_3} | n \rangle \langle n | c_{j,d_{\beta},\sigma_2}^{\dagger} \bm{T}_{\bm{M}_2\bm{M}_1}^{eff} c_{i,d_{\alpha},\sigma_1} | GS \rangle,\label{Eq-S20}
\end{equation}
where $\langle LS |$ and $| GS \rangle$ denote the final state and the ground state, respectively. The effective hopping integral contains both constant terms and spin operators; we separate these by factoring out the constants, denoting $\bm{T}_{\bm{M}_2\bm{M}_1}^{eff} = \sum_{\gamma} T_{\bm{M}_2\bm{M}_1,\gamma}^{eff} \tilde{\sigma}_{\gamma}$. The effective hopping integral may contain multiple terms; for notational simplicity, we use $\tilde{\sigma}_{\gamma}$ to represent them, where $\gamma = 0, x, y, z$. The expression can then be written as:
\begin{equation}
    \begin{aligned}
    \langle LS | H_s | GS \rangle &= - \sum_{\gamma, \gamma'} \frac{T_{\bm{M}_2\bm{M}_1,\gamma'}^{eff\dagger} T_{\bm{M}_2\bm{M}_1,\gamma}^{eff}}{U + 3J_H} \langle LS | c_{i,d_{\alpha},\sigma_4}^{\dagger} [\tilde{\sigma}_{\gamma'}]_{\sigma_4 \sigma_3} c_{j,d_{\beta},\sigma_3} c_{j,d_{\beta},\sigma_2}^{\dagger} [\tilde{\sigma}_{\gamma}]_{\sigma_2 \sigma_1} c_{i,d_{\alpha},\sigma_1} | GS \rangle \\
    &= \sum_{\gamma, \gamma'} \frac{T_{\bm{M}_2\bm{M}_1,\gamma'}^{eff\dagger} T_{\bm{M}_2\bm{M}_1,\gamma}^{eff}}{U + 3J_H} \langle LS | [\tilde{\sigma}_{\gamma'}]_{\sigma_4 \sigma_3} [\tilde{\sigma}_{\gamma}]_{\sigma_2 \sigma_1} c_{j,d_{\beta},\sigma_2}^{\dagger} c_{j,d_{\beta},\sigma_3} c_{i,d_{\alpha},\sigma_4}^{\dagger} c_{i,d_{\alpha},\sigma_1} | GS \rangle \\
    &= \sum_{\gamma, \gamma'} \frac{T_{\bm{M}_2\bm{M}_1,\gamma'}^{eff\dagger} T_{\bm{M}_2\bm{M}_1,\gamma}^{eff}}{U + 3J_H} \langle LS | ([\tilde{\bm{\sigma}}]_{\sigma_4 \sigma_3} \cdot \tilde{\bm{s}}_{j,\beta}) ([\tilde{\bm{\sigma}}]_{\sigma_2 \sigma_1} \cdot \tilde{\bm{s}}_{i,\alpha}) | GS \rangle.
\end{aligned}\label{Eq-S21}
\end{equation}

During this process, spin-independent density operator terms have been omitted. Here, $\tilde{\bm{s}}_{i,\alpha}$ and $\tilde{\bm{s}}_{j,\beta}$ are the spin operators for a single hole on the respective orbitals, derived from $c_{i,d_{\alpha},\tilde{\sigma}_4}^{\dagger} c_{i,d_{\alpha},\tilde{\sigma}_1}$ and $c_{j,d_{\beta},\tilde{\sigma}_2}^{\dagger} c_{j,d_{\beta},\tilde{\sigma}_3}$, respectively. According to the equivalent operator theorem~[\textcolor{green}{41,42}], in the high-spin $S = 2$ case, their relation to the total spin can be expressed as:
\begin{equation}
    \tilde{\bm{s}}_{i,\alpha} = \frac{1}{4}\tilde{\bm{S}}_i, \quad \tilde{\bm{s}}_{j,\beta} = \frac{1}{4}\tilde{\bm{S}}_j.
    \label{Eq-S22}
\end{equation}

In the calculation of Eq.~(\ref{Eq-S20}), the Pauli matrix combination $[\tilde{\sigma}_{\gamma'}]_{\tilde{\sigma}_4\tilde{\sigma}_3}[\tilde{\sigma}_{\gamma}]_{\tilde{\sigma}_2\tilde{\sigma}_1}$ inherently selects specific spin configurations; therefore, the spin operator calculations can be determined based on these Pauli matrices. We illustrate this with $[\tilde{\sigma}_0]_{\tilde{\sigma}_4\tilde{\sigma}_3}[\tilde{\sigma}_0]_{\tilde{\sigma}_2\tilde{\sigma}_1}$ as a concrete example:
\begin{equation}
    \begin{aligned}
    [\tilde{\sigma}_0]_{\tilde{\sigma}_4\tilde{\sigma}_3}[\tilde{\sigma}_0]_{\tilde{\sigma}_2\tilde{\sigma}_1} &\, ([\tilde{\bm{\sigma}}]_{\tilde{\sigma}_1\tilde{\sigma}_4} \cdot \tilde{\bm{s}}_{i,\alpha})([\tilde{\bm{\sigma}}]_{\tilde{\sigma}_3\tilde{\sigma}_2} \cdot \tilde{\bm{s}}_{j,\beta}) \\
    &= [\tilde{\sigma}_0]_{\uparrow\uparrow}[\tilde{\sigma}_0]_{\uparrow\uparrow}([\tilde{\bm{\sigma}}]_{\uparrow\uparrow} \cdot \tilde{\bm{s}}_{i,\alpha})([\tilde{\bm{\sigma}}]_{\uparrow\uparrow} \cdot \tilde{\bm{s}}_{j,\beta}) \\
    &\quad + [\tilde{\sigma}_0]_{\downarrow\downarrow}[\tilde{\sigma}_0]_{\downarrow\downarrow}([\tilde{\bm{\sigma}}]_{\downarrow\downarrow} \cdot \tilde{\bm{s}}_{i,\alpha})([\tilde{\bm{\sigma}}]_{\downarrow\downarrow} \cdot \tilde{\bm{s}}_{j,\beta}) \\
    &\quad + [\tilde{\sigma}_0]_{\uparrow\uparrow}[\tilde{\sigma}_0]_{\downarrow\downarrow}([\tilde{\bm{\sigma}}]_{\downarrow\uparrow} \cdot \tilde{\bm{s}}_{i,\alpha})([\tilde{\bm{\sigma}}]_{\uparrow\downarrow} \cdot \tilde{\bm{s}}_{j,\beta}) \\
    &\quad + [\tilde{\sigma}_0]_{\downarrow\downarrow}[\tilde{\sigma}_0]_{\uparrow\uparrow}([\tilde{\bm{\sigma}}]_{\uparrow\downarrow} \cdot \tilde{\bm{s}}_{i,\alpha})([\tilde{\bm{\sigma}}]_{\downarrow\uparrow} \cdot \tilde{\bm{s}}_{j,\beta}) \\
    &= 2\,\tilde{\bm{s}}_{i,\alpha} \cdot \tilde{\bm{s}}_{j,\beta} = \frac{1}{8}\,\tilde{\bm{S}}_i \cdot \tilde{\bm{S}}_j = \frac{1}{8}\,\bm{S}_i \cdot \bm{S}_j,
    \end{aligned}
    \label{Eq-S23}
\end{equation}
where the magnetic order selected by $[\tilde{\sigma}_0]_{\tilde{\sigma}_4\tilde{\sigma}_3}[\tilde{\sigma}_0]_{\tilde{\sigma}_2\tilde{\sigma}_1}$ is solved and summed. The product of spin and Pauli vector operators is computed by expanding the dot product as $\tilde{\sigma}_x \tilde{s}_x + \tilde{\sigma}_y \tilde{s}_y + \tilde{\sigma}_z \tilde{s}_z$ and substituting the Pauli matrices according to the spin configuration. As shown, $[\tilde{\sigma}_0]_{\tilde{\sigma}_4\tilde{\sigma}_3}[\tilde{\sigma}_0]_{\tilde{\sigma}_2\tilde{\sigma}_1}$ indicates that, in the absence of spin flips, the superexchange contributes to the Heisenberg interaction. Here, the dot product in the local coordinate system can be directly transformed into the global coordinate system. The results for different possible Pauli matrix combinations are presented below:
\begin{equation}
    \begin{aligned}
    [\tilde{\sigma}_x]_{\tilde{\sigma}_4\tilde{\sigma}_3}[\tilde{\sigma}_x]_{\tilde{\sigma}_2\tilde{\sigma}_1}([\tilde{\bm{\sigma}}]_{\tilde{\sigma}_1\tilde{\sigma}_4} \cdot \tilde{\bm{s}}_{i,\alpha})([\tilde{\bm{\sigma}}]_{\tilde{\sigma}_3\tilde{\sigma}_2} \cdot \tilde{\bm{s}}_{j,\beta}) &= \frac{1}{8}(\tilde{S}_i^x \tilde{S}_j^x - \tilde{S}_i^y \tilde{S}_j^y - \tilde{S}_i^z \tilde{S}_j^z), \\
    [\tilde{\sigma}_x]_{\tilde{\sigma}_4\tilde{\sigma}_3}[\tilde{\sigma}_y]_{\tilde{\sigma}_2\tilde{\sigma}_1}([\tilde{\bm{\sigma}}]_{\tilde{\sigma}_1\tilde{\sigma}_4} \cdot \tilde{\bm{s}}_{i,\alpha})([\tilde{\bm{\sigma}}]_{\tilde{\sigma}_3\tilde{\sigma}_2} \cdot \tilde{\bm{s}}_{j,\beta}) &= \frac{1}{8}(\tilde{S}_i^x \tilde{S}_j^y + \tilde{S}_i^y \tilde{S}_j^x), \\
    [\tilde{\sigma}_y]_{\tilde{\sigma}_4\tilde{\sigma}_3}[\tilde{\sigma}_y]_{\tilde{\sigma}_2\tilde{\sigma}_1}([\tilde{\bm{\sigma}}]_{\tilde{\sigma}_1\tilde{\sigma}_4} \cdot \tilde{\bm{s}}_{i,\alpha})([\tilde{\bm{\sigma}}]_{\tilde{\sigma}_3\tilde{\sigma}_2} \cdot \tilde{\bm{s}}_{j,\beta}) &= \frac{1}{8}(-\tilde{S}_i^x \tilde{S}_j^x + \tilde{S}_i^y \tilde{S}_j^y - \tilde{S}_i^z \tilde{S}_j^z), \\
    [\tilde{\sigma}_x]_{\tilde{\sigma}_4\tilde{\sigma}_3}[\tilde{\sigma}_z]_{\tilde{\sigma}_2\tilde{\sigma}_1}([\tilde{\bm{\sigma}}]_{\tilde{\sigma}_1\tilde{\sigma}_4} \cdot \tilde{\bm{s}}_{i,\alpha})([\tilde{\bm{\sigma}}]_{\tilde{\sigma}_3\tilde{\sigma}_2} \cdot \tilde{\bm{s}}_{j,\beta}) &= \frac{1}{8}(\tilde{S}_i^x \tilde{S}_j^z + \tilde{S}_i^z \tilde{S}_j^x), \\
    [\tilde{\sigma}_y]_{\tilde{\sigma}_4\tilde{\sigma}_3}[\tilde{\sigma}_z]_{\tilde{\sigma}_2\tilde{\sigma}_1}([\tilde{\bm{\sigma}}]_{\tilde{\sigma}_1\tilde{\sigma}_4} \cdot \tilde{\bm{s}}_{i,\alpha})([\tilde{\bm{\sigma}}]_{\tilde{\sigma}_3\tilde{\sigma}_2} \cdot \tilde{\bm{s}}_{j,\beta}) &= \frac{1}{8}(\tilde{S}_i^y \tilde{S}_j^z + \tilde{S}_i^z \tilde{S}_j^y), \\
    [\tilde{\sigma}_z]_{\tilde{\sigma}_4\tilde{\sigma}_3}[\tilde{\sigma}_z]_{\tilde{\sigma}_2\tilde{\sigma}_1}([\tilde{\bm{\sigma}}]_{\tilde{\sigma}_1\tilde{\sigma}_4} \cdot \tilde{\bm{s}}_{i,\alpha})([\tilde{\bm{\sigma}}]_{\tilde{\sigma}_3\tilde{\sigma}_2} \cdot \tilde{\bm{s}}_{j,\beta}) &= \frac{1}{8}(-\tilde{S}_i^x \tilde{S}_j^x - \tilde{S}_i^y \tilde{S}_j^y + \tilde{S}_i^z \tilde{S}_j^z),
\end{aligned}
\label{Eq-S24}
\end{equation}
swapping the positions of the two Pauli matrices of $[\tilde{\sigma}_{\gamma'}]_{\tilde{\sigma}_4\tilde{\sigma}_3}[\tilde{\sigma}_{\gamma}]_{\tilde{\sigma}_2\tilde{\sigma}_1}$ does not affect the result in Eq.~\ref{Eq-S24}.

Substituting the spin operator results and effective hopping integrals into the calculation yields, for the 1NN:
\begin{equation}
    \begin{aligned}
    H_{s_1}^{t_2-t_2} &= \frac{1}{4(U+3J_H)}\Bigl[8t_1^2 t_2^2 + 4t_3^2 t_4^2 + (t_3^2 + t_4^2)^2\Bigr] A^2 \,(\bm{S}_i \cdot \bm{S}_j) \\
    &\quad - \frac{1}{8(U+3J_H)}\Bigl[(t_1 t_3 + t_2 t_4)^2 + (t_1 t_4 + t_2 t_3)^2\Bigr] B^2 \,(\tilde{S}_i^z \tilde{S}_j^z) \\
    &\quad - \frac{1}{4(U+3J_H)}(t_1 t_3 + t_2 t_4)(t_1 t_4 + t_2 t_3) B^2 \,(\tilde{S}_i^x \tilde{S}_j^y + \tilde{S}_i^y \tilde{S}_j^x),
\end{aligned}
\label{Eq-S25}
\end{equation}
and for the 2NN:
\begin{equation}
\begin{aligned}
    H_{s_2}^{t_2-t_2} &= \frac{1}{16(U + 3J_H)} \left\{ 4\left[ t_3^2 t_4^2 + 2t_1^2 t_2^2 + \frac{1}{2}t_4^4 + \frac{1}{2}t_3^4 \right] A^2 - \frac{1}{2}(t_2^2 - t_1^2)^2 B^2 \right\} (\bm{S}_i \cdot \bm{S}_j) \\
    &\quad - \frac{1}{16(U + 3J_H)} \left[ (t_1^2 + t_2^2)(t_3^2 + t_4^2) - (t_2^2 - t_1^2)^2 \right] B^2 (\tilde{S}_i^z \tilde{S}_j^z) \\
    &\quad - \frac{1}{8(U + 3J_H)} t_1 t_2 (t_3^2 + t_4^2)^2 B^2 (\tilde{S}_i^x \tilde{S}_j^y + \tilde{S}_i^y \tilde{S}_j^x) \\
    &\quad + \frac{1}{16(U + 3J_H)} (t_1^2 - t_2^2)(t_3^2 + t_4^2) B^2 (\tilde{S}_i^x \tilde{S}_j^x - \tilde{S}_i^y \tilde{S}_j^y).
\end{aligned}
    \label{Eq-S26}
\end{equation}

However, the above results are expressed in the local coordinate system; the spin operators must now be rotated back to the global coordinate system. Based on the previously defined rotation matrix and the bond-direction definitions in the Kitaev-like bond-dependent model, the following relations are obtained:
\begin{equation}
    \begin{aligned}
        S^{\alpha} &= \cos\theta \tilde{S}^x - \sin\theta \tilde{S}^y \\
        S^{\beta} &= -\sin\theta \tilde{S}^x + \cos\theta \tilde{S}^y \\
        \quad S^{\gamma} &= \tilde{S}^z
    \end{aligned},
    \label{Eq-S27}
\end{equation}
where $S^{\alpha}, S^{\beta}$, and $S^{\gamma}$ correspond to the bond directions on the Fe-Te-Fe(-Te) plane in the Kitaev-like model, as shown in Fig.~\ref{FIG-s3}. Through this transformation, the correspondence between the spin components in the local coordinate system and those in the global coordinate system is given by:
\begin{equation}
    \tilde{S}_i^z \tilde{S}_j^z = S_i^{\gamma} S_j^{\gamma}, \quad \tilde{S}_i^x \tilde{S}_j^y + \tilde{S}_i^y \tilde{S}_j^x = (S_i^{\alpha} S_j^{\beta} + S_i^{\beta} S_j^{\alpha}) + \frac{1}{3}\bm{S}_i \cdot \bm{S}_j - \frac{1}{3} S_i^{\gamma} S_j^{\gamma},
    \label{Eq-S28}
\end{equation}
where the Kitaev-like interaction term is extracted from $\tilde{S}_i^z \tilde{S}_j^z$, the off-diagonal interaction term is extracted from $\tilde{S}_i^x \tilde{S}_j^y + \tilde{S}_i^y \tilde{S}_j^x$, and these also contain partial Heisenberg and Kitaev-like interaction contributions. This demonstrates that the Heisenberg interaction in our Hamiltonian originates not only from spin-conserving processes but also receives contributions from spin-flip processes. This is also evident from the expression for the second-nearest neighbors in Eq.~(\ref{Eq-S26}), where the Heisenberg interaction arising from spin flips is not solely from the $\tilde{S}_i^x \tilde{S}_j^y + \tilde{S}_i^y \tilde{S}_j^x$ term. For different planes, the results should be identical under the local coordinate system defined in Fig.~\ref{FIG-s3}, because the correspondence between $\tilde{x},\tilde{y},\tilde{z}$ and $\alpha,\beta,\gamma$ is the same under our coordinate selection. For the additional $(\tilde{S}_i^x \tilde{S}_j^x - \tilde{S}_i^y \tilde{S}_j^y)$ term in the 2NN case, which is independent of the bond direction, this term vanishes upon transformation to the global coordinate system and summation over different coordinate selections, and thus does not appear in the total Hamiltonian.

Substituting Eq.~(\ref{Eq-S28}) into Eqs.~(\ref{Eq-S25}) and~(\ref{Eq-S26}) yields the Hamiltonian expressions in the global coordinate system. For the 1NN:
\begin{equation}
\begin{aligned}
    H_{s_1}^{t_2-t_2} &= \frac{1}{4(U + 3J_H)} \left\{ \left[ 8t_1^2 t_2^2 + 4t_3^2 t_4^2 + (t_3^2 + t_4^2)^2 \right] A^2 - \frac{1}{3}(t_1 t_3 + t_2 t_4)(t_1 t_4 + t_2 t_3) B^2 \right\} (\bm{S}_i \cdot \bm{S}_j) \\
    &\quad - \frac{1}{8(U + 3J_H)} \left[ (t_1 t_3 + t_2 t_4)^2 + (t_1 t_4 + t_2 t_3)^2 - \frac{2}{3}(t_1 t_3 + t_2 t_4)(t_1 t_4 + t_2 t_3) \right] B^2 (S_i^{\gamma} S_j^{\gamma}) \\
    &\quad - \frac{1}{4(U + 3J_H)} (t_1 t_3 + t_2 t_4)(t_1 t_4 + t_2 t_3) B^2 (S_i^{\alpha} S_j^{\beta} + S_i^{\beta} S_j^{\alpha}),
\end{aligned}
    \label{Eq-S29}
\end{equation}
and for the second-nearest neighbors:
\begin{equation}
\begin{aligned}
    H_{s_2}^{t_2-t_2} &= \frac{1}{16(U + 3J_H)} \left\{ 2\left[ 4t_1^2 t_2^2 + (t_3^2 + t_4^2)^2 \right] A^2 - \left[ \frac{1}{2}(t_2^2 - t_1^2)^2 + \frac{2}{3}t_1 t_2 (t_3^2 + t_4^2) \right] B^2 \right\} (\bm{S}_i \cdot \bm{S}_j) \\
    &\quad - \frac{1}{16(U + 3J_H)} \left[ (t_1^2 + t_2^2)(t_3^2 + t_4^2) - (t_2^2 - t_1^2)^2 - \frac{2}{3}t_1 t_2 (t_3^2 + t_4^2) \right] B^2 (S_i^{\gamma} S_j^{\gamma}) \\
    &\quad - \frac{1}{8(U + 3J_H)} t_1 t_2 (t_3^2 + t_4^2)^2 B^2 (S_i^{\alpha} S_j^{\beta} + S_i^{\beta} S_j^{\alpha}).
\end{aligned}
    \label{Eq-S30}
\end{equation}

Here, the Heisenberg interaction coefficient corresponds to the coefficient of the $\bm{S}_i \cdot \bm{S}_j$ term, the Kitaev-like interaction coefficient corresponds to the coefficient of the $S_i^{\gamma} S_j^{\gamma}$ term, and the off-diagonal interaction coefficient corresponds to the coefficient of the $S_i^{\alpha} S_j^{\beta} + S_i^{\beta} S_j^{\alpha}$ term, clearly verifying the Kitaev-like bond-dependent model.

For the 2NN, inversion symmetry is broken, leading to the emergence of Dzyaloshinskii-Moriya (DM) interactions under SOC. In the $t_2$-$t_2$ exchange, the DM interaction originates from the $d_{xy}$-$d_{xy}$ coupling. Due to the symmetry breaking between $\text{Fe}_1$ and $\text{Fe}_2$, the Pauli matrix combinations $[\tilde{\sigma}_0]_{\tilde{\sigma}_4\tilde{\sigma}_3}[\tilde{\sigma}_{\gamma}]_{\tilde{\sigma}_2\tilde{\sigma}_1}$ and $[\tilde{\sigma}_{\gamma}]_{\tilde{\sigma}_4\tilde{\sigma}_3}[\tilde{\sigma}_0]_{\tilde{\sigma}_2\tilde{\sigma}_1}$ do not cancel across different pathways but are instead retained. These combinations correspond to the DM interaction on different planes, namely:
\begin{equation}
    \begin{aligned}
    [\tilde{\sigma}_0]_{\tilde{\sigma}_4\tilde{\sigma}_3}[\tilde{\sigma}_x]_{\tilde{\sigma}_2\tilde{\sigma}_1}
    ([\tilde{\bm{\sigma}}]_{\tilde{\sigma}_1\tilde{\sigma}_4} \cdot \tilde{\bm{s}}_{i,\alpha})
    ([\tilde{\bm{\sigma}}]_{\tilde{\sigma}_3\tilde{\sigma}_2} \cdot \tilde{\bm{s}}_{j,\beta})
    &= \frac{i}{8}(\tilde{S}_i^y \tilde{S}_j^z - \tilde{S}_i^z \tilde{S}_j^y), \\
    [\tilde{\sigma}_0]_{\tilde{\sigma}_4\tilde{\sigma}_3}[\tilde{\sigma}_y]_{\tilde{\sigma}_2\tilde{\sigma}_1}
    ([\tilde{\bm{\sigma}}]_{\tilde{\sigma}_1\tilde{\sigma}_4} \cdot \tilde{\bm{s}}_{i,\alpha})
    ([\tilde{\bm{\sigma}}]_{\tilde{\sigma}_3\tilde{\sigma}_2} \cdot \tilde{\bm{s}}_{j,\beta})
    &= \frac{i}{8}(\tilde{S}_i^z \tilde{S}_j^x - \tilde{S}_i^x \tilde{S}_j^z), \\
    [\tilde{\sigma}_0]_{\tilde{\sigma}_4\tilde{\sigma}_3}[\tilde{\sigma}_z]_{\tilde{\sigma}_2\tilde{\sigma}_1}
    ([\tilde{\bm{\sigma}}]_{\tilde{\sigma}_1\tilde{\sigma}_4} \cdot \tilde{\bm{s}}_{i,\alpha})
    ([\tilde{\bm{\sigma}}]_{\tilde{\sigma}_3\tilde{\sigma}_2} \cdot \tilde{\bm{s}}_{j,\beta})
    &= \frac{i}{8}(\tilde{S}_i^x \tilde{S}_j^y - \tilde{S}_i^y \tilde{S}_j^x),
    \end{aligned}
    \label{Eq-S31}
\end{equation}
where $[\tilde{\sigma}_0]_{\tilde{\sigma}_4\tilde{\sigma}_3}[\tilde{\sigma}_{\gamma}]_{\tilde{\sigma}_2\tilde{\sigma}_1}$ and 
$[\tilde{\sigma}_{\gamma}]_{\tilde{\sigma}_4\tilde{\sigma}_3}[\tilde{\sigma}_0]_{\tilde{\sigma}_2\tilde{\sigma}_1}$ yield opposite results. The result in the local coordinate system is:
\begin{equation}
    H_D = -\frac{1}{4(U + 3J_H)} t_1 t_2 (t_2^2 - t_1^2) A B \,
    (\tilde{S}_i^x \tilde{S}_j^y - \tilde{S}_i^y \tilde{S}_j^x),
    \label{Eq-S32}
\end{equation}
which corresponds to the following result in the global coordinate system:
\begin{equation}
    H_D = -\frac{1}{4(U + 3J_H)} t_1 t_2 (t_2^2 - t_1^2)
    \bigl[(\mathbf{z} \times \mathbf{d}_{ij}) \cdot (\bm{S}_i \times \bm{S}_j)\bigr].
    \label{Eq-S33}
\end{equation}

In the global Hamiltonian, this is expressed as $\bm{D}_{ij} = D_2 (\mathbf{z} \times \mathbf{d}_{ij})$, where $D_2$ denotes the magnitude of the DM interaction parameter for the 2NN, $\mathbf{z}$ is the unit vector pointing from the Te atom toward the Fe layer, and $\mathbf{d}_{ij}$ is the unit vector pointing from $\text{Fe}_i$ toward $\text{Fe}_j$.

\subsubsection{\texorpdfstring{$t_2$–$e$ superexchange process}{t2–e superexchange process}}\label{Subsubsec-S2.4.2}
We next consider the hopping process in which a hole moves from the $t_2$ orbital of $\mathrm{Fe}_1$ to the $e$ orbital of $\mathrm{Fe}_2$. In this case, however, an orbital degeneracy arises. In the ideal tetrahedral lattice considered here, the $d_{z^2}$($d_{3z^2-r^2}$) and $d_{x^2-y^2}$ orbitals in the $e$ manifold are degenerate. Therefore, in the hole representation $d^4(t_2^3e^1)$, the single hole occupying the $e$ level retains an orbital degree of freedom. To account for this degree of freedom, we introduce an independent orbital pseudospin $\tau$. The occupations of the $d_{z^2}$ and $d_{x^2-y^2}$ orbitals can be written as
\begin{equation}
|d_{z^2}\rangle = |\tau^z = \frac{1}{2}\rangle, \quad |d_{x^2-y^2}\rangle = |\tau^z = -\frac{1}{2}\rangle, \quad \tau^z = \frac{1}{2}(n_z - n_x),
\label{Eq-S34}
\end{equation}
where $\tau^z$ denotes the $z$ component of the orbital pseudospin, and $n_z$ and $n_x$ are the occupation numbers of the $d_{z^2}$ and $d_{x^2-y^2}$ orbitals, respectively. The orbital occupation can then be identified from the following pseudospin states:
\begin{equation}
    \begin{aligned}
|d_{z^2}\rangle\langle d_{z^2}| &= \frac{1}{2} + \tau^z, \\
|d_{x^2-y^2}\rangle\langle d_{x^2-y^2}| &= \frac{1}{2} - \tau^z.
\end{aligned}
\label{Eq-S35}
\end{equation}

In the $t_2$-$e$ superexchange process, several intermediate states appear. We briefly discuss them within the Kanamori Hamiltonian. There are three relevant classes of intermediate states. In the first case, a hole hops from the $t_2$ orbital of $\mathrm{Fe}_1$ to an empty $e$ orbital of $\mathrm{Fe}_2$, with its spin parallel to the spin already present on $\mathrm{Fe}_2$. The corresponding energy difference between the intermediate state and the ground state is $\Delta E_{t_2-e}^1 = U' - J_H + \Delta_c$. In the second case, the hole hops from the $t_2$ orbital of $\mathrm{Fe}_1$ to an already occupied $e$ orbital of $\mathrm{Fe}_2$. The $e$ manifold then contains one doubly occupied orbital and one empty orbital, so that the pair-hopping term in the Kanamori interaction becomes relevant. The corresponding intermediate-state energies are $\Delta E_{t_2-e}^2 = U + 4J_H + \Delta_c$, $\Delta E_{t_2-e}^{2'} = U + 2J_H + \Delta_c$, which correspond to the antisymmetric and symmetric orbital states, respectively, each carrying a weight of $1/2$. In the third case, the hole hops from the $t_2$ orbital of $\mathrm{Fe}_1$ to an empty $e$ orbital of $\mathrm{Fe}_2$, but with spin antiparallel to the original spin. This gives rise to high-spin and low-spin intermediate multiplets with energies $\Delta E_{t_2-e}^3 = U' - J_H + \Delta_c$, $\Delta E_{t_2-e}^{3'} = U' + 4J_H + \Delta_c$, with weights $1/5$ and $4/5$, respectively.
\begin{figure}[!ht]
\centering
\includegraphics[width=0.85\columnwidth, clip]{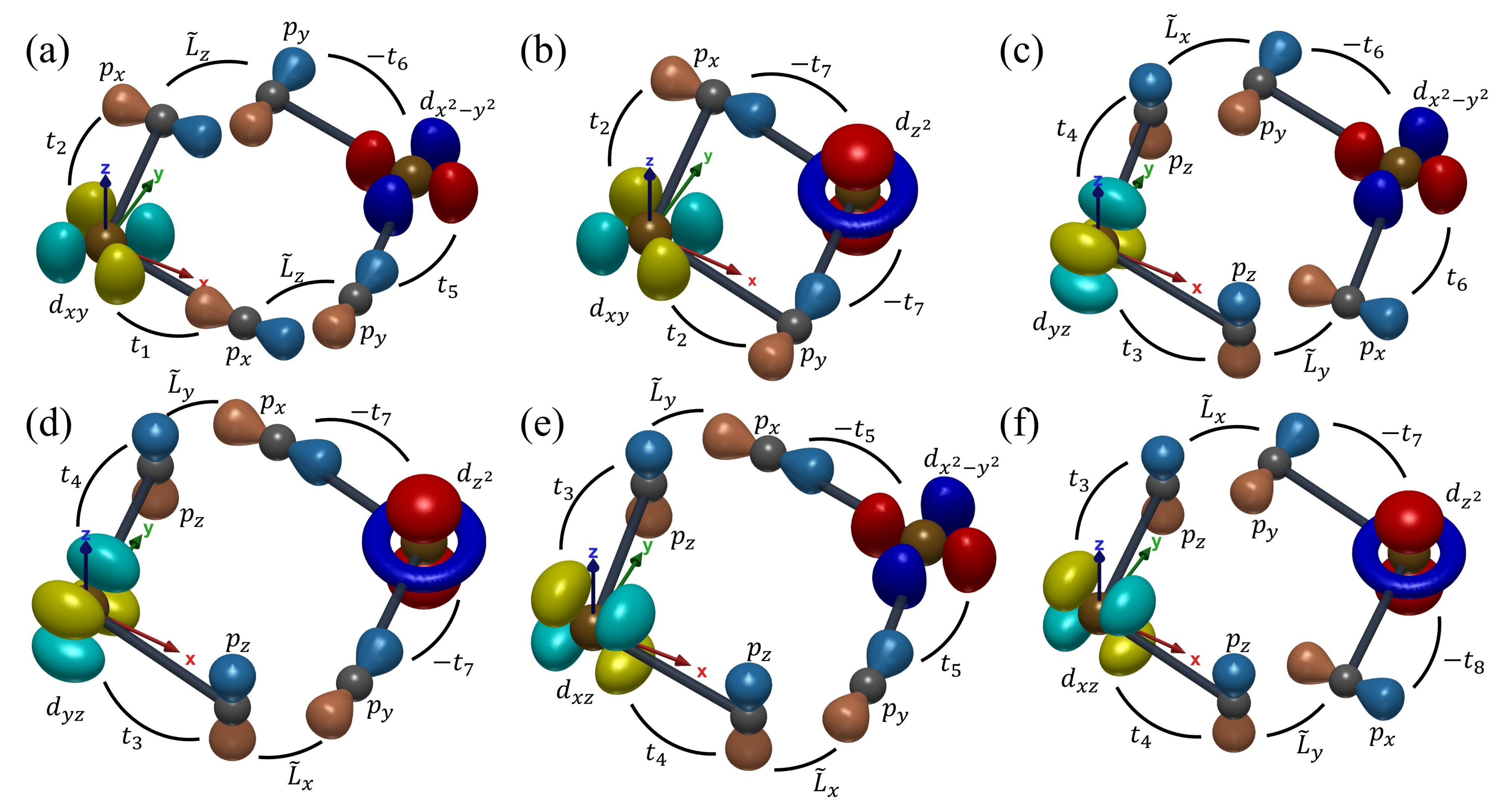}\\
\caption{Schematic of the $t_2$-$e$ virtual hopping processes in the hole representation. Brown and dark-gray spheres denote Fe and Te atoms, respectively. Only representative virtual hopping paths that contribute to the 1NN $t_2$-$e$ effective hopping integrals are shown.
  }\label{FIG-s5}
\end{figure}  

Since the evaluation of the $t_2$-$e$ superexchange processes proceeds in close analogy to the $t_2$-$t_2$ case, apart from the inclusion of the orbital pseudospin and the modified intermediate-state energies, we do not repeat the detailed derivation here. Representative hopping paths are shown in Fig.~\ref{FIG-s5}. Below we present only the resulting effective hopping integrals and the corresponding effective spin Hamiltonians.

The 1NN effective hopping integrals for the $t_2-e$ channel are
\begin{equation}
T_{M_2 M_1}^{\text{eff}}(t_2 \otimes e) = 
\begin{pmatrix}
i(t_1 t_5 - t_2 t_6) B \widetilde{\sigma}_z & -2(t_1 t_8 + t_2 t_7) A \widetilde{\sigma}_0 \\[8pt]
i \dfrac{1}{2}(t_3 t_6 - t_4 t_5) B \widetilde{\sigma}_y + i \dfrac{1}{2}(t_4 t_6 - t_3 t_5) B \widetilde{\sigma}_x & -i \dfrac{1}{2}(t_3 t_8 + t_4 t_7) B \widetilde{\sigma}_y + i \dfrac{1}{2}(t_3 t_7 + t_4 t_8) B \widetilde{\sigma}_x \\[8pt]
i \dfrac{1}{2}(t_4 t_6 - t_3 t_5) B \widetilde{\sigma}_y + i \dfrac{1}{2}(t_3 t_6 - t_4 t_5) B \widetilde{\sigma}_x & i \dfrac{1}{2}(t_3 t_8 + t_4 t_7) B \widetilde{\sigma}_x - i \dfrac{1}{2}(t_3 t_7 + t_4 t_8) B \widetilde{\sigma}_y
\end{pmatrix}.
\label{Eq-S36}
\end{equation}

The 2NN effective hopping integrals for the $t_2$-$e$ channel are
\begin{equation}
T_{M_2' M_1}^{\text{eff}}(t_2 \otimes e) = 
\begin{pmatrix}
(t_2 t_5 + t_1 t_6) A \widetilde{\sigma}_0 + i \dfrac{1}{2}(t_2 t_6 - t_1 t_5) B \widetilde{\sigma}_z & (t_2 t_7 + t_1 t_8) A \widetilde{\sigma}_0 + i \dfrac{1}{2}(t_2 t_8 - t_1 t_7) B \widetilde{\sigma}_z \\[8pt]
i \dfrac{1}{2} t_4 t_5 B \widetilde{\sigma}_y - i \dfrac{1}{2} t_4 t_6 B \widetilde{\sigma}_x & i \dfrac{1}{2} t_4 t_7 B \widetilde{\sigma}_y - i \dfrac{1}{2} t_4 t_8 B \widetilde{\sigma}_x \\[8pt]
i \dfrac{1}{2} t_3 t_5 B \widetilde{\sigma}_y - i \dfrac{1}{2} t_3 t_6 B \widetilde{\sigma}_x & i \dfrac{1}{2} t_3 t_7 B \widetilde{\sigma}_y - i \dfrac{1}{2} t_3 t_8 B \widetilde{\sigma}_x
\end{pmatrix}.
\label{Eq-S37}
\end{equation}

Substituting these effective hopping integrals into the superexchange expression gives the effective spin Hamiltonian in the local coordinate frame. After rotating back to the global coordinate frame, the 1NN effective spin Hamiltonian becomes
\begin{equation}
\begin{aligned}
H_{s_1}^{t_2-e} &= \frac{1}{8} \left\{ \left[ \frac{1}{3}(t_3 t_6 - t_4 t_5)(t_4 t_6 - t_3 t_5) - (t_1 t_5 - t_2 t_6) \right] B^2 \Delta_{t_2-e}^1 \right. \\
&\quad + \left. \left[ 4(t_1 t_8 + t_2 t_7)^2 A^2 - \frac{1}{3}(t_3 t_8 + t_4 t_7)(t_3 t_7 + t_4 t_8) B^2 \right] \Delta_{t_2-e}^2 \right\} (\mathbf{S}_i \cdot \mathbf{S}_j) \\
&\quad + \frac{1}{16} \left\{ \left[ 2(t_1 t_5 - t_2 t_6)^2 - (t_3 t_6 - t_4 t_5)^2 - (t_4 t_6 - t_3 t_5)^2 + \frac{2}{3}(t_3 t_6 - t_4 t_5)(t_4 t_6 - t_3 t_5) \right] \Delta_{t_2-e}^1 \right. \\
&\quad - \left. \left[ (t_3 t_8 + t_4 t_7)^2 + (t_3 t_7 + t_4 t_8)^2 - \frac{2}{3}(t_3 t_8 + t_4 t_7)(t_3 t_7 + t_4 t_8) \right] \Delta_{t_2-e}^2 \right\} B^2 S_i^\gamma S_j^\gamma \\
&\quad + \frac{1}{8} \left[ (t_3 t_6 - t_4 t_5)(t_4 t_6 - t_3 t_5) \Delta_{t_2-e}^1 - (t_3 t_8 + t_4 t_7)(t_3 t_7 + t_4 t_8) \Delta_{t_2-e}^2 \right] B^2 (S_i^\alpha S_j^\beta + S_i^\beta S_j^\alpha).
\end{aligned}
\label{Eq-S38}
\end{equation}

The corresponding 2NN effective spin Hamiltonian is
\begin{equation}
\begin{aligned}
H_{s_2}^{t_2-e} &= \frac{1}{16} \left\{ \left[ 2(t_2 t_5 + t_1 t_6)^2 A^2 - \frac{1}{2}(t_2 t_6 - t_1 t_5) B^2 - \frac{1}{3}(t_3^2 + t_4^2) t_5 t_6 B^2 \right] \Delta_{t_2-e}^1 \right. \\
&\quad + \left. \left[ 2(t_2 t_7 + t_1 t_8)^2 A^2 - \frac{1}{2}(t_2 t_8 - t_1 t_7) B^2 - \frac{1}{3}(t_3^2 + t_4^2) t_7 t_8 B^2 \right] \Delta_{t_2-e}^2 \right\} (\mathbf{S}_i \cdot \mathbf{S}_j) \\
&\quad + \frac{1}{16} \left\{ \left[ (t_2 t_6 - t_1 t_5)^2 - \frac{1}{2}(t_3^2 + t_4^2)(t_5^2 + t_6^2) + \frac{1}{3}(t_3^2 + t_4^2) t_5 t_6 \right] B^2 \Delta_{t_2-e}^1 \right. \\
&\quad + \left. \left[ (t_2 t_8 - t_1 t_7)^2 - \frac{1}{2}(t_3^2 + t_4^2)(t_7^2 + t_8^2) + \frac{1}{3}(t_3^2 + t_4^2) t_7 t_8 \right] B^2 \Delta_{t_2-e}^2 \right\} S_i^\gamma S_j^\gamma \\
&\quad - \frac{1}{16}(t_3^2 + t_4^2)(t_5 t_6 \Delta_{t_2-e}^1 + t_7 t_8 \Delta_{t_2-e}^2) (S_i^\alpha S_j^\beta + S_i^\beta S_j^\alpha) \\
&\quad - \frac{1}{8} \left[ (t_2 t_5 + t_1 t_6)(t_2 t_6 - t_1 t_5) \Delta_{t_2-e}^1 + (t_2 t_7 + t_1 t_8)(t_2 t_8 - t_1 t_7) \Delta_{t_2-e}^2 \right] AB \left[ (\mathbf{z} \times \mathbf{d}_{ij}) \cdot (\mathbf{S}_i \times \mathbf{S}_j) \right].
\end{aligned}
\label{Eq-S39}
\end{equation}

In Eqs.~(\ref{Eq-S38}) and~(\ref{Eq-S39}), we have incorporated the orbital degree of freedom and the multiple intermediate states into $\Delta_{t_2-e}^1$ and $\Delta_{t_2-e}^2$. Their explicit forms are
\begin{equation}
    \begin{aligned}
\Delta_{t_2-e}^1 &= \frac{\frac{1}{2} + \tau_j^z}{U' - J_H + \Delta_c} + \frac{(\frac{1}{2} - \tau_j^z)(U + 3J_H + \Delta_c)}{(U + 4J_H + \Delta_c)(U + 2J_H + \Delta_c)} + \frac{(\frac{1}{2} + \tau_j^z)(U' + \Delta_c)}{(U' + 4J_H + \Delta_c)(U' - J_H + \Delta_c)}, \\
\Delta_{t_2-e}^2 &= \frac{\frac{1}{2} - \tau_j^z}{U' - J_H + \Delta_c} + \frac{(\frac{1}{2} + \tau_j^z)(U + 3J_H + \Delta_c)}{(U + 4J_H + \Delta_c)(U + 2J_H + \Delta_c)} + \frac{(\frac{1}{2} - \tau_j^z)(U' + \Delta_c)}{(U' + 4J_H + \Delta_c)(U' - J_H + \Delta_c)}.
\end{aligned}
\label{Eq-S40}
\end{equation}
The expressions above give the effective hopping integrals and effective spin Hamiltonians arising from the $t_2$-$e$ superexchange processes in the hole representation. They show that the $t_2$-$e$ channel contributes not only to the Heisenberg interaction, but also to the Kitaev-like interaction, the off-diagonal $\Gamma$ interaction, and the DM interaction.

\subsubsection{\texorpdfstring{$e$-$t_2$ superexchange process}{e-t2 superexchange process}}\label{Subsubsec-S2.4.3}
We next consider the hole hopping process from the $e$ orbital of $\mathrm{Fe}_1$ to the $t_2$ orbital of $\mathrm{Fe}_2$. The orbital degree of freedom must again be taken into account. In this case, however, only a single intermediate state exists: the hole in the $e$ orbital of $\mathrm{Fe}_1$ hops to the $t_2$ orbital of $\mathrm{Fe}_2$. The energy difference between this intermediate state and the ground state is $\Delta E_{e-t_2} = U + 3J_H - \Delta_c$. We again omit the detailed derivation; representative virtual hopping paths are shown in Fig.~\ref{FIG-s5}, and only the resulting effective hopping integrals and effective spin Hamiltonians are presented below.
The 1NN effective hopping integral for the $e$-$t_2$ channel is
\begin{equation}
    \begin{matrix}
T_{M_2 M_1}^{\text{eff}}(e \otimes t_2) = 
\begin{pmatrix}
i(t_2 t_6 - t_1 t_5) B \widetilde{\sigma}_z & i\dfrac{1}{2}(t_4 t_5 - t_3 t_6)B\widetilde{\sigma}_y + i\dfrac{1}{2}(t_3 t_5 - t_4 t_6)B\widetilde{\sigma}_x & T_{M_2 M_1, 13}^{\text{eff}} \\
-2(t_1 t_8 + t_2 t_7) A \widetilde{\sigma}_0 & i\dfrac{1}{2}(t_3 t_8 + t_4 t_7)B\widetilde{\sigma}_y - i\dfrac{1}{2}(t_3 t_7 + t_4 t_8)B\widetilde{\sigma}_x & T_{M_2 M_1, 23}^{\text{eff}}
\end{pmatrix}, \\
T_{M_2 M_1, 13}^{\text{eff}}(e \otimes t_2) = i\frac{1}{2}(t_3 t_5 - t_4 t_6) B \widetilde{\sigma}_y + i\frac{1}{2}(t_4 t_5 - t_3 t_6) B \widetilde{\sigma}_x, \\
T_{M_2 M_1, 23}^{\text{eff}}(e \otimes t_2) = -i\frac{1}{2}(t_3 t_8 + t_4 t_7) B \widetilde{\sigma}_x + i\frac{1}{2}(t_3 t_7 + t_4 t_8) B \widetilde{\sigma}_y.
\end{matrix}
\label{Eq-S41}
\end{equation}

The 2NN effective hopping integral for the $e$-$t_2$ channel is
\begin{equation}
\begin{matrix}
T_{M_2' M_1}^{\text{eff}}(e \otimes t_2) = 
\begin{pmatrix}
-(t_1 t_5 + t_2 t_6) A \widetilde{\sigma}_0 - i\dfrac{1}{2}(t_1 t_6 - t_2 t_5) B \widetilde{\sigma}_z & i\dfrac{1}{2} t_3 t_6 B \widetilde{\sigma}_y - i\dfrac{1}{2} t_3 t_5 B \widetilde{\sigma}_x & T_{M_2' M_1, 13}^{\text{eff}} \\
(t_1 t_8 + t_2 t_7) A \widetilde{\sigma}_0 + i\dfrac{1}{2}(t_2 t_8 - t_1 t_7) B \widetilde{\sigma}_z & -i\dfrac{1}{2} t_3 t_8 B \widetilde{\sigma}_y + i\dfrac{1}{2} t_3 t_7 B \widetilde{\sigma}_x & T_{M_2' M_1, 23}^{\text{eff}}
\end{pmatrix}, \\
T_{M_2' M_1, 13}^{\text{eff}}(e \otimes t_2) = i\frac{1}{2} t_4 t_6 B \widetilde{\sigma}_y - i\frac{1}{2} t_4 t_5 B \widetilde{\sigma}_x, \\
T_{M_2' M_1, 23}^{\text{eff}}(e \otimes t_2) = -i\frac{1}{2} t_4 t_8 B \widetilde{\sigma}_y + i\frac{1}{2} t_4 t_7 B \widetilde{\sigma}_x.
\end{matrix}
\label{Eq-S42}
\end{equation}

Substituting these effective hopping integrals into the superexchange expression yields the effective spin Hamiltonian for the $e$-$t_2$ channel. The 1NN effective spin Hamiltonian is
\begin{equation}
\begin{aligned}
H_{s_1}^{e-t_2} &= \left\{ \left[ \frac{1}{3}(t_3 t_6 - t_4 t_5)(t_4 t_6 - t_3 t_5) - (t_1 t_5 - t_2 t_6)^2 \right] B^2 (1 - \tau_i^z) \right.  \\
&\quad \left. + \left[ 4(t_1 t_8 + t_2 t_7)^2 A^2 - \frac{1}{3}(t_3 t_8 + t_4 t_7)(t_3 t_7 + t_4 t_8) B^2 \right] (\frac{1}{2} + \tau_i^z) \right\} \frac{(\mathbf{S}_i \cdot \mathbf{S}_j)}{8(U + 3J_H - \Delta_c)}  \\
&\quad + \left\{ \left[ 2(t_1 t_5 - t_2 t_6)^2 - (t_3 t_6 - t_4 t_5)^2 - (t_4 t_6 - t_3 t_5)^2 - \frac{2}{3}(t_3 t_6 - t_4 t_5)(t_4 t_6 - t_3 t_5) \right] (\frac{1}{2} - \tau_i^z) \right.  \\
&\quad \left. - \left[ (t_3 t_8 + t_4 t_7)^2 + (t_3 t_7 + t_4 t_8)^2 - \frac{2}{3}(t_3 t_8 + t_4 t_7)(t_3 t_7 + t_4 t_8) \right] (\frac{1}{2} + \tau_i^z) \right\} \frac{B^2 S_i^\gamma S_j^\gamma}{16(U + 3J_H - \Delta_c)}  \\
&\quad + \left[ (t_3 t_6 - t_4 t_5)(t_4 t_6 - t_3 t_5) (\frac{1}{2} - \tau_i^z) - (t_3 t_8 + t_4 t_7)(t_3 t_7 + t_4 t_8) (\frac{1}{2} + \tau_i^z) \right] \frac{B^2 (S_i^\alpha S_j^\beta + S_i^\beta S_j^\alpha)}{8(U + 3J_H - \Delta_c)}.
\end{aligned}
\label{Eq-S43}
\end{equation}

The corresponding 2NN effective spin Hamiltonian is
\begin{equation}
\begin{aligned}
H_{s_2}^{e-t_2} &= \left\{ \left[ 2(t_1 t_5 + t_2 t_6)^2 A^2 - \frac{1}{2}(t_1 t_6 - t_2 t_5)^2 B^2 - \frac{1}{3}(t_3^2 + t_4^2)t_5 t_6 B^2 \right] (\frac{1}{2} - \tau_i^z) \right.  \\
&\quad \left. + \left[ 2(t_1 t_8 + t_2 t_7)^2 A^2 - \frac{1}{2}(t_2 t_8 - t_1 t_7)^2 B^2 - \frac{1}{3}(t_3^2 + t_4^2)t_7 t_8 B^2 \right] (\frac{1}{2} + \tau_i^z) \right\} \frac{\mathbf{S}_i \cdot \mathbf{S}_j}{16(U + 3J_H - \Delta_c)}  \\
&\quad + \left\{ \left[ (t_1 t_6 - t_2 t_5)^2 - \frac{1}{2}(t_3^2 + t_4^2)(t_5^2 + t_6^2) + \frac{1}{3}(t_3^2 + t_4^2) t_5 t_6 \right] (\frac{1}{2} - \tau_i^z) \right.  \\
&\quad \left. + \left[ (t_2 t_8 - t_1 t_7)^2 - \frac{1}{2}(t_3^2 + t_4^2)(t_7^2 + t_8^2) + \frac{1}{3}(t_3^2 + t_4^2)t_7 t_8 \right] (\frac{1}{2} + \tau_i^z) \right\} \frac{B^2 S_i^\gamma S_j^\gamma}{16(U + 3J_H - \Delta_c)}  \\
&\quad - (t_3^2 + t_4^2) \left[ t_5 t_6 (\frac{1}{2} - \tau_i^z) + t_7 t_8 (\frac{1}{2} + \tau_i^z) \right] \frac{B^2 (S_i^\alpha S_j^\beta + S_i^\beta S_j^\alpha)}{16(U + 3J_H - \Delta_c)}  \\
&\quad - \left[ (t_1 t_5 + t_2 t_6)(t_1 t_6 - t_2 t_5) (\frac{1}{2} - \tau_i^z) + (t_1 t_8 + t_2 t_7)(t_2 t_8 - t_1 t_7) (\frac{1}{2} + \tau_i^z) \right] \frac{AB [(\mathbf{z} \times \mathbf{d}_{ij}) \cdot (\mathbf{S}_i \times \mathbf{S}_j)]}{8(U + 3J_H - \Delta_c)}.
\end{aligned}
\label{Eq-S44}
\end{equation}

The expressions above give the effective hopping integrals and effective spin Hamiltonians arising from the $e$-$t_2$ superexchange processes in the hole representation. As the result, the $e$-$t_2$ channel contributes to the Heisenberg interaction, the Kitaev-like interaction, the off-diagonal $\Gamma$ interaction, and the DM interaction.

\subsubsection{\texorpdfstring{$e$-$e$ superexchange process}{e-e superexchange process}}\label{Subsubsec-S2.4.4}
The final contribution arises from the superexchange process between the $e$ orbitals of $\mathrm{Fe}_1$ and $\mathrm{Fe}_2$. Since the orbital degrees of freedom on both $\mathrm{Fe}$ sites must be retained and several intermediate states are involved, the resulting parameters of effective spin Hamiltonian is more intricate than those obtained for other channels.

\begin{figure}[!ht]
\centering
\includegraphics[width=0.65\columnwidth, clip]{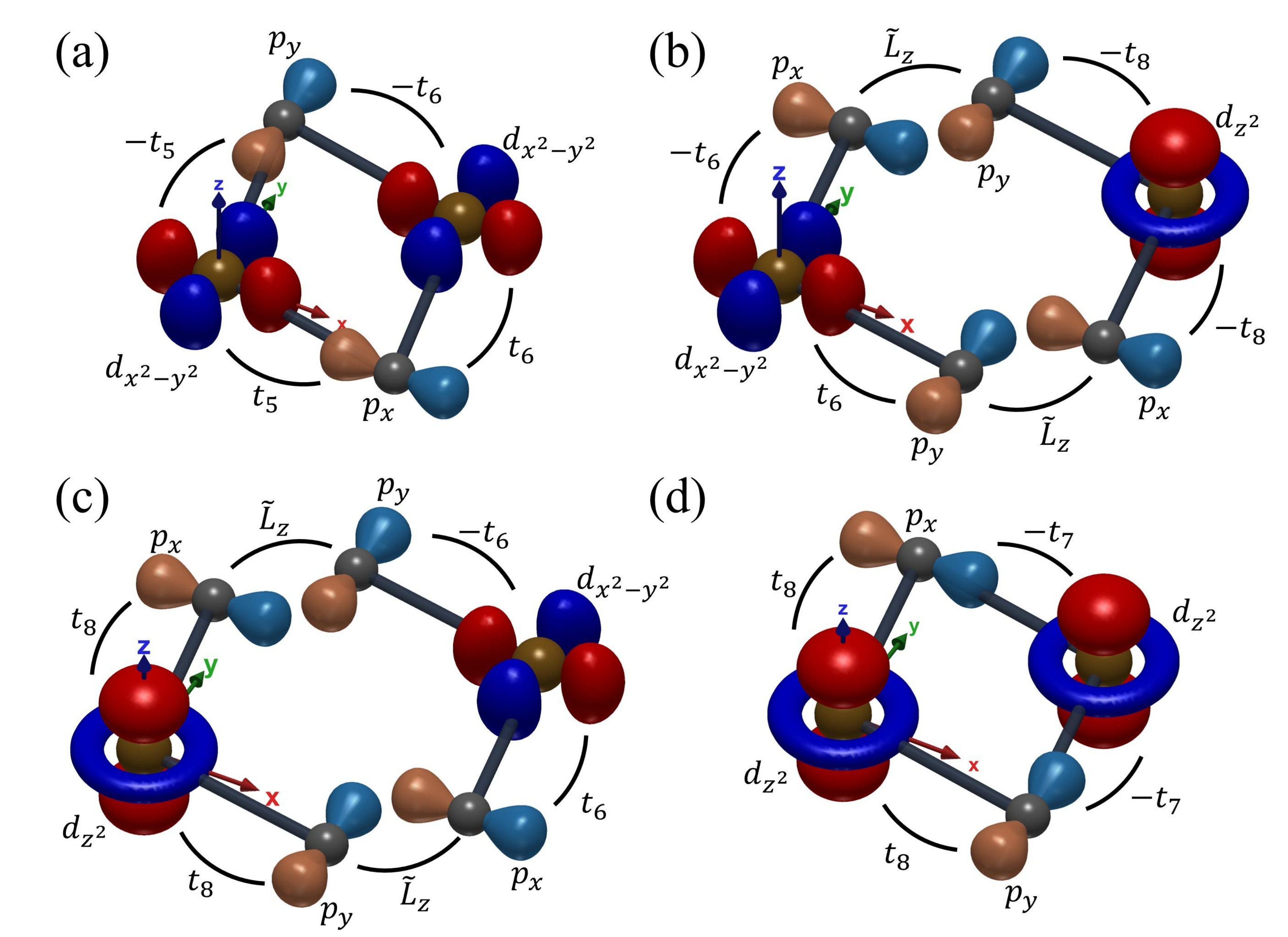 }\\
\caption{Schematic of the $e$-$e$ virtual hopping processes in the hole representation. Brown and dark-gray spheres denote $\mathrm{Fe}$ and $\mathrm{Te}$ atoms, respectively. Only representative virtual hopping paths that contribute to the 1NN $e$-$e$ effective hopping integrals are shown.}
\label{FIG-s6}
\end{figure}

We first specify the relevant intermediate states. Again, three classes of intermediate states occur. In the first case, a hole hops from an $e$ orbital of $\mathrm{Fe}_1$ to an empty $e$ orbital of $\mathrm{Fe}_2$, with its spin parallel to the spin of the hole initially occupying the other $e$ orbital on $\mathrm{Fe}_2$. The corresponding excitation energy is $\Delta E_{e-e}^1 = U' - J_H$. In the second case, a hole hops from an $e$ orbital of $\mathrm{Fe}_1$ to an already occupied $e$ orbital of $\mathrm{Fe}_2$. This process generates a doubly occupied orbital and an empty orbital within the $e$ manifold. The pair-hopping term in the Kanamori interaction then mixes these configurations, producing antisymmetric and symmetric orbital states with excitation energies $\Delta E_{e-e}^2 = U + 4J_H, \Delta E_{e-e}^{2'} = U + 2J_H$, respectively, each with weight $1/2$. In the third case, a hole hops from an $e$ orbital of $\mathrm{Fe}_1$ to an empty $e$ orbital of $\mathrm{Fe}_2$ with spin antiparallel to the original spin. This process gives rise to high-spin and low-spin intermediate multiplets, with excitation energies $\Delta E_{e-e}^3 = U' - J_H, \Delta E_{e-e}^{3'} = U' + 4J_H$ and weights $1/5$ and $4/5$, respectively. The corresponding $e$-$e$ virtual hopping processes are illustrated in Fig.~\ref{FIG-s6}.
From the $e$-$e$ virtual hopping paths, the effective hopping integrals can be obtained. For 1NN pairs, the result is
\begin{equation}
T_{M_2 M_1}^{\text{eff}}(e \otimes e) = 
\begin{pmatrix}
4t_5 t_6 A \sigma_0 & -i(t_5 t_7 - t_6 t_8) B \sigma_z \\
i(t_5 t_7 - t_6 t_8) B \sigma_z & -4t_7 t_8 A \sigma_0
\end{pmatrix}.
\label{Eq-S45}
\end{equation}

For 2NN pairs, the effective hopping integral is
\begin{equation}
\begin{aligned}
T_{M'_2 M_1}^{\text{eff}}(e \otimes e) = \begin{pmatrix}
-2t_5 t_6 A \sigma_0 + i \dfrac{1}{2}(t_5^2 - t_6^2) B \sigma_z & -(t_5 t_8 + t_6 t_7) A \sigma_0 + i \dfrac{1}{2}(t_5 t_7 - t_6 t_8) B \sigma_z \\[8pt]
(t_5 t_8 + t_6 t_7) A \sigma_0 + i \dfrac{1}{2}(t_6 t_8 - t_5 t_7) B \sigma_z & 2t_7 t_8 A \sigma_0 + i \dfrac{1}{2}(t_8^2 - t_7^2) B \sigma_z
\end{pmatrix}.
\end{aligned}
\label{Eq-S46}
\end{equation}

Substituting the effective hopping integrals into the superexchange expression yields the 1NN effective spin Hamiltonian,
\begin{equation}
\begin{aligned}
H_{s_1}^{e-e} &= \frac{1}{8} \left[ 16(t_5^2 t_6^2 \Delta_{e-e}^1 + t_7^2 t_8^2 \Delta_{e-e}^4) A^2 + (t_5 t_7 - t_6 t_8)^2 (\Delta_{e-e}^2 + \Delta_{e-e}^3) B^2 \right] (\mathbf{S}_i \cdot \mathbf{S}_j) \\
&\quad - \frac{1}{4} (t_5 t_7 - t_6 t_8)^2 (\Delta_{e-e}^2 + \Delta_{e-e}^3) B^2 S_i^\gamma S_j^\gamma.
\end{aligned}
\label{Eq-S47}
\end{equation}

Similarly, the 2NN effective spin Hamiltonian is
\begin{equation}
\begin{aligned}
H_{s_2}^{e-e} &= \frac{1}{16} \left\{ \left[ 8t_5^2 t_6^2 A^2 + \frac{1}{2}(t_5^2 - t_6^2)^2 B^2 \right] \Delta_{e-e}^1 + \left[ 2(t_5 t_8 + t_6 t_7)^2 A^2 + \frac{1}{2}(t_5 t_7 - t_6 t_8)^2 B^2 \right] \Delta_{e-e}^2 \right. \\
&\quad + \left[ 2(t_5 t_8 + t_6 t_7)^2 A^2 + \frac{1}{2}(t_5 t_7 - t_6 t_8)^2 B^2 \right] \Delta_{e-e}^3 \\
&\quad + \left. \left[ 8t_7^2 t_8^2 A^2 + \frac{1}{2}(t_8^2 - t_7^2)^2 B^2 \right] \Delta_{e-e}^4 \right\} (\mathbf{S}_i \cdot \mathbf{S}_j) \\
&\quad - \frac{1}{16} \left[ (t_5^2 - t_6^2)^2 \Delta_{e-e}^1 + (t_5 t_7 - t_6 t_8)^2 (\Delta_{e-e}^2 + \Delta_{e-e}^3) + (t_8^2 - t_7^2)^2 \Delta_{e-e}^4 \right] B^2 S_i^\gamma S_j^\gamma \\
&\quad + \frac{1}{8} \left[ 2(t_5^2 - t_6^2) t_5 t_6 \Delta_{e-e}^1 + (t_5 t_8 + t_6 t_7)(t_5 t_7 - t_6 t_8) (\Delta_{e-e}^2 + \Delta_{e-e}^3) \right. \\
&\quad - \left. 2(t_8^2 - t_7^2) t_7 t_8 \Delta_{e-e}^4 \right] AB [(\mathbf{z} \times \mathbf{d}_{ij}) \cdot (\mathbf{S}_i \times \mathbf{S}_j)].
\end{aligned}
\label{Eq-S48}
\end{equation}

In Eqs.~(\ref{Eq-S47}) and ~(\ref{Eq-S48}), the orbital degrees of freedom of the $e$ orbitals on both $\mathrm{Fe}$ sites and the multiple intermediate states in the virtual hopping processes have been incorporated into $\Delta_{e-e}^1$, $\Delta_{e-e}^2$, $\Delta_{e-e}^3$, and $\Delta_{e-e}^4$. Their explicit forms are
\begin{equation}
\begin{aligned}
\Delta_{e-e}^1 &= \frac{(\frac{1}{2} - \tau_i^z)(\frac{1}{2} + \tau_j^z)}{U' - J_H} + \frac{(\frac{1}{2} - \tau_i^z)(\frac{1}{2} - \tau_j^z)(U + 3J_H)}{(U + 2J_H)(U + 4J_H)} + \frac{(\frac{1}{2} - \tau_i^z)(\frac{1}{2} + \tau_j^z) U'}{(U' - J_H)(U' + 4J_H)}, \\
\Delta_{e-e}^2 &= \frac{(\frac{1}{2} - \tau_i^z)(\frac{1}{2} - \tau_j^z)}{U' - J_H} + \frac{(\frac{1}{2} - \tau_i^z)(\frac{1}{2} + \tau_j^z)(U + 3J_H)}{(U + 2J_H)(U + 4J_H)} + \frac{(\frac{1}{2} - \tau_i^z)(\frac{1}{2} - \tau_j^z) U'}{(U' - J_H)(U' + 4J_H)}, \\
\Delta_{e-e}^3 &= \frac{(\frac{1}{2} + \tau_i^z)(\frac{1}{2} + \tau_j^z)}{U' - J_H} + \frac{(\frac{1}{2} + \tau_i^z)(\frac{1}{2} - \tau_j^z)(U + 3J_H)}{(U + 2J_H)(U + 4J_H)} + \frac{(\frac{1}{2} + \tau_i^z)(\frac{1}{2} + \tau_j^z) U'}{(U' - J_H)(U' + 4J_H)}, \\
\Delta_{e-e}^4 &= \frac{(\frac{1}{2} + \tau_i^z)(\frac{1}{2} - \tau_j^z)}{U' - J_H} + \frac{(\frac{1}{2} + \tau_i^z)(\frac{1}{2} + \tau_j^z)(U + 3J_H)}{(U + 2J_H)(U + 4J_H)} + \frac{(\frac{1}{2} + \tau_i^z)(\frac{1}{2} - \tau_j^z) U'}{(U' - J_H)(U' + 4J_H)}.
\end{aligned}
\label{Eq-S49}
\end{equation}

The above results constitute the effective spin Hamiltonian generated by the $e$-$e$ superexchange channel. This channel contributes to the Heisenberg interaction, the Kitaev-like interaction, and the DM interaction, while it does not generate an off-diagonal interaction $\Gamma$.

\subsection{Effective spin Hamiltonian of FeX}\label{Subsec-S2.5}
In Sec.~\ref{Subsec-S2.4}, we presented the superexchange processes between Fe and its 1NN and 2NN Fe sites mediated by the coordinating ligands X (X=Te, Se), together with the resulting effective spin Hamiltonians. As shown there, the 1NN interactions contain three types of magnetic couplings, namely, the Heisenberg, Kitaev-like, and off-diagonal exchange interactions, whereas the 2NN interactions additionally host a Dzyaloshinskii--Moriya (DM) term. The effective Hamiltonian for monolayer FeX can therefore be written as
\begin{equation}
H = \sum_{\langle i,j \rangle_{1,2}} \frac{1}{2} \left[ J_{ij}(\mathbf{S}_i \cdot \mathbf{S}_j) + K_{ij} S_i^\gamma S_j^\gamma + \Gamma_{ij}(S_i^\alpha S_j^\beta + S_i^\beta S_j^\alpha) \right] + \sum_{\langle i,j \rangle_2} \frac{1}{2} D_{ij} \left[ (\mathbf{z} \times \mathbf{d}_{ij}) \cdot (\mathbf{S}_i \times \mathbf{S}_j) \right],
\label{Eq-S50}
\end{equation}
where the terms represent the Heisenberg interaction $J_{ij}$, the Kitaev-like interaction $K_{ij}$, the off-diagonal interaction $\Gamma_{ij}$, and the DM interaction $D_{ij}$ with their respective coupling coefficients, incorporating a factor of $1/2$ to avoid double counting. The $\langle i, j \rangle$ represent Fe atom neighbor pairs, and $S$ is the normalized spin operator. The local spin quantization axes $\{\alpha, \beta, \gamma\}$ are the spin component directions of the planar basis in Fig.~(\textcolor{red}{1})(b),(d) and (e). Unlike the edge-sharing octahedral Kitaev interaction, $\alpha$ and $\beta$ form an angle of $109.47^\circ$ and are not perpendicular. $\mathbf{i}, \mathbf{j}, \mathbf{k}$ are the basis vectors of the Cartesian coordinate system.

The Kitaev-like and off-diagonal interactions originate from spin-flip virtual hopping processes~[\textcolor{green}{27}] in which the SOC of the ligand atoms participates in both the $\mathrm{Fe}_i \to \mathrm{Fe}_j$ and $\mathrm{Fe}_j \to \mathrm{Fe}_i$ hopping pathways. In contrast, the DM interaction arises from superexchange processes where SOC is active in only one of the two virtual hopping steps. The Heisenberg interaction receives contributions from both spin-conserving processes (without SOC) and spin-flip processes.

Furthermore, the contributions from different orbital channels vary. In the local coordinate frame, the $t_2$-$t_2$, $t_2$-$e$, and $e$-$t_2$ superexchange channels all contribute to the Heisenberg, Kitaev-like, off-diagonal, and DM interactions, whereas the $e$-$e$ channel does not contribute to the off-diagonal interaction.

When discussing magnetic anisotropy, studies often consider single-ion anisotropy (SIA). Previous investigations of FeX systems have largely been restricted to SIA as the sole source of magnetic anisotropy~[\textcolor{green}{27,37}]. To provide a complete description, we incorporate SIA into the Hamiltonian, which is expressed in its full form as
\begin{equation}
\begin{aligned}
H &= \sum_{\langle i,j \rangle_{1,2}} \frac{1}{2} \left[ J_{ij}(\mathbf{S}_i \cdot \mathbf{S}_j) + K_{ij} S_i^\gamma S_j^\gamma + \Gamma_{ij}(S_i^\alpha S_j^\beta + S_i^\beta S_j^\alpha) \right] \\
&\quad + \sum_{\langle i,j \rangle_2} \frac{1}{2} D_{ij} \left[ (\mathbf{z} \times \mathbf{d}_{ij}) \cdot (\mathbf{S}_i \times \mathbf{S}_j) \right] + \sum_i A_k [1 - (\mathbf{S}_i \cdot \mathbf{k})^2],
\end{aligned}
\label{Eq-S51}
\end{equation}
where the SIA term is denoted by $A_k$. In the following, the anisotropy parameters in the Hamiltonian will be extracted via density functional theory (DFT) calculations.

In the effective spin Hamiltonian, since the various anisotropic interactions and the Heisenberg interaction fundamentally arise from combinations of different spin components, perturbation theory allows us to cast the Hamiltonian in the general tensor form:
\begin{equation}
H_{s}(i,j) = (\tilde{S}_i^\mu)^T M_{ij}^{\mu\nu} \tilde{S}_j^\nu,
\label{Eq-S52}
\end{equation}
where $M_{ij}^{\mu\nu}$ represents the exchange tensor of the spin Hamiltonian between magnetic atoms $i$ and $j$. However, for bond-dependent Kitaev-like interactions in the 2D edge-sharing tetrahedral structure, the bond directions are not mutually orthogonal, preventing the use of a single unified Cartesian coordinate system. Having derived the spin-exchange expressions in the local coordinate frame via perturbation theory, we can explicitly write the exchange tensor for the 1NN Fe pairs as
\begin{equation}
\widetilde{M}_1 = 
\begin{pmatrix}
J_1 - \dfrac{1}{3}\Gamma_1 & \Gamma_1 & \dfrac{\sqrt{6}}{2}\Gamma'_1 \\[8pt]
\Gamma_1 & J_1 - \dfrac{1}{3}\Gamma_1 & \dfrac{\sqrt{6}}{2}\Gamma'_1 \\[8pt]
\dfrac{\sqrt{6}}{2}\Gamma'_1 & \dfrac{\sqrt{6}}{2}\Gamma'_1 & J_1 + K_1
\end{pmatrix},
\label{Eq-S53}
\end{equation}

The appearance of $\Gamma_1$ in the diagonal components stems from the fact that the Heisenberg parameters receive contributions not only from $\widetilde{S}_i^x \widetilde{S}_j^x + \widetilde{S}_i^y \widetilde{S}_j^y + \widetilde{S}_i^z \widetilde{S}_j^z$ in the local frame, but also from off-diagonal terms introduced through our bond-dependent coordinate transformation, as shown in Eq.~(\ref{Eq-S28}). The off-diagonal component $\Gamma'$ vanishes, as proved in Sec. ~\ref{Subsubsec-S3.3.4} (i.e., its corresponding coefficient is zero).

For the 2NNs, the exchange tensor can be expressed as
\begin{equation}
\widetilde{M}_2 = 
\begin{pmatrix}
J_2 - \dfrac{1}{3}\Gamma_2 & \Gamma_2 + D_2 & \dfrac{\sqrt{6}}{2}\Gamma'_2 \\[8pt]
\Gamma_2 - D_2 & J_2 - \dfrac{1}{3}\Gamma_2 & \dfrac{\sqrt{6}}{2}\Gamma'_2 \\[8pt]
\dfrac{\sqrt{6}}{2}\Gamma'_2 & \dfrac{\sqrt{6}}{2}\Gamma'_2 & J_2 + K_2
\end{pmatrix},
\label{Eq-S54}
\end{equation}
where $D_2$ denotes the DM interaction parameter, and its sign depends on the convention adopted in matrices under global coordinates. Furthermore, since no secondary off-diagonal interactions are present for either 1NN or 2NN pairs, $\Gamma'_1 = \Gamma'_2 = 0$.

Since these exchange tensors are defined in the local coordinate system, the exchange tensors for equivalent neighbor bonds are identical in the designated local frames. Upon mapping to the global coordinate frame, they transform according to the specific bond directions to yield the corresponding global Hamiltonians.

\newpage
\section{DFT calculation scheme and validation}\label{Sec-S3}
A major obstacle to generalizing the Kitaev interaction to broader lattice structures is the lack of a concise and universal density functional theory (DFT) scheme capable of disentangling complex magnetic interactions within a lattice. To address this, our group employs an ``Energy Mapping + MAE'' approach~[\textcolor{green}{21,22,23}]. By calculating the angle-dependent MAE for various collinear magnetic orders, we obtain MAE curves that isolate contributions solely from Kitaev(-like) interactions, off-diagonal interactions, and single-ion anisotropy (SIA). Since the functional forms of the MAE curves and the corresponding interaction expressions differ across magnetic configurations, considering a sufficient number of collinear magnetic orders allows for the separation of distinct interaction parameters. Below, we present the detailed formalism of this method applied to the two-dimensional edge-sharing tetrahedral structure, along with the MAE results for $\mathrm{FeX}$ ($X=\mathrm{Te}, \mathrm{Se}$).

\subsection{Energy mapping + MAE method}\label{Subsec-S3.1}
In this method, the first step is to select appropriate collinear magnetic orders based on the specific lattice structure. For the two-dimensional monolayer FeX system, we adopt three commonly studied magnetic configurations, as illustrated in Fig.~\ref{FIG-s7}: ferromagnetic (FM) order, N\'{e}el-type antiferromagnetic order (commonly abbreviated as GAFM in square lattices), and bicollinear antiferromagnetic (BAFM) order. The computed MAE includes both in-plane and out-of-plane components, expressed as:
\begin{equation}
H_{MAE}^{in} = H_{MAE}^{\theta=90^\circ, \varphi} - H_{MAE}^{\theta=90^\circ, \varphi=0^\circ}, \quad 
H_{MAE}^{out} = H_{MAE}^{\theta, \varphi=0} - H_{MAE}^{\theta=90^\circ, \varphi=0^\circ},
\label{Eq-S55}
\end{equation}
where the in-plane term corresponds to the MAE relative to the $[100]$ direction as a function of the in-plane angle $\varphi$, and the out-of-plane term corresponds to the MAE relative to the $[100]$ direction as a function of the out-of-plane angle $\theta$. The definitions of $\theta$ and $\varphi$ are provided in Fig.~\ref{FIG-s7}.

Consequently, in our angle-dependent MAE calculations, the contribution from the isotropic Heisenberg interaction remains constant and thus does not appear explicitly. Furthermore, since the magnetic orders considered are strictly collinear (parallel or antiparallel), the Dzyaloshinskii--Moriya (DM) interaction vanishes identically. Therefore, the MAE for different magnetic orders contains only the pure contributions from Kitaev-like interactions, off-diagonal interactions, and SIA. This can be expressed as:
\begin{equation}
H_{MAE} = H_K + H_\Gamma + H_{SIA},
\label{Eq-S56}
\end{equation}
where the bond-dependent Kitaev-like interaction is collectively denoted as $H_{Kit} = H_K + H_\Gamma$ in subsequent discussions. To derive the MAE expressions for different magnetic configurations, we represent the spin directions in spherical coordinates as:
\begin{equation}
\mathbf{S} = S_0 (\sin\theta \cos\varphi \mathbf{i} + \sin\theta \sin\varphi \mathbf{j} + \cos\theta \mathbf{k}),
\label{Eq-S57}
\end{equation}
where the opposite magnetic moment direction undergoes the following transformation: $\theta \to (\pi - \theta)$, $\varphi \to (\varphi + \pi)$.

According to the Hamiltonian, the SIA energy acts solely along the direction perpendicular to the plane and is independent of the spin arrangement; thus, its contribution to the MAE is identical across all magnetic orders and can be written as:
\begin{equation}
H_{SIA}^{FM} = H_{SIA}^{GAFM} = H_{SIA}^{BAFM} = A_k \cos^2\theta.
\label{Eq-S58}
\end{equation}

In contrast, the bond-dependent Kitaev-like interaction varies with the magnetic configuration and exhibits angular dependence in both the in-plane and out-of-plane directions. For the FM order, this contribution can be written as (with analogous expressions for other magnetic orders):
\begin{equation}
H_{Kit}^{FM}(\theta, \varphi) = H_K^{FM}(\theta, \varphi) + H_\Gamma^{FM}(\theta, \varphi),
\end{equation}
where the expression above represents the FM case; analogous expressions apply to other magnetic configurations. Consequently, the final expression for the total MAE (taking FM as an example) becomes:
\begin{equation}
H_{MAE}^{FM}(\theta, \varphi) = H_{Kit}^{FM}(\theta, \varphi) + H_{SIA}(\theta),
\end{equation}
where the forms for other magnetic configurations are similar. Thus, by analyzing the MAE results across different magnetic orders, we can first extract the SIA contribution and subsequently determine the parameters for the bond-dependent Kitaev-like interactions by incorporating multiple magnetic configurations.

Therefore, a simple criterion to determine whether anisotropic magnetic interactions beyond SIA exist in the system is to calculate the out-of-plane MAE for both FM and G-AFM orders. If the trends of the MAE curves differ between these two orders, interactions beyond SIA must be present. Conversely, if the trends are identical, further calculations involving additional magnetic orders are required for confirmation.

\subsection{Contributions of Kitaev-like and Off-diagonal Interactions to the MAE}\label{Subsec-S3.2}
The effective spin Hamiltonian for the Kitaev-like and off-diagonal interactions in two-dimensional monolayer FeX is given by Eq.~(\ref{Eq-S51}), while the bond directions between different atoms are listed in Table~\ref{Table-s1}. We can therefore expand the bond-dependent Hamiltonian into a form expressed in terms of the distinct bonds:
\begin{equation}
\begin{aligned}
H_{Kit} &= \sum_{|m|=|n|=1} \frac{1}{2} \left[ K_1 S_{00}^{w(p+2)-(p+3)} S_{mn}^{w(p+2)-(p+3)} + \Gamma_1 (S_{00}^{\bar{w}_p} S_{mn}^{w_{p+1}} + S_{00}^{\bar{w}_{p+1}} S_{mn}^{w_p}) \right] \\
&\quad + \sum_{|m| \text{ or } |n|=2} \frac{1}{2} \left[ K_2 S_{00}^{w(p+1)-(p+3)} S_{mn}^{w(p+1)-(p+3)} + \Gamma_2 S_{00}^{\bar{w}_p} S_{mn}^{\bar{w}_{p+2}} \right],
\end{aligned}
\label{Eq-S61}
\end{equation}
where the spin subscript $00$ denotes the spin of the central atom under consideration, and the subscript $mn$ denotes the spin of the atom located at $\mathbf{R} = m\mathbf{i} + n\mathbf{j}$ relative to the central atom. Different values of $mn$ correspond to different bond labels $p$, with the specific correspondence detailed in Table~\ref{Table-s1}.

Based on the above bond-dependent Hamiltonian, we can derive the respective contributions of the Kitaev-like and off-diagonal interactions to the MAE under different magnetic orders, as presented below.

For the 1NN, the angular dependence of the Kitaev-like interaction under different magnetic orders is:
\begin{equation}
\begin{aligned}
H_{K1}^{FM} &= \frac{1}{2} K_1 S_0^2 (1 + \cos^2\theta), \\
H_{K1}^{GAFM} &= -\frac{1}{2} K_1 S_0^2 (1 + \cos^2\theta), \\
H_{K1}^{BAFM} &= \sqrt{2} K_1 S_0^2 \sin\theta \cos\theta \cos\varphi.
\end{aligned}
\label{Eq-S62}
\end{equation}

The corresponding result for the 1NN off-diagonal interaction is:
\begin{equation}
\begin{aligned}
H_{\Gamma1}^{FM} &= \frac{4}{3} \Gamma_1 S_0^2 \cos^2\theta, \\
H_{\Gamma1}^{GAFM} &= -\frac{4}{3} \Gamma_1 S_0^2 \cos^2\theta, \\
H_{\Gamma1}^{BAFM} &= -\frac{4\sqrt{2}}{3} \Gamma_1 S_0^2 \sin\theta \cos\theta \cos\varphi.
\end{aligned}
\label{Eq-S63}
\end{equation}

For the 2NN, the Kitaev-like interaction result is:
\begin{equation}
\begin{aligned}
H_{K2}^{FM} &= K_2 S_0^2 (1 - \cos^2\theta), \\
H_{K2}^{GAFM} &= K_2 S_0^2 (1 - \cos^2\theta), \\
H_{K2}^{BAFM} &= K_2 S_0^2 (1 - \cos^2\theta)(1 - 2\sin^2\theta).
\end{aligned}
\label{Eq-S64}
\end{equation}

The 2NN off-diagonal interaction result is:
\begin{equation}
\begin{aligned}
H_{\Gamma2}^{FM} &= \frac{4}{3} \Gamma_2 S_0^2 (2\cos^2\theta - 1), \\
H_{\Gamma2}^{GAFM} &= \frac{4}{3} \Gamma_2 S_0^2 (2\cos^2\theta - 1), \\
H_{\Gamma2}^{BAFM} &= \frac{2}{3} \Gamma_2 S_0^2 (1 - \cos^2\theta)(2\cos^2\varphi - 1).
\end{aligned}
\label{Eq-S65}
\end{equation}

The above expressions describe the bond-dependent Kitaev-like interactions for spins oriented at arbitrary angles in space. However, our goal is to obtain the energy variation relative to the $[100]$ axis. We therefore process these results separately to obtain the out-of-plane and in-plane components. For the out-of-plane component, we set $\varphi = 0^\circ$ and subtract the result at $\theta = 90^\circ$. For the in-plane component, we set $\theta = 90^\circ$ and subtract the result at $\varphi = 0^\circ$. After applying these operations and summing the in-plane and out-of-plane expressions of the individual interaction terms, we obtain the total in-plane and out-of-plane MAE expressions.

The out-of-plane MAE is given by:
\begin{equation}
\begin{aligned}
H_{MAEout}^{FM} &= (\frac{1}{2}K_1 + \frac{4}{3}\Gamma_1 - K_2 + \frac{8}{3}\Gamma_2 + A_k) S_0^2 \cos^2\theta, \\
H_{MAEout}^{GAFM} &= (-\frac{1}{2}K_1 - \frac{4}{3}\Gamma_1 - K_2 + \frac{8}{3}\Gamma_2 + A_k) S_0^2 \cos^2\theta, \\
H_{MAEout}^{BAFM} &= (\sqrt{2}K_1 - \frac{4\sqrt{2}}{3}\Gamma_1) S_0^2 \sin\theta \cos\theta + (-K_2 - \frac{4}{3}\Gamma_2 + A_k) S_0^2 \cos^2\theta.
\end{aligned}
\label{Eq-S66}
\end{equation}

The in-plane MAE is given by:
\begin{equation}
\begin{aligned}
H_{MAEin}^{FM} &= 0, \\
H_{MAEin}^{GAFM} &= 0, \\
H_{MAEin}^{BAFM} &= (-2K_2 - \frac{8}{3}\Gamma_2) S_0^2 \sin^2\varphi.
\end{aligned}
\label{Eq-S67}
\end{equation}

\begin{figure}[!ht]
\centering
\includegraphics[width=0.7\columnwidth, clip]{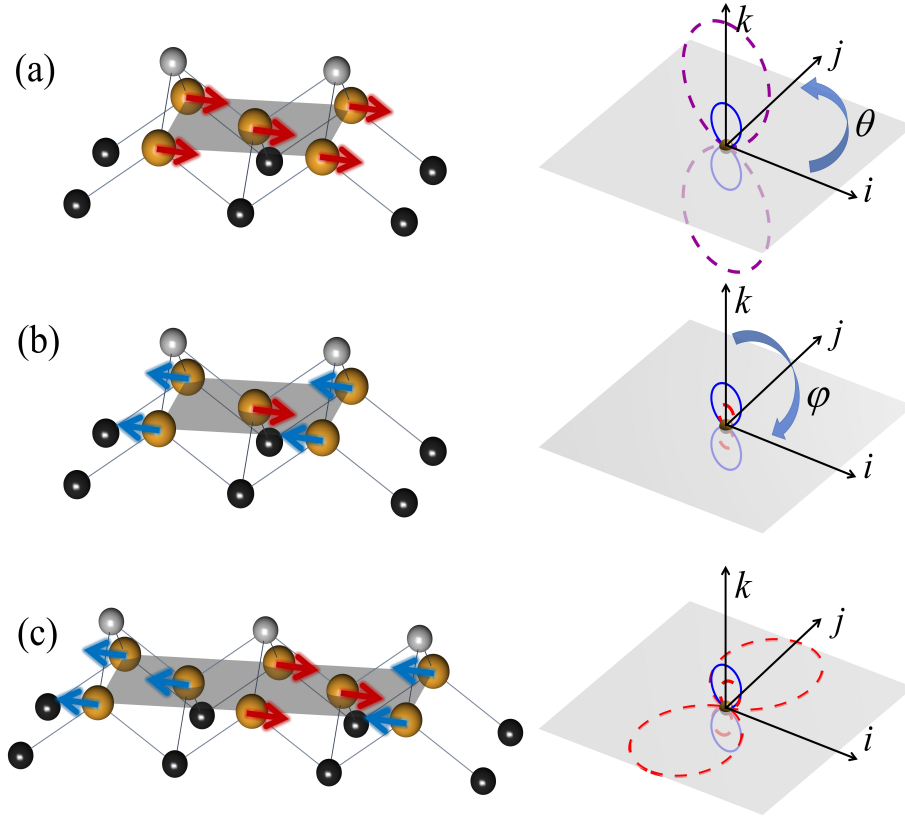}\\
\caption{Magnetic orders and corresponding anisotropy landscapes in monolayer FeX Schematic of Possible Magnetic Orders in Monolayer FeX and SIA($\theta$), MAE($\theta/\varphi$). (a) Ferromagnetic order (FM), (b) Neel antiferromagnetic order (N\'{e}el AFM), (c) Bilinear antiferromagnetic order (BAFM). Red and blue arrows indicate opposite magnetic moment directions. The right side shows the polar plots of SIA($\theta$) and MAE($\theta/\varphi$) for each magnetic order (all parameters set to 1), divided into components within the Fe plane (gray plane) and perpendicular to the Fe layer (out-of-plane). The in-plane component uses $\mathbf{i}$ as the polar axis, and the out-of-plane component uses $\mathbf{k}$ as the polar axis; distance from the origin represents the magnitude of the effect in that direction. Blue solid lines represent SIA (positive values), while purple and red dashed lines represent positive and negative values of MAE($\theta/\varphi$), respectively.}
\label{FIG-s7}
\end{figure}

Based on the above in-plane and out-of-plane MAE expressions, we first set all interaction parameters to unity for a preliminary analysis, with the results shown in Fig~\ref{FIG-s7}.

Our analytical derivation reveals a critical decoupling mechanism: while the SIA energy is determined solely by the local ligand field and thus remains robust against magnetic ordering, the bond-dependent Kitaev-like energy fluctuates significantly with the magnetic structure. We visualize this contrast in Fig.~\ref{FIG-s7} by plotting the theoretical angular dependece for FM, N\'{e}el AFM, and BAFM orders (with all parameters set to 1). As shown by the blue curves, SIA is identical across different magnetic orders and only has an out-of-plane component. However, the total MAE exhibits drastic variations driven by bond-dependent terms: they amplify the anisotropy in the FM state [Fig.~\ref{FIG-s7}(a)], effectively cancel the SIA in the GAFM state [Fig.~\ref{FIG-s7}(b)], and induce a complex in-plane anisotropy in the BAFM [Fig.~\ref{FIG-s7}(c)]. The differences in MAE among these three magnetic configurations intuitively and concisely express the existence of Kitaev-like interactions beyond SIA in the model.

In our Kitaev-like bond-dependent model, the MAE consists of the 1NN Kitaev-like term and off-diagonal term, the 2NN Kitaev-like term and off-diagonal term, and SIA. For specific FeX systems, we can obtain the parameters for Kitaev-like and SIA interactions by fitting the theoretically derived formulas to DFT calculation results using the least squares method. Substituting these parameters into the formulas to generate MAE curves, self-consistency with calculation results would verify the correctness of the model.

\subsection{DFT calculation results and analysis}\label{Subsec-S3.3]}
After theoretically verifying that our Kitaev-like bond-dependent model, combined with the Energy Mapping + MAE scheme, can successfully disentangle the individual interaction parameters, we performed the corresponding DFT calculations. These include the electronic structure of $\mathrm{FeX}$ and the MAE under different magnetic orders. Some of these results have already been presented in Fig.~\textcolor{red}{3} of the main text and will not be repeated here. In this section, we describe our computational methods, analyze the electronic structure, and present the results obtained when the secondary off-diagonal interaction $\Gamma'$ is taken into account.
\subsubsection{DFT Computational Methods}\label{Subsubsec-S3.3.1}
We employed first-principles calculations under the generalized Bloch condition using the Vienna Ab initio Simulation Package (VASP)~[\textcolor{green}{36}] based on DFT pseudopotential plane-wave method. The Perdew-Burke-Ernzerhof (PBE) functional~[\textcolor{green}{37}] was used. VASP solves the Kohn-Sham equations self-consistently using plane-wave basis functions and calculates forces and tensors via wavefunctions~[\textcolor{green}{38}]. To study monolayer $\mathrm{FeTe}$ and $\mathrm{FeSe}$ properties effectively, we included a vacuum layer of $20\text{ \AA}$ to avoid periodic interactions. The force convergence criterion was set to $5\times10^{-3}\text{ eV/\AA}$, and the plane-wave cutoff energy was chosen as $400\text{ eV}$. The self-consistent electronic step convergence criterion was $1.0\times10^{-6}\text{ eV}$. Calculations used the $\text{GGA}+U$ method~[\textcolor{green}{39}] with an effective on-site Coulomb interaction parameter $U = 0.5\text{ eV}$. To ensure convergence, $k$-point meshes in the reciprocal space of the primitive cell were set to $18\times18\times1$ for $\mathrm{FeTe}$ and $\mathrm{FeSe}$. For $2\times1$ supercells, $k$-points were $9\times18\times1$, and for $\sqrt{2}\times\sqrt{2}$ supercells, they were $12\times12\times1$. Our calculations also included SOC as implemented by Hobbs et al. in VASP~[\textcolor{green}{40}].
\subsubsection{Electronic Structure of FeTe}\label{Subsubsec-S3.3.2}
\begin{figure}[!ht]
\centering
\includegraphics[width=0.95\columnwidth, clip]{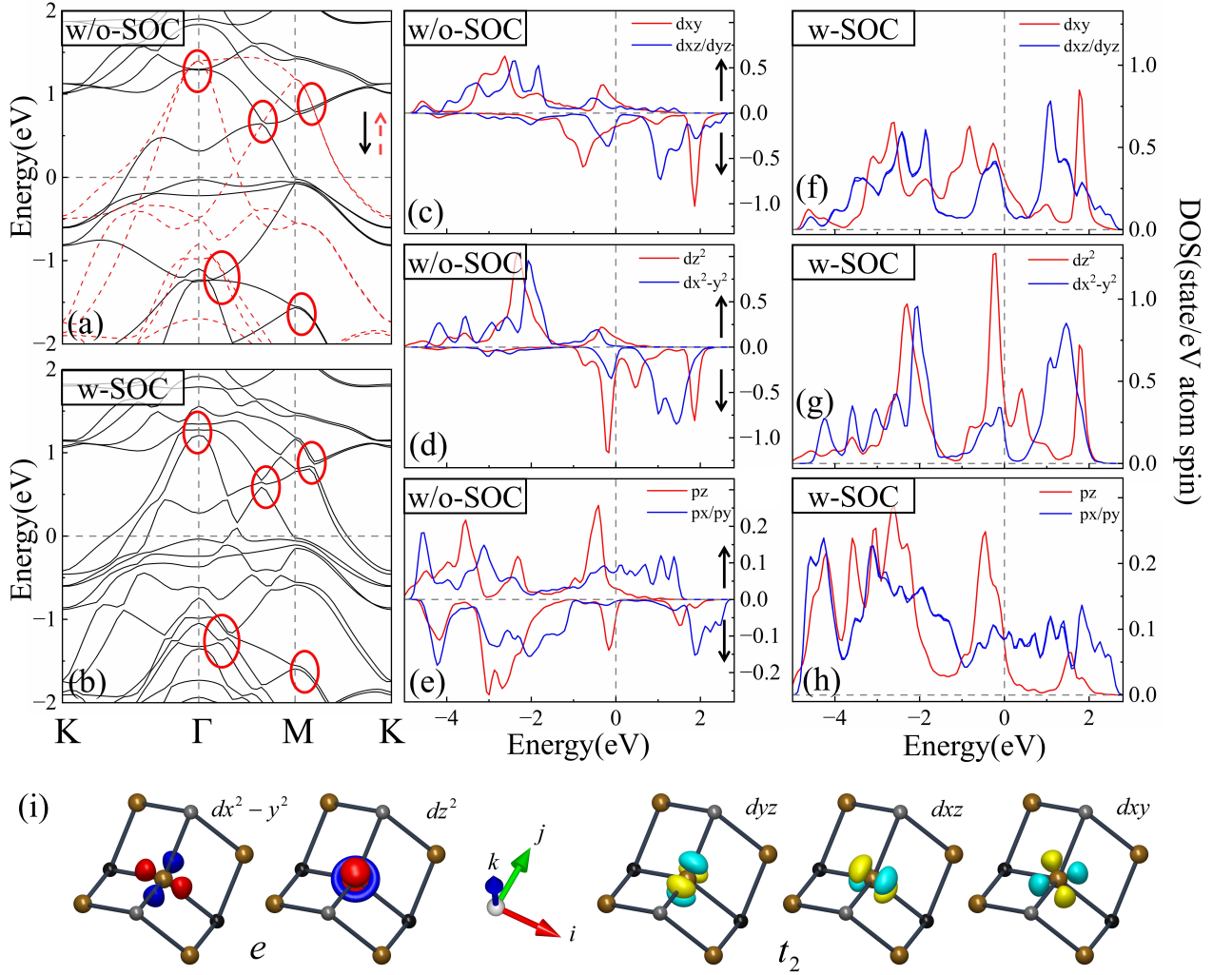}\\
\caption{Electronic structure of monolayer $\mathrm{FeTe}$ with and without SOC, along with schematic illustrations of orbital interactions. (a) and (b) show the band structures without and with SOC, respectively. The red ellipses indicate the lifting of band degeneracies induced by SOC. (c)-(e) display the projected density of states (PDOS) onto $\mathrm{Fe}$ $3d$ orbitals and $\mathrm{Te}$ $5p$ orbitals in the absence of SOC. (f)-(h) present the corresponding PDOS when SOC is included. (i) Schematic of the five-fold degenerate $3d$ orbitals of the $\mathrm{Fe}$ atom.}
\label{FIG-s8}
\end{figure}

The significance of SOC is primarily manifested in the electronic structure. As illustrated in Fig.~\ref{FIG-s8}, in the ferromagnetic ground state, the Fermi level crosses electronic states in both cases with and without SOC, indicating metallic behavior. Upon inclusion of SOC, band crossings that appear in the absence of SOC are observed to open gaps (see red ellipses in Figs.~\ref{FIG-s8}(a) and (b)), demonstrating that SOC plays a crucial role in this system. This observation provides the physical foundation for the emergence of Kitaev-like interactions.

In the 2D tetrahedral coordination environment, the $\mathrm{Fe}$ atom is situated at the center of the crystal field formed by the surrounding $\mathrm{Te}$ atoms. The five degenerate $d$ orbitals split into a higher-energy triply degenerate $t_2$ manifold and a lower-energy doubly degenerate $e$ manifold (see Fig.~\ref{FIG-s8}(i)). The density of states shown in Figs.~\ref{FIG-s8}(c)-(h) reveals that, overall, the $e$ orbitals exhibit relatively localized character, whereas the $t_2$ orbitals are more spatially extended, indicating that the $t_2$ states predominantly mediate the interatomic bonding interactions. However, owing to the two-dimensional nature of $\mathrm{FeTe}$, the $t_{2g}$ and $e$ states are not perfectly degenerate.

The main peak of $\mathrm{Fe}$ $d_{xy}$ is located well below the Fermi level, reflecting partial direct interactions between first nearest-neighbor $\mathrm{Fe}$ atoms. The SOC-induced effects, however, are primarily mediated through indirect exchange interactions via the $\mathrm{Te}$ atoms. The $\mathrm{Fe}$ $d_{xz}$ and $d_{yz}$ orbitals, together with the $\mathrm{Te}$ $p_x$ and $p_y$ states, exhibit relatively extended distributions near and below the Fermi level, characterizing the indirect $\mathrm{Fe}\text{-}\mathrm{Te}\text{-}\mathrm{Fe}$ superexchange pathways. Specifically, the $d_{xz}$ orbitals of $\mathrm{Fe}_0$ and the $d_{yz}$ orbitals of $\mathrm{Fe}_1$ mediate the first nearest-neighbor Kitaev-like interactions through the $\mathrm{Te}$ $p$ orbitals, while the $d_{xz}$ orbitals of $\mathrm{Fe}_0$ and the $d_{xz}$ orbitals of $\mathrm{Fe}_2$ mediate the second nearest-neighbor Kitaev-like interactions via the same pathway (see Fig.~\textcolor{red}{2}(a)). The presence of these bond-dependent Kitaev-like interactions is further corroborated by first-principles DFT calculations.
\subsubsection{Electronic Structure of FeSe}\label{Subsubsec-S3.3.3}
\begin{figure}[!ht]
\centering
\includegraphics[width=0.95\columnwidth, clip]{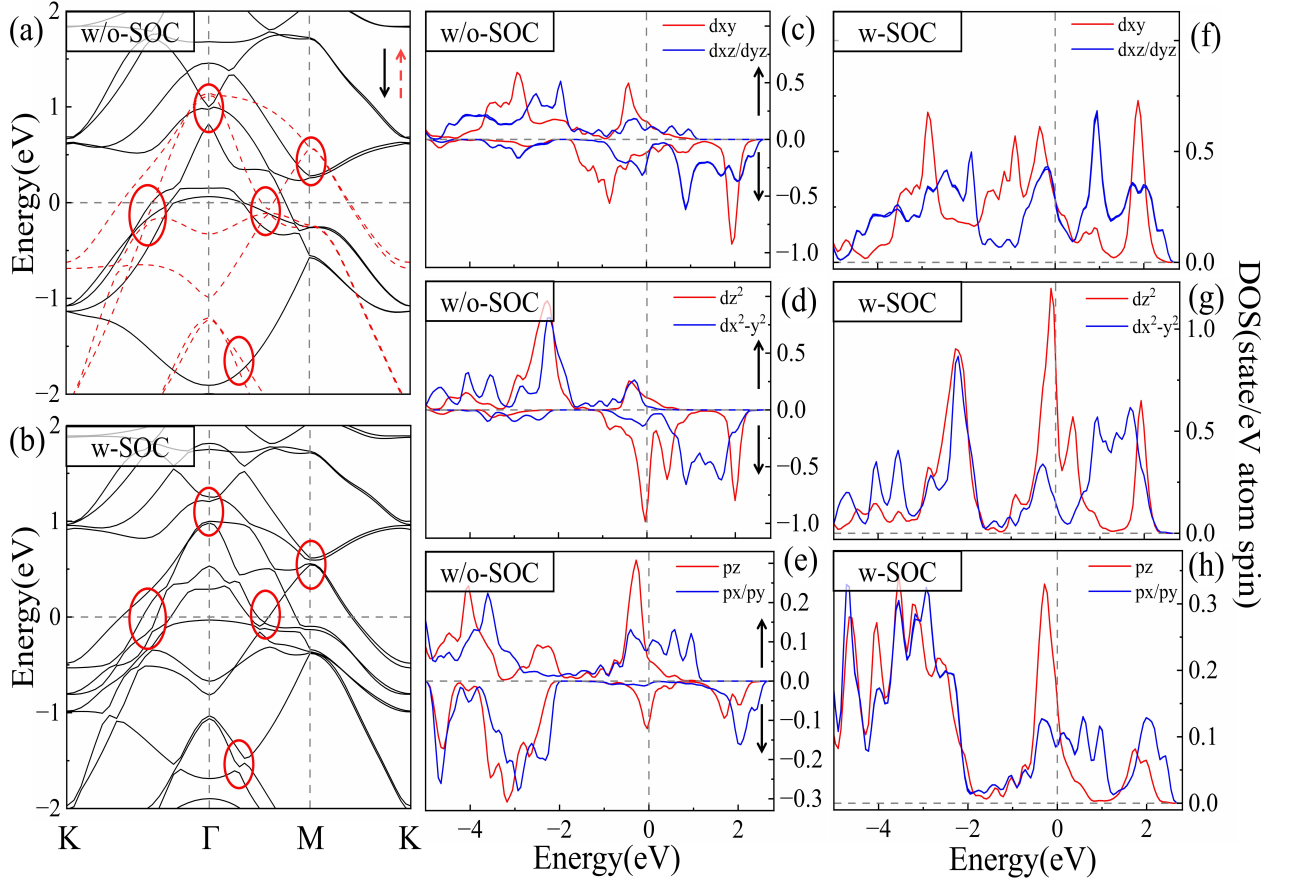}\\
\caption{Electronic structure of monolayer $\mathrm{FeSe}$ with and without SOC, along with schematic illustrations of orbital interactions. (a) and (b) show the band structures without and with SOC, respectively. The red ellipses indicate the lifting of band degeneracies induced by SOC. (c)-(e) display the PDOS onto $\mathrm{Fe}$ $3d$ orbitals and $\mathrm{Se}$ $4p$ orbitals in the absence of SOC. (f)-(h) present the corresponding PDOS when SOC is included.}
\label{FIG-s9}
\end{figure}
The electronic structure of monolayer $\mathrm{FeSe}$ exhibits similar characteristics to that of monolayer $\mathrm{FeTe}$. Upon inclusion of SOC, band gaps emerge at the crossing points that appear in the band structure calculated without SOC. The $\mathrm{Fe}$ atom is situated at the center of the tetrahedral crystal field formed by the surrounding $\mathrm{Se}$ atoms. Under the influence of this coordination field, the Kitaev-like interactions in $\mathrm{FeSe}$ are also mediated through the $p$ orbitals of the $\mathrm{Se}$ atoms.
\subsubsection{Discussion of the Off-diagonal Interaction $\Gamma'$}\label{Subsubsec-S3.3.4}
In studies of bond-dependent interactions, an additional off-diagonal interaction, commonly denoted by $\Gamma'$, often appears. Similar to $\Gamma$, it generally arises from lattice distortions in octahedrally coordinated systems~[\textcolor{green}{11,23,27}]. The distinction between $\Gamma$ and $\Gamma'$ lies in the bond components involved: $\Gamma$ describes the coupling between the $\alpha$ and $\beta$ components, whereas $\Gamma'$ describes the coupling between the $\alpha$ and $\gamma$ components or between the $\beta$ and $\gamma$ components. If the relevant symmetry is absent, a further parameter $\Gamma''$ may be introduced to distinguish these contributions. In the two-dimensional edge-sharing tetrahedral system, the $\alpha$ and $\beta$ bonds are symmetry-equivalent in the ideal structure. We therefore introduce a single parameter $\Gamma'$ to describe these interactions. The corresponding Hamiltonian can be written as
\begin{equation}
H_{\Gamma'} = \sum_{\langle i,j \rangle} \frac{1}{2} \Gamma'_{ij} \bigl( S_i^\alpha S_j^\gamma + S_i^\gamma S_j^\alpha + S_i^\beta S_j^\gamma + S_i^\gamma S_j^\beta \bigr),
\label{Eq-S68}
\end{equation}

We next examine whether the off-diagonal interaction $\Gamma'$ exists in $\mathrm{FeX}$ in the absence of lattice distortions, using both perturbation theory and DFT calculations. According to Eq.~(\ref{Eq-S27}), the effective spin interaction associated with $\Gamma'$ in the local coordinate frame takes the form
\begin{equation}
\begin{aligned}
S_i^\alpha S_j^\gamma + S_i^\gamma S_j^\alpha &= \cos\theta\,(\widetilde{S}_i^x \widetilde{S}_j^z + \widetilde{S}_i^z \widetilde{S}_j^x) - \sin\theta\,(\widetilde{S}_i^y \widetilde{S}_j^z + \widetilde{S}_i^z \widetilde{S}_j^y), \\
S_i^\beta S_j^\gamma + S_i^\gamma S_j^\beta &= -\sin\theta\,(\widetilde{S}_i^x \widetilde{S}_j^z + \widetilde{S}_i^z \widetilde{S}_j^x) + \cos\theta\,(\widetilde{S}_i^y \widetilde{S}_j^z + \widetilde{S}_i^z \widetilde{S}_j^y), \\
S_i^\alpha S_j^\gamma + S_i^\gamma S_j^\alpha + S_i^\beta S_j^\gamma + S_i^\gamma S_j^\beta &= \frac{\sqrt{6}}{3}\,(\widetilde{S}_i^x \widetilde{S}_j^z + \widetilde{S}_i^z \widetilde{S}_j^x + \widetilde{S}_i^y \widetilde{S}_j^z + \widetilde{S}_i^z \widetilde{S}_j^y).
\end{aligned}
\label{Eq-S69}
\end{equation}

For this off-diagonal interaction, the $\alpha,\gamma$ and $\beta,\gamma$ components cannot be separated and have identical coefficients. We therefore describe them using a single parameter $\Gamma'$. The corresponding cross terms between spin components do not arise in the theoretical derivation presented in Sec.~\ref{Sec-S2}. We therefore conclude that, in the absence of lattice distortions, the $\Gamma'$ interaction is absent from the model considered here.

We further verify this conclusion using DFT calculations. Starting from the spin Hamiltonian in Eq.~(\ref{Eq-S68}), one can derive the MAE resulting from a finite $\Gamma'$ for different magnetic configurations:
\begin{equation}
\begin{aligned}
H_{\Gamma'_1}^{FM} = H_{\Gamma'_1}^{GAFM} &= 0, &
H_{\Gamma'_1}^{BAFM} &= -\frac{2\sqrt{3}}{3}\Gamma'_1 S_0^2\sin\theta\cos\theta\sin\varphi, \\
H_{\Gamma'_2}^{FM} = H_{\Gamma'_2}^{GAFM} &= 0, &
H_{\Gamma'_2}^{BAFM} &= \frac{4\sqrt{6}}{3}\Gamma'_2 S_0^2\sin^2\theta\sin\varphi\cos\varphi.
\end{aligned}
\label{Eq-S70}
\end{equation}

The expression shows that the contribution of $\Gamma'$ vanishes for both the 1NN and 2NN interactions in the FM and GAFM configurations. This cancellation results from symmetry: the contributions of the $\Gamma'$ interactions on different bonds exactly cancel in the total MAE. In the BAFM configuration, however, $\Gamma'$ gives a finite contribution to the MAE because the magnetic order lowers the corresponding symmetry.

\begin{figure}[!ht]
\centering
\includegraphics[width=0.5\columnwidth, clip]{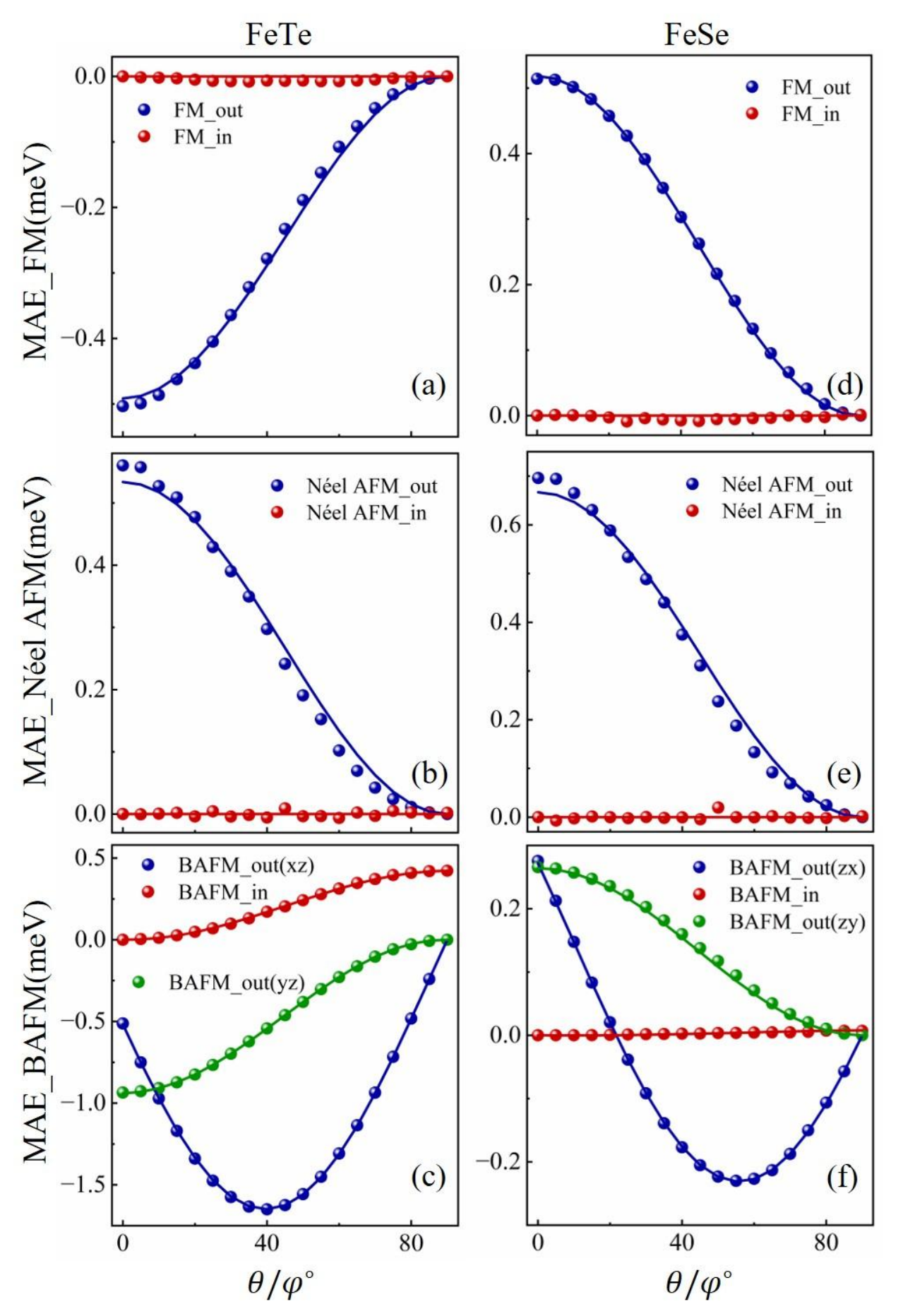 }\\
\caption{Calculated and fitted MAE curves for monolayer $\mathrm{FeTe}$ and $\mathrm{FeSe}$ based on the Kitaev-like model. The discrete points denote the calculated values, with one data point obtained every $5^\circ$. The solid curves are obtained by substituting the fitted interaction parameters into the analytical expressions. The blue, red, and green curves correspond to the out-of-plane rotation in the $zx$ plane ($\varphi = 0^\circ$), the in-plane rotation in the $xy$ plane ($\theta = 90^\circ$), and the out-of-plane rotation in the $zy$ plane ($\varphi = 90^\circ$), respectively, for the different magnetic orders. Panels (a)-(c) show the results for $\mathrm{FeTe}$, while panels (d)-(e) show the results for $\mathrm{FeSe}$.}
\label{FIG-s10}
\end{figure}

The extraction of the 1NN parameter $\Gamma'_1$ from the MAE therefore requires an additional calculation. As seen from Eq.~(\ref{Eq-S70}), for the originally defined in-plane rotation, $[100]\to[010]$ with $\theta=90^\circ$, and the out-of-plane rotation, $[001]\to[010]$ with $\varphi=90^\circ$, the BAFM MAE remains angle-independent and is therefore insensitive to $\Gamma'_1$. To determine its contribution, we consider an additional out-of-plane rotation, $[001]\to[010]$, corresponding to $\varphi=90^\circ$. In this geometry, the out-of-plane MAE contains a contribution from $\Gamma'_1$ and is given by
\begin{equation}
E_{\mathrm{MAEper2}}^{BAFM} = -\frac{2}{\sqrt{3}}\Gamma'_1 S_0^2\sin\theta\cos\theta + \bigl(K_2 + \tfrac{2}{3}\Gamma_2 + A_k\bigr)S_0^2\cos\theta^2.
\end{equation}
Similarly, the in-plane MAE contains a contribution from $\Gamma'_2$ and should be modified as
\begin{equation}
E_{\mathrm{MAEin}}^{BAFM} = \bigl(-2K_2 - \tfrac{4}{3}\Gamma_2\bigr)S_0^2\sin^2\varphi + \frac{4\sqrt{6}}{3}\Gamma'_2 S_0^2\sin\varphi\cos\varphi.
\end{equation}
Refitting the calculated MAE data gives the interaction parameters for $\mathrm{FeTe}$ and $\mathrm{FeSe}$ listed in Table~\ref{Table-s2}. All values are given in meV/Fe, and the results are compared with those obtained previously.

\begin{table}[htbp]
  \centering
  \caption{Interaction parameters of $\mathrm{FeTe}$ and $\mathrm{FeSe}$ obtained after including the off-diagonal interaction $\Gamma'$.}
  \label{Table-s2}
   \begin{ruledtabular}
 \label{Table-s2}
    \begin{tabular}{ c c c c c c c c }
      Name & $K_1$ & $\Gamma_1$ & $\Gamma_1'$ & $K_2$ & $\Gamma_2$ & $\Gamma_2'$ & $A_k$ \\
      \hline
      FeTe & -1.6296 & 0.2266 & -0.0136 & -0.3907 & 0.1346 & -0.0030 & -0.7286 \\
      FeTe(Previous) & -1.6282 & 0.2261 & -- & -0.3901 & 0.1352 & -- & -0.7295 \\
      \hline
      \hline
      FeSe & -0.3712 & 0.0836 & -0.0090 & -0.1106 & 0.0801 & -0.0006 & 0.2679 \\
      FeSe(Previous) & -0.3717 & 0.0837 & -- & -0.1096 & 0.0799 & -- & 0.2699 \\
    \end{tabular}
     \end{ruledtabular}
\end{table}

Here, ``Previous'' denotes the results obtained from the original fitting procedure without including $\Gamma'$. The comparison shows that, even after incorporating the out-of-plane $zy$-plane MAE of the BAFM configuration, the extracted interaction parameters remain nearly unchanged, with relative variations below 1\%. Furthermore, both $\Gamma'_1$ and $\Gamma'_2$ are nearly zero in $\mathrm{FeTe}$ and $\mathrm{FeSe}$ and are much smaller than the other interaction parameters. We therefore conclude that the contribution of the $\Gamma'$ interaction in monolayer $\mathrm{FeTe}$ and $\mathrm{FeSe}$ is negligible and may be safely omitted.

To further substantiate this conclusion, we compare the calculated MAE with the analytical expressions evaluated using the fitted parameters, as shown in Fig.~\ref{FIG-s10}.

The fitted curves agree very well with the DFT results for all interaction parameters. We also performed fits using the parameters obtained previously without including $\Gamma'$. The resulting curves are nearly indistinguishable from those obtained after including $\Gamma'$, and are therefore not shown. This indicates that including $\Gamma'$ leads to overfitting and further supports the validity of our model, as well as the conclusion that any $\Gamma'$ interaction in this structure is extremely weak. Combined with the perturbative derivation, which shows that the spin-component terms associated with $\Gamma'$ do not arise in the two-dimensional edge-sharing tetrahedral structure, these results demonstrate that the ideal $\mathrm{FeX}$ structure does not support a $\Gamma'$ off-diagonal interaction in the absence of lattice distortions.

\newpage
\section{Studies of the Dzyaloshinskii-Moriya interaction}\label{Sec-S4}
In Sec.~\ref{Sec-S2}, perturbation theory shows that the Dzyaloshinskii--Moriya (DM) interaction arises between 2NN atoms due to broken inversion symmetry. However, because the Energy Mapping + MAE scheme employs only collinear magnetic configurations, the DM interaction parameter cannot be extracted directly within that framework. To address this limitation, we designed two specific noncollinear magnetic configurations and performed DFT calculations to determine the DM interaction parameters in two-dimensional monolayer $\text{Fe}X$ ($X=\text{Te}$, Se). We also discuss the influence of the DM interaction on the ground-state magnetic ordering.
\subsection{Extraction of the DM interaction parameter}\label{Subsec-S4.1}
\begin{figure}[!ht]
\centering
\includegraphics[width=0.85\columnwidth, clip]{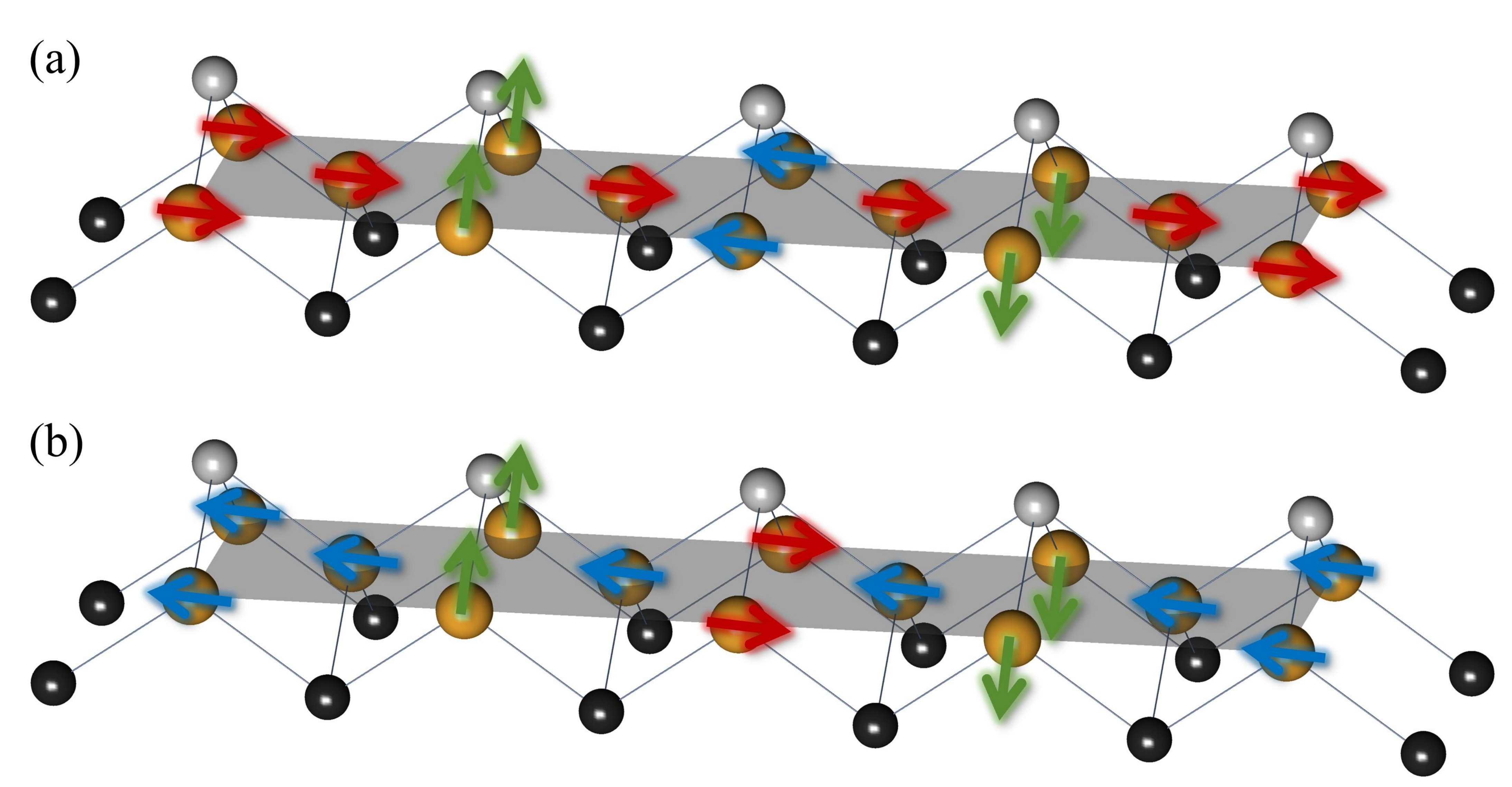 }\\
\caption{Magnetic configurations designed for verifying the DM interaction. (a) and (b) show the two respective configurations. The arrows indicate the orientations of the magnetic moments at each site: red arrows denote the (1,0,0) direction, blue arrows denote the ($-1$,0,0) direction, and green arrows denote the (0,0,1) and (0,0,$-1$) directions, respectively.}
\label{FIG-s11}
\end{figure}

Multiple magnetic interactions coexist in $\text{Fe}X$, including the Heisenberg interaction, biquadratic Heisenberg exchange, single-ion anisotropy, and Kitaev-like and off-diagonal interactions. To extract the pure DM interaction, the optimal approach is to construct two magnetic configurations that alter exclusively the DM interaction while keeping the contributions of all other magnetic interactions identical between the two. The energy difference between these configurations thus isolates the pure DM contribution. The magnetic configurations designed for this purpose are illustrated in Fig.~\ref{FIG-s11}.

The two configurations shown in Fig.~\ref{FIG-s11} involve noncollinear spin arrangements. Through analytical derivation, we verify that the Heisenberg interaction vanishes in both configurations, while the biquadratic exchange and the SIA contribute equally per atomic site in the two structures. Furthermore, the total energies of the Kitaev-like and off-diagonal interactions are identical in both configurations, and the contributions from atoms with spins along (0,0,1) and (0,0,$-1$) cancel each other exactly between the two configurations. Consequently, the energy difference contains only the pure DM contribution, which can be written as
\begin{equation}
    E_{b-a} = 8D_2,
    \label{Eq-S73}
\end{equation}

From our DFT calculations, the energy difference between configurations (b) and (a) in Fig.~\ref{FIG-s11} is found to be $E_{b-a} = -60.493\,\text{meV}$ for FeTe, yielding a DM interaction coefficient of $D_2 = -7.56\,\text{meV/Fe}$. For FeSe, the energy difference is $E_{b-a} = -39.833\,\text{meV}$, corresponding to a DM interaction coefficient of $D_2 = -4.98\,\text{meV/Fe}$. These values indicate that the DM interaction is relatively robust. Below, we discuss in detail its effects on the spin configuration.
\subsection{Ground-state spin configuration under the DM interaction}\label{Subsec-S4.2}
In the perturbative derivation of Sec.~\ref{Sec-S2}, the effective spin Hamiltonian for the 2NN DM interaction is given by
\begin{equation}
    H_{DMI} = \frac{1}{2} \mathbf{D}_{ij} \cdot (\mathbf{S}_i \times \mathbf{S}_j),
    \label{Eq-S74}
\end{equation}
where $\mathbf{D}_{ij} = D_2(\mathbf{z} \times \mathbf{d}_{ij})$, $D_2$ denotes the magnitude of the 2NN DM interaction parameter, $\mathbf{z}$ is the unit vector pointing from the Te ligand toward the Fe layer, and $\mathbf{d}_{ij}$ is the unit vector from $\text{Fe}_i$ to $\text{Fe}_j$. Thus, the DM interaction depends on the ligand position relative to the Fe plane, the relative Fe-Fe geometry, and the spin configuration.

\begin{figure}[!ht]
\centering
\includegraphics[width=0.85\columnwidth, clip]{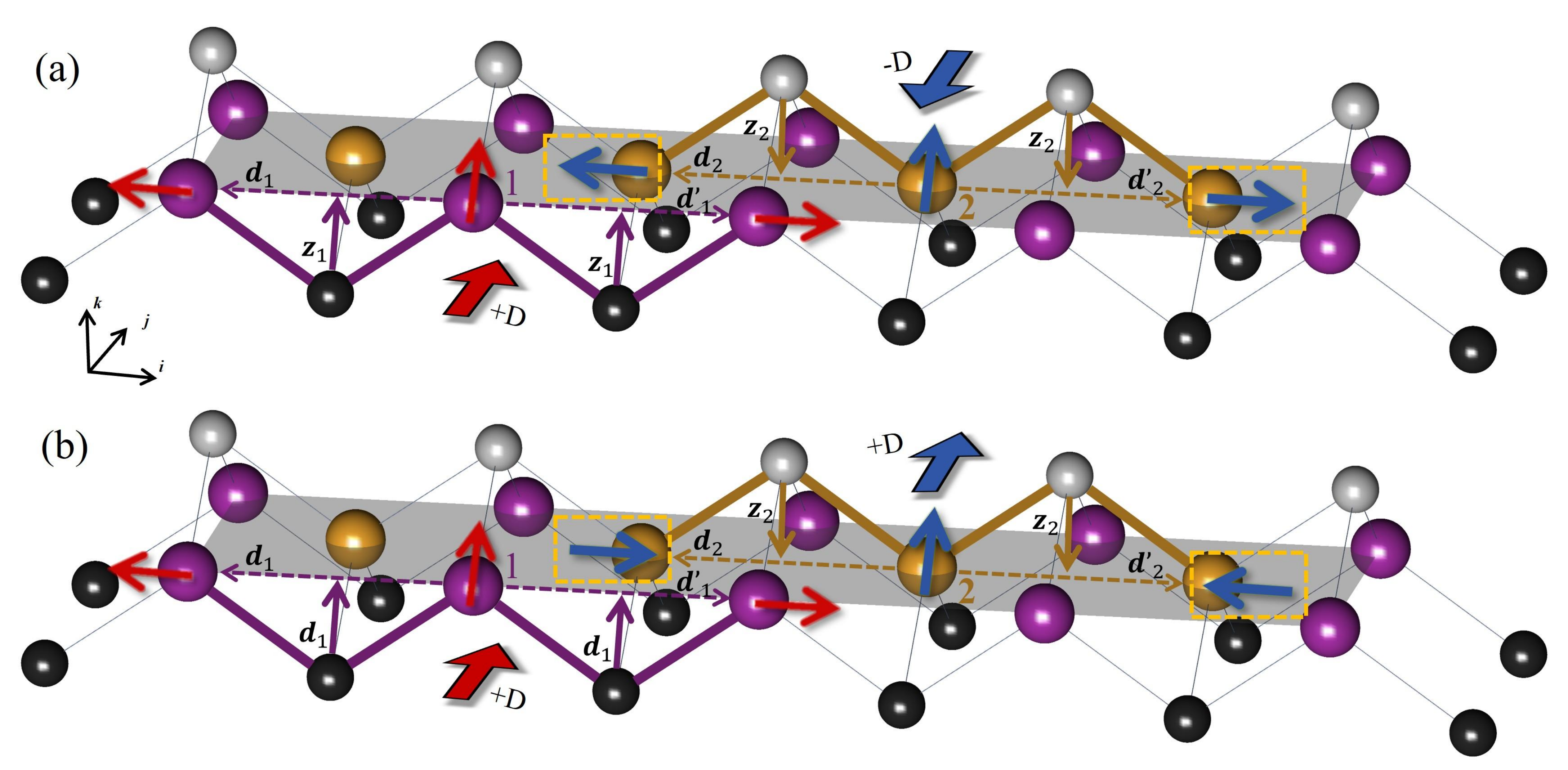 }\\
\caption{DM interaction representations within different sublattices. Purple and brown spheres with arrows denote the two sublattices of Fe atoms and their corresponding parameters, respectively; gray and black spheres represent the upper and lower Te atoms, respectively. Red and blue arrows indicate the spin directions and the signs of the DM interaction for different sublattices. (a) When both sublattices adopt the same spin arrangement (both clockwise or both counterclockwise), the local DM interaction contributions are opposite, causing cancellation globally, so that only local DM interaction survives. (b) When the two sublattices adopt opposite spin arrangements, the DM interaction contributions add constructively both locally and globally.}
\label{FIG-s12}
\end{figure}

In monolayer FeTe the DM interaction exists only between 2NNs and therefore acts independently on the two sublattices (the 1NNs within each sublattice coincide with the 2NNs of the full lattice), as illustrated in Fig.~\ref{FIG-s12}.

\begin{figure}[!ht]
\centering
\includegraphics[width=0.85\columnwidth, clip]{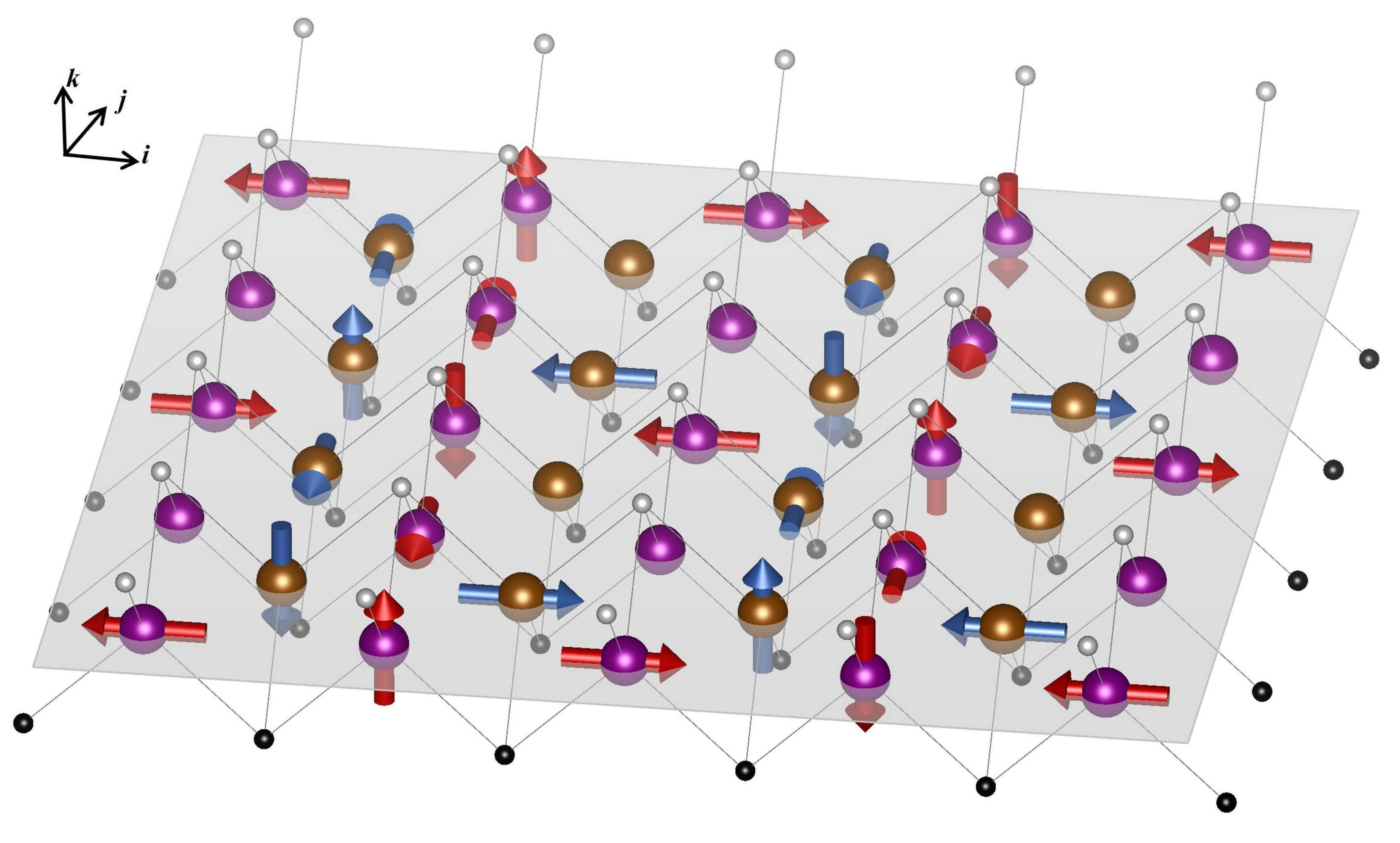 }\\
\caption{Single Fe-atom chain arrangements along the $i$ and $j$ axes under pure DM interaction ($D < 0$). Red and blue arrows correspond to the spin arrangements of the purple and brown sublattices, respectively. Along the $j$ direction, the spins are oriented along the $k$ directions; atoms without spin labels represent sites where any spin configuration yields zero DM interaction energy.}
\label{FIG-s13}
\end{figure}

To define the spin-rotation direction, we adopt the following convention: if atoms are arranged along the $i$-axis, the perspective is taken along the positive $j$-axis direction; if arranged along the $j$-axis, the perspective is along the positive $i$-axis direction. The physics of Fig.~\ref{FIG-s12} can then be interpreted as follows. When both sublattices follow the same spin sequence (both clockwise or both counterclockwise), the DM interaction appears locally within each sublattice but cancels globally, yielding no net DM interaction, as shown in Fig.~\ref{FIG-s12}(a). Conversely, when they follow opposite sequences (one clockwise, one counterclockwise), the DM interaction is present both locally and globally, as shown in Fig.~\ref{FIG-s12}(b). The above describes the DM interaction rule for Fe-Fe chains along the $i$ direction. Assuming the arrangement in Fig.~\ref{FIG-s12}(b), i.e., the purple sublattice is arranged clockwise and the brown sublattice counterclockwise, then along the $j$ direction the spins should be arranged counterclockwise and clockwise, respectively, to manifest global DM interaction, as depicted in Fig.~\ref{FIG-s13}.

In the two-dimensional configuration of Fig.~\ref{FIG-s13}, we identify the lowest-energy arrangement for a single-atom cross-chain along the $i$ and $j$ axes under pure DM interaction (with $D < 0$; the DFT extraction scheme and results are presented below). For the purple sublattice, the spin arrangements along the positive $i$ and $j$ axes are both clockwise; for the brown sublattice, both are counterclockwise. We now evaluate the total energy. In this configuration, the spins along the $k$-axis contribute $1/4$ of the total, with a per-site DM interaction energy of $2D$; the spins along the $i$ or $j$ axes contribute $1/2$, with a per-site contribution of $D$; and the remaining $1/4$ of sites yield zero DM interaction regardless of spin arrangement. Thus, the average DM interaction energy per atom is $D$, which is not the globally lowest-energy configuration.

\begin{figure}[!ht]
\centering
\includegraphics[width=0.85\columnwidth, clip]{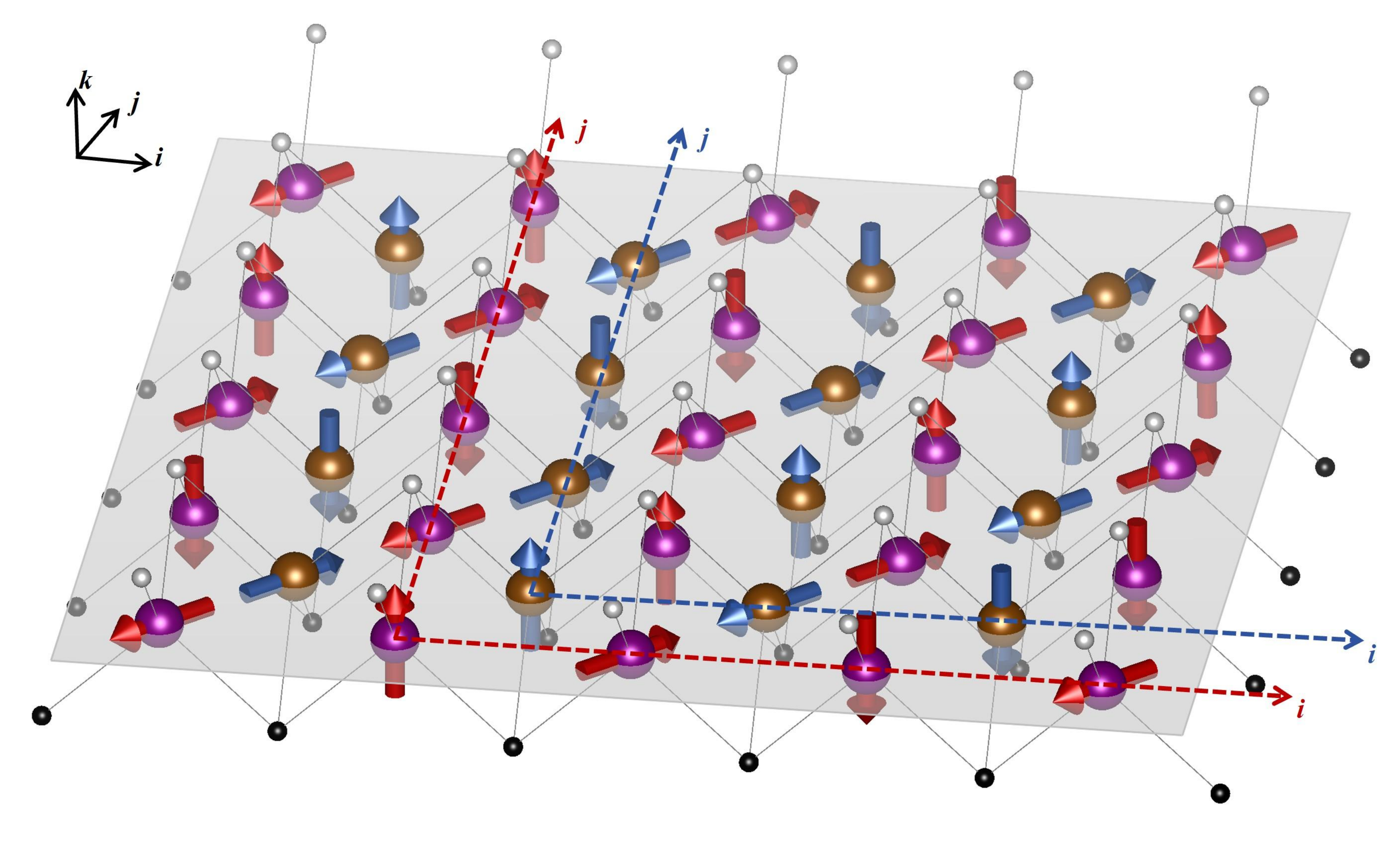 }\\
\caption{Ground-state spin arrangement of monolayer $\text{Fe}X$ ($X=\text{Te}$, Se) under pure DMI. The spin directions are restricted to $(0,0,1)$, $(0,0,-1)$, $(1,1,0)$ and $(1,-1,0)$.}
\label{FIG-s14}
\end{figure}

Next, we determine the ground state by employing the spin-spiral method. We assume the spin at site $(m, n)$ forms a spiral wave and represent it as a unit vector:
\begin{equation}
    \mathbf{S}_{mn} = (A \sin(q_x m + q_y n), B \sin(q_x m + q_y n), C \cos(q_x m + q_y n)),
    \label{Eq-S75}
\end{equation}
where $\mathbf{q} = (q_x, q_y)$ denotes the wavevector components, $\mathbf{R} = m\mathbf{i} + n\mathbf{j}$ is the lattice position, and the parameters satisfy $A^2 + B^2 = C = 1$. In this framework, we represent the interaction between neighboring spins in a bond-dependent form. Only two types of 2NN bonds exist: those oriented along the $i$ direction are designated as x-bonds, and those along the $j$ direction as y-bonds. The energies of the x and y bonds can therefore be expressed as
\begin{equation}
    \begin{aligned}
        E_x =& D \{C \cos(q_x m + q_y n) A \sin[q_x(m + 1) + q_y n] \\
        &- A \sin(q_x m + q_y n) C \cos[q_x(m + 1) + q_y n]\} = DA \sin(q_x), \\
        E_y =& D \{B \sin(q_x m + q_y n) C \cos[q_x m + q_y(n + 1)] \\
        &- C \cos(q_x m + q_y n) B \sin[q_x m + q_y(n + 1)]\} = DB \sin(-q_y),
    \end{aligned}
    \label{Eq-S76}
\end{equation}
where $E_x$ and $E_y$ depend only on $q_x$ and $q_y$, not on the absolute atomic positions. Summing over all bonds yields the total energy
\begin{equation}
    E_{\text{Bond}} = E_x + E_y = D[A \sin(q_x) - B \sin(q_y)],
    \label{Eq-S77}
\end{equation}
Minimization of $E$ with respect to $q_x$ and $q_y$ gives the lowest-energy state when $A = B = \frac{1}{\sqrt{2}}, q_x = \frac{\pi}{2}, q_y = -\frac{\pi}{2}$, which is one of several degenerate solutions. The corresponding spin configuration for the purple sublattice is
\begin{equation}
    \mathbf{S}_{mn} = \left( \frac{1}{\sqrt{2}} \sin \frac{\pi}{2}(m - n), \frac{1}{\sqrt{2}} \sin \frac{\pi}{2}(m - n), \cos \frac{\pi}{2}(m - n) \right),
    \label{Eq-S78}
\end{equation}
For the brown sublattice, the spin configuration is obtained by reversing the chirality (e.g., changing the sign of the $A$ or $B$ component), yielding the ground-state arrangement shown in Fig.~\ref{FIG-s14}.

As shown in Fig.~\ref{FIG-s14}, the spin arrangement along the Fe-Fe chains in the $i$ and $j$ directions corresponds to a rotation within the $(1\bar{1}0)$ plane. At this configuration, the average DMI energy per site is $\sqrt{2}D$, which is the global minimum for pure DM interaction. Conversely, if the spiral propagates along the $(1\bar{1}0)$ direction, the energy is equivalent and also minimal. So it selects the chirality and propagation direction of the spin rotation.

In quadratically symmetric lattices, multiple symmetry-equivalent spiral wavevectors exist. Under the combined effects of external magnetic fields, magnetic anisotropy, strain, or higher-order exchange interactions, these single-q spiral states may compete or superpose, potentially forming multi-q noncollinear magnetic structures such as square skyrmions, antiskyrmions, or bimerons~[\textcolor{green}{54,55}]. However, the stability of such structures and their corresponding phase diagrams require further verification through Monte Carlo simulations, spin dynamics, or energy-barrier calculations based on the complete spin Hamiltonian.~[\textcolor{green}{56,57}] Consequently, monolayer $\text{Fe}X$ is expected to serve as a promising platform for studying DM-interaction-driven two-dimensional noncollinear magnetism, as well as the external-field or strain-controlled topological spin textures.

\newpage
\section{Orbital superexchange paths in the global coordinate frame}\label{Sec-S5}
In the preceding analysis, the orbital components were defined in the local coordinate frame. To elucidate the overall nature of the bond-dependent Kitaev-like interaction, we now analyze the corresponding superexchange pathways in the global coordinate frame. For this purpose, we also provide the Slater--Koster expressions in the global frame to facilitate the analysis of the hopping integrals between different orbitals. Since the symbol $t$ was already used for the hopping integrals in the local coordinate frame, we denote by $S$ the hopping integrals containing both $V_{pd\sigma}$ and $V_{pd\pi}$ contributions, and by $P$ those containing only the $V_{pd\pi}$ contribution. The hopping matrix from the Fe $d$ orbitals to the Te $p$ orbitals is then written as
\begin{equation}
\mathbf{T}_{X_1 M_0} = 
\begin{pmatrix}
0 & P_3 & 0 \\
0 & -P_4 & 0 \\
S_3 & 0 & S_1 \\
S_1 & 0 & S_2 \\
P_1 & 0 & -P_2
\end{pmatrix}, \quad
\mathbf{T}_{X_2 M_0}  = 
\begin{pmatrix}
P_3 & 0 & 0 \\
0 & -S_3 & S_1 \\
P_4 & 0 & 0 \\
0 & -S_1 & S_2 \\
0 & P_1 & P_2
\end{pmatrix},\quad
\mathbf{T}_{X_1 M_2} = 
\begin{pmatrix}
0 & -P_3 & 0 \\
0 & -P_4 & 0 \\
S_3 & 0 & -S_1 \\
-S_1 & 0 & S_2 \\
-P_1 & 0 & -P_2
\end{pmatrix}.
\label{Eq-S79}
\end{equation}

The individual hopping components are given by
\begin{equation}
\begin{aligned}
S_1 &= \frac{\sqrt{2}}{3} V_{pd\sigma} &+ \frac{\sqrt{6}}{9} V_{pd\pi}, \\
S_2 &= -\frac{1}{3} V_{pd\sigma} &+ \frac{2\sqrt{3}}{9} V_{pd\pi}, \\
S_3 &= -\frac{2}{3} V_{pd\sigma} &+ \frac{\sqrt{3}}{9} V_{pd\pi}, \\
P_1 &= &-\frac{\sqrt{2}}{3} V_{pd\pi}, \\
P_2 &= &+\frac{2}{3} V_{pd\pi}, \\
P_3 &= &+\frac{\sqrt{6}}{3} V_{pd\pi}, \\
P_4 &= &+\frac{\sqrt{3}}{3} V_{pd\pi},
\end{aligned}
\label{Eq-S80}
\end{equation}
where $V_{pd\sigma}$ and $V_{pd\pi}$ have the same meanings as in the local coordinate frame. The orbital bases in the global coordinate frame are defined as $(d_{xy}, d_{yz}, d_{xz}, d_{x^2-y^2}, d_{3z^2-r^2})$ and $(p_x, p_y, p_z)$.

The exchange pathways of the $d_{xz}$ and $d_{yz}$ orbitals along the $x$ and $y$ directions are symmetry-related. Likewise, the exchange pathways involving the $d_{x^2-y^2}$ and $d_{z^2}$ orbitals have the same structure and differ only in their hopping amplitudes. We therefore present only representative superexchange processes below. The global-frame Slater--Koster hopping matrix further shows that the $\sigma$-bond channel appears only in the hopping pathways involving the $d_{xz}$, $d_{yz}$, and $d_{z^2}$ orbitals. Accordingly, Fig.~\textcolor{red}{2} of the main text focuses on the superexchange processes involving these three orbitals. We next briefly discuss several representative processes in the complete superexchange pathways. The $\sigma$- and $\pi$-bond channels will not be distinguished separately here; the corresponding hopping integrals can be obtained from Eq.~(\ref{Eq-S80}).
\begin{figure}[!ht]
\centering
\includegraphics[width=0.85\columnwidth, clip]{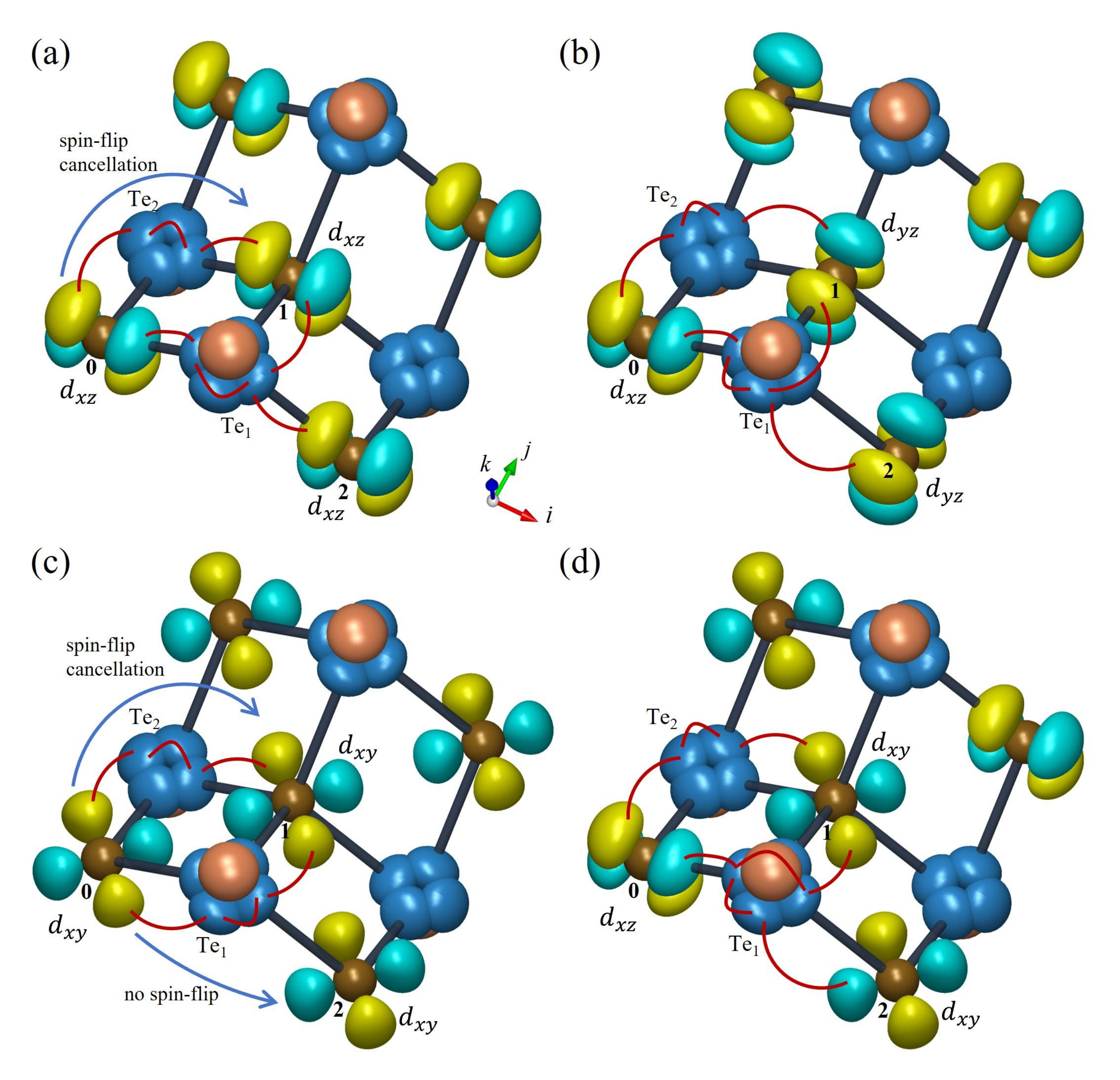 }\\
\caption{Representative spin-flip superexchange pathways between $t_2$ orbitals. The labels $0$, $1$ and $2$ denote the central Fe atom, 1NN Fe atom, and 2NN Fe atom, respectively. The paths connecting different atoms represent hopping processes, whereas the paths involving the same Te atom represent transformations between Te $p$ orbitals induced by SOC. In the schematic representation of the Te orbitals, the blue regions denote hybridized $p_x/p_z$ or $p_y/p_z$ orbitals, while the brown regions denote $p_z$ orbitals. (a) $d_{xz}$-$d_{xz}$. (b) $d_{xz}$-$d_{yz}$. (c) $d_{xy}$-$d_{xy}$. (d) $d_{xz}$-$d_{xy}$.}
\label{FIG-s15}
\end{figure}

As shown in Fig.~\ref{FIG-s15}(a), the $d_{xz}$-$d_{xz}$ channel contains spin-flip processes in the superexchange pathway. For the 1NN pair, a hole hops from the $\text{Fe}_0$ $d_{xz}$ orbital to the hybridized $\text{Te}_1$ $p_x/p_z$ orbital, is projected onto the $\text{Te}_1$ $p_x$ orbital through SOC, and subsequently hops to the Fe $d_{xz}$ orbital of the neighboring Fe atom. Among these pathways, only the $d_{xz}$-$p_z$-$p_x$-$d_{xz}$ path contributes to spin-flip hopping. The corresponding process through $\text{Te}_2$ is analogous, giving the spin-flip path $d_{xz}$-$p_x$-$p_z$-$d_{xz}$. Because the pathways through $\text{Te}_1$ and $\text{Te}_2$ are symmetry-related, their hopping integrals are identical. However, the SOC matrix elements associated with the two spin-flip processes are $\langle p_z | H_{SOC} | p_x \rangle$ and $\langle p_x | H_{SOC} | p_z \rangle$, respectively, and their contributions cancel each other. Thus, although spin-flip pathways are present for the 1NN pair, they do not produce a net bond-dependent Kitaev-like interaction. For the 2NN pair, this cancellation does not occur because the relevant symmetry is broken and the virtual hopping proceeds through only one coordinating atom. The corresponding pathways are $d_{xz}$-$p_x$-$p_z$-$d_{xz}$ and $d_{xz}$-$p_z$-$p_x$-$d_{xz}$. Although these pathways appear analogous to those discussed above, their hopping integrals are different, as specified in Eq.~(\ref{Eq-S80}). Consequently, the two pathways yield the same net contribution and contribute equally to the spin-flip exchange.

The $d_{xz} \text{-} d_{yz}$ and $d_{xz} \text{-} d_{yz}$ pathways shown in Figs.~\ref{FIG-s15}(b) and (d), respectively, correspond to conventional spin-flip processes and will not be discussed further here. The Slater--Koster hopping integrals in Eq.~(\ref{Eq-S80}) can be used to identify the specific spin-flip superexchange pathways.

For the $d_{xy} \text{-} d_{xy}$ channel shown in Fig.~\ref{FIG-s15}(c), the 1NN pathway is analogous to the $d_{xz} \text{-} d_{xz}$ case: spin-flip pathways exist, but their contributions cancel in the total exchange process. For the 2NN pair, no spin-flip hopping pathway exists, and this channel therefore does not contribute to the Kitaev-like interaction.
\begin{figure}[!ht]
\centering
\includegraphics[width=0.85\columnwidth, clip]{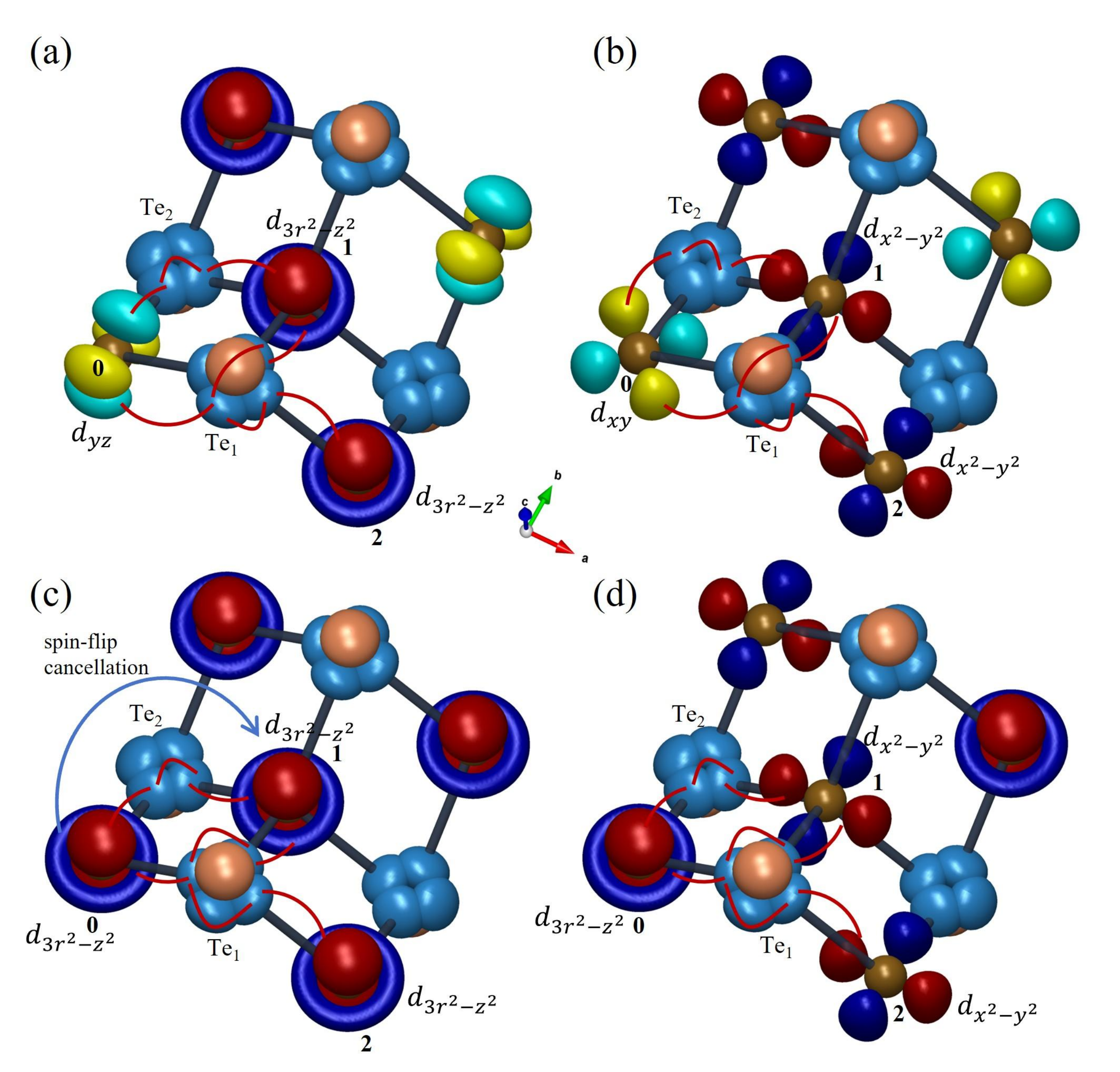 }\\
\caption{Representative spin-flip superexchange pathways involving the $e$ orbitals. The labels $0$, $1$, and $2$ denote the central Fe atom, 1NN Fe atom, and 2NN Fe atom, respectively. The paths connecting different atoms represent hopping processes, whereas the paths involving the same Te atom represent transformations between Te $p$ orbitals induced by SOC. In the schematic representation of the Te orbitals, the blue regions denote hybridized $p_x/p_z$ or $p_y/p_z$ orbitals, while the brown regions denote $p_z$ orbitals. (a) $d_{yz} - d_{z^2}$. (b) $d_{xy} - d_{x^2-y^2}$. (c) $d_{z^2} - d_{z^2}$. (d) $d_{z^2} - d_{x^2-y^2}$.}
\label{FIG-s16}
\end{figure}

Fig.~\ref{FIG-s16} shows representative superexchange pathways involving the $e$ orbitals. Similar to the $t_2 \text{-} t_2$ channels, spin-flip pathways exist between the same orbitals for 1NN pairs, but their contributions cancel in the total exchange, as illustrated in Fig.~\ref{FIG-s16}(c). By contrast, interorbital exchange pathways generally contain nonvanishing spin-flip contributions.

\end{document}